\documentclass[article,a4paper,nojss]{jss}
\usepackage[]{graphicx}
\usepackage[]{color}
\makeatletter
\def\maxwidth{ %
  \ifdim\Gin@nat@width>\linewidth
    \linewidth
  \else
    \Gin@nat@width
  \fi
}
\makeatother

\usepackage{amssymb,amsmath}
\usepackage{bm}
\usepackage{bbm}
\usepackage[margin=1in]{geometry}
\usepackage{setspace}
\usepackage{multirow}
\usepackage{array}
\usepackage[normalem]{ulem}
\usepackage{amsthm}
\usepackage[printwatermark]{xwatermark}
\usepackage{xcolor}
\usepackage{lipsum}

\DeclareSymbolFont{matha}{OML}{txmi}{m}{it}
\DeclareMathSymbol{\varv}{\mathord}{matha}{118}

\renewcommand{\tt} {\texttt}

\newcommand{\zerob} {{\bf 0}}

\newcommand{\expect} {\E}

\newcommand{\thetab} {{\boldsymbol{\theta}}}
\newcommand{\alphab} {{\boldsymbol{\alpha}}}

\newcommand{\nub} {{\boldsymbol{\nub}}}

\newcommand{\deltab} {{\boldsymbol{\delta}}}
\newcommand{\xib} {{\boldsymbol{\xi}}}

\newcommand{\varthetab} {{\boldsymbol{\vartheta}}}

\newcommand{\intd} {\textrm{d}}
\newcommand{\phib} {\boldsymbol{\phi}}

\newcommand{\etab} {\boldsymbol{\eta}}

\newcommand{\Sigmamat} {{\bm \Sigma}}

\newcommand{\Omegab} {{\bm \Omega}}

\newcommand{\Lambdamat} {\mathbf{\Lambda}}

\newcommand{\Pimat} {{\bm \Pi}}
\newcommand{\Amat} {\textbf{A}}

\newcommand{\Dmat} {\textbf{D}}

\newcommand{\Smat} {\textbf{S}}
\newcommand{\Tmat} {\textbf{T}}

\newcommand{\Cmat} {\mathbf{C}}

\newcommand{\Kmat} {\textbf{K}}

\newcommand{\Zmat} {\textbf{Z}}

\newcommand{\Imat} {\textbf{I}}

\newcommand{\Vmat} {\textbf{V}}

\newcommand{\hvec} {\textbf{h}}

\newcommand{\svec} {\textbf{s}}
\newcommand{\tvec} {\textbf{t}}

\newcommand{\muvec} {\boldsymbol{\mu}}

\newcommand{\upsilonb} {\boldsymbol {\upsilon}}

\newcommand{\NS} {\textit{NS}}
\newcommand{\RS} {\textit{RS}}
\newcommand{\RMSPE} {\textit{RMSPE}}
\newcommand{\SNR} {\textit{SNR}}
\newcommand{\FRK} {\textit{FRK}}
\newcommand{\SPDE} {\textit{SPDE}}
\newcommand{\gstat} {\textit{gstat}}
\newcommand{\LTK} {\textit{LTK}}

\renewcommand{\zerob}{\mathbf{0}}

\newcommand{\Yvec}{\mathbf{Y}}
\newcommand{\Wvec}{\mathbf{W}}

\newcommand{\Zvec}{\mathbf{Z}}
\newcommand{\epsilonb}{\boldsymbol{\epsilon}}

\newcommand{\cov}{\mathrm{\COV}}

\newcommand{\var}{\mathrm{\VAR}}

\newcommand{\tr}{\mathrm{tr}}
\newcommand{\diag}{\mathrm{diag}}

\newcommand{\Gau}{\mathrm{Gau}}

\DeclareMathOperator*{\argmax}{arg\,max}

\author{Andrew Zammit-Mangion\\University of Wollongong \And
        Noel Cressie\\University of Wollongong}
\title{\pkg{FRK}: An \proglang{R} Package for Spatial and Spatio-Temporal Prediction with Large Datasets}
\Plainauthor{Andrew Zammit-Mangion, Noel Cressie} 
\Plaintitle{FRK: An R package for spatial and spatio-temporal prediction with large datasets} 
\Shorttitle{\pkg{FRK}: An \proglang{R} package for spatial and spatio-temporal prediction with large datasets} 
\Abstract{
  \pkg{FRK} is an \proglang{R} software package for spatial/spatio-temporal modelling and prediction with large datasets. It facilitates optimal spatial prediction (kriging) on the most commonly used manifolds (in Euclidean space and on the surface of the sphere), for both spatial and spatio-temporal fields. It differs from many of the packages for spatial modelling and prediction by avoiding stationary and isotropic covariance and variogram models, instead constructing a spatial random effects (SRE) model on a fine-resolution discretised spatial domain. The discrete element is known as a basic areal unit (BAU), whose introduction in the software leads to several practical advantages. The software can be used to (i)  integrate multiple observations with different supports with relative ease; (ii) obtain exact predictions at millions of prediction locations (without conditional simulation); and (iii) distinguish between measurement error and fine-scale variation at the resolution of the BAU, thereby allowing for reliable uncertainty quantification. The temporal component is included by adding another dimension. A key component of the SRE model is the specification of spatial or spatio-temporal basis functions; in the package, they can be generated automatically or by the user. The package also offers automatic BAU construction, an expectation-maximisation (EM) algorithm for parameter estimation, and functionality for prediction over any user-specified polygons or BAUs. Use of the package is illustrated on several spatial and spatio-temporal datasets, and its predictions and the model it implements are extensively compared to others commonly used for spatial prediction and modelling.

}
\Plainkeywords{Fixed rank kriging, basic areal units, EM algorithm, spatial prediction, spatial random effects model} 

\Address{
  Andrew Zammit-Mangion\\
  National Institute for Applied Statistics Research Australia (NIASRA)\\
  School of Mathematics and Applied Statistics\\
  University of Wollongong\\
  Wollongong, Australia\\
  E-mail: \email{azm@uow.edu.au}\\
  URL: \url{https://andrewzm.wordpress.com}
}

\IfFileExists{upquote.sty}{\usepackage{upquote}}{}
\begin{document}


\section{Introduction}\label{sec:intro}

Fixed rank kriging (FRK) is a spatial/spatio-temporal modelling and prediction framework that is scaleable, works well with large datasets, and can deal easily with data that have different spatial supports. FRK hinges on the use of a spatial random effects (SRE) model, in which a spatially correlated mean-zero random process is decomposed using a linear combination of spatial basis functions with random coefficients plus a term that captures the random process' fine-scale variation. Dimensionality reduction through a relatively small number of basis functions ensures computationally efficient prediction, while the reconstructed spatial process is, in general, non-stationary.  The SRE model has a spatial covariance function that is always nonnegative-definite and, because any (possibly non-orthogonal) basis functions can be used, it can be constructed so as to approximate standard families of covariance functions \citep{Kang_2011}.  For a detailed treatment of FRK, see \cite{Cressie_2006,Cressie_2008}, \cite{Shi_2007}, and \cite{Nguyen_2012}.

There are numerous \proglang{R}  \citep{R} packages available for modelling and prediction with spatial or spatio-temporal data,\footnote{see \url{https://cran.r-project.org/web/views/Spatial.html}.} although relatively few of these make use of a model with spatial basis functions. A few variants of FRK have been developed to date, and the one that comes closest to the present software is \pkg{LatticeKrig} \citep{Nychka_2015,LatticeKrig}. \pkg{LatticeKrig} implements what we call a \emph{LatticeKrig} model, which is made up of Wendland basis functions (that have compact support)  decomposing a spatially correlated process. LatticeKrig models use a Markov assumption to construct a precision matrix  (the matrix $\Kmat^{-1}$ in Section \ref{sec:SREModel}) to describe the dependence between the coefficients of these basis functions. This, in turn, results in efficient computations and the potential use of a large number ($>10,000$) of basis functions. LatticeKrig models do not cater for what we term fine-scale-process variation and, instead, the finest scale of the process is limited to the finest resolution of the basis functions used.

The package \pkg{INLA} \citep{Lindgren_2015} is a general-purpose package for model fitting and prediction. One advantage of \pkg{INLA} is that it contains functionality for fitting Gaussian processes that have covariance functions from the Mat{\'e}rn class \citep[see][for details on software interface]{Lindgren_2015} by approximating a stochastic partial differential equation (SPDE) using a Gaussian Markov random field (GMRF). Specifically, the process is decomposed using basis functions that are triangular `tent' functions, and the coefficients of these basis functions are normally distributed with a sparse precision matrix. Thus, these models, which we term \emph{SPDE--GMRF} models, share many of the features of LatticeKrig models. A key advantage of \pkg{INLA} over \pkg{LatticeKrig} is that once the spatial or spatio-temporal model is constructed, one has access to all the approximate-inference machinery and likelihood models available within the package.

\cite{Kang_2011} develop Bayesian FRK; they keep the spatial basis functions fixed and put a prior distribution on $\Kmat$. The predictive-process approach of \citet{Banerjee_2008} can also be seen as a type of Bayesian FRK, where the basis functions are constructed from the postulated covariance function of the spatial random effects and hence depend on parameters \citep[see][for an equivalence argument]{Katzfuss_2014}. An \proglang{R} package that implements predictive processes is \pkg{spBayes} \citep{Finley_2007}. It allows for multivariate spatial or spatio-temporal processes, and Bayesian inference is carried out using Markov chain Monte Carlo (MCMC), thus allowing for a variety of likelihood models. Because the implied basis functions are constructed based on a parametric covariance model, a prior distribution on parameters results in new basis functions generated at each MCMC iteration. Since this can slow down the computation, the number of knots used in predictive processes is usually chosen to be small, which has the effect of limiting their ability to model finer scales.

Our software package \pkg{FRK} differs from spatial prediction packages currently available by constructing an SRE model on a discretised domain, where the discrete element is known as a basic areal unit \citep[BAU; see, e.g.,][]{Nguyen_2012}. The BAU can be viewed as the smallest spatial area or spatio-temporal volume that can be resolved by the process and, to reflect this, the process itself is assumed to be piecewise constant over the set of BAUs. The BAUs serve many purposes in {\bf FRK}: They define a fine grid over which to do numerical integrations for change-of-support problems; a fine lattice of discrete points over which to predict (although \pkg{FRK} implements functions to predict over any arbitrary user-defined polygons); and a set of bins within which to average large spatio-temporal datasets, if so desired, for computational efficiency. BAUs do not need to be square or all equal in size, but they do  need to be `small,' in the sense that they should be able to reconstruct the (undiscretised) process with minimal error.


In the standard `flavour' of FRK \citep{Cressie_2008}, which we term \emph{vanilla} FRK (FRK-V), there is an explicit reliance on multi-resolution basis functions to give complex non-stationary spatial patterns at the cost of not imposing any structure on $\Kmat$, the covariance matrix of the basis function weights. This can result in identifiability issues and hence in over-fitting the data when $\Kmat$ is estimated using standard likelihood methods  \citep[e.g.,][]{Nguyen_2014}, especially in regions of data paucity. Therefore, in \pkg{FRK} we also implement a model (FRK-M) where a parametric structure is imposed on $\Kmat$ \citep[e.g.,][]{Stein_2008,Nychka_2015}. The main aim of the package \pkg{FRK} is to facilitate spatial and spatio-temporal analysis and prediction for large datasets, where multiple observatons come with different spatial supports. We see that in `big data' scenarios, lack of consideration of fine-scale variation may lead to over-confident predictions, irrespective of the number of basis functions adopted. 

In Section 2, we describe the modelling, estimation, and prediction approach we adopt in \pkg{FRK}. In Section 3, we discuss further details of the package and provide a simple example on the classic \code{meuse} dataset. In Section 4, we evaluate the SRE model implemented in \pkg{FRK} in controlled cases, against LatticeKrig models and SPDE--GMRF models through use of the packages \pkg{LatticeKrig} and \pkg{INLA}. In Section 5, we show its capability to deal with change-of-support issues and anisotropic processes.  In Section 6, we show how to use \pkg{FRK} with spatio-temporal data and illustrate its use on the modelling and prediction of column-averaged carbon dioxide on the globe from remote sensing data produced by NASA's OCO-2 mission. The spatio-temporal dataset contains millions of observations. Finally, Section 7 discusses future work.

\section[Outline of FRK: Modelling, estimation and prediction]{Outline of {\pkg FRK}: Modelling, estimation and prediction} \label{sec:theory}

In this section we present the theory behind the operations implemented in \pkg{FRK}. In Section \ref{sec:SREModel} we introduce the SRE model, in Section \ref{sec:estimation} we discuss the EM algorithm for parameter estimation, and in Section \ref{sec:prediction} we present the spatial prediction equations.

\subsection{The SRE model} \label{sec:SREModel}

Denote the spatial process of interest as $\{Y(\svec) : \svec \in D\}$, where $\svec$ indexes the location of $Y(\svec)$ in our domain of interest $D$. In what follows, we assume that $D$ is a spatial domain but extensions to spatio-temporal domains are natural within the framework (Section \ref{sec:ST}). Consider the classical spatial statistical model,
\begin{equation*}
Y(\svec) = \tvec(\svec)^\top\alphab + \upsilon(\svec) + \xi(\svec); \quad \svec \in D,
\end{equation*}
where, for $\svec \in D$, $\tvec(\svec)$ is a vector of spatially referenced covariates, $\alphab$ is a vector of regression coefficients, $\upsilon(\svec)$ is a small-scale, spatially correlated random effect, and $\xi(\svec)$ is a fine-scale random effect that is `almost' spatially uncorrelated. It is natural to let $\E(\upsilon(\cdot)) = \E(\xi(\cdot)) = 0$. Define $\lambda(\cdot) \equiv \upsilon(\cdot) + \xi(\cdot)$, so that $\expect(\lambda(\cdot)) = 0$. It is the structure of the process $\upsilon(\cdot)$ in terms of a linear combination of a fixed number of spatial basis functions that defines the SRE model for $\lambda(\cdot)$:
$$
\lambda(\svec) = \sum_{l=1}^r \phi_l(\svec)\eta_l + \xi(\svec);\quad \svec \in D,
$$
\noindent where $\etab \equiv (\eta_1,\dots,\eta_r)^\top$ is an $r$-variate random vector, and $\phib(\cdot) \equiv (\phi_1(\cdot),\dots,\phi_r(\cdot))^\top$ is an $r$-dimensional vector of pre-specified spatial basis functions. Sometimes, $\phib(\cdot)$ contains basis functions of multiple resolutions (e.g., wavelets), they may or may not be orthogonal, and they may or may not have compact support. The basis functions chosen should be able to adequately reconstruct realisations of $Y(\cdot)$; an empirical spectral-based approach that can ensure this is discussed in \cite{Zammit_2012}.

In order to cater for different observation supports $\{B_j\}$ (defined below), it is convenient to assume a discretised domain of interest $D^G \equiv \{A_i \subset D: i = 1,\dots,N\}$ that is made up of $N$ small, non-overlapping basic areal units or BAUs \citep{Nguyen_2012}, and $D = \bigcup_{i=1}^N A_i$. The set $D^G$ of BAUs is a discretisation, or `tiling,' of the original domain $D$, and typically $N \gg r$. The process $\{Y(\svec): \svec \in D\}$ is then averaged over the BAUs, giving the vector $\Yvec = (Y_i : i = 1,\dots,N)^\top$, where

\begin{equation}\label{eq:Yi0}
Y_i \equiv \frac{1}{|A_i|}\int_{A_i} Y(\svec) \intd \svec; \quad i = 1,\dots,N,
\end{equation}

\noindent and $N$ is the number of BAUs. At this BAU level,
\begin{equation} \label{eq:Yi1}
Y_i = \tvec_i^\top\alphab + \upsilon_i + \xi_i,
\end{equation}
\noindent where for $i = 1,\dots, N,$ $\tvec_i \equiv \frac{1}{|A_i|}\int_{A_i} \tvec(\svec) \intd \svec$, $\upsilon_i\equiv \frac{1}{|A_i|}\int_{A_i} \upsilon(\svec) \intd \svec$, and $\xi_i$ is specified below. The SRE model specifies that the small-scale random variation is $\upsilon(\cdot) = \phib(\cdot)^\top\etab,$ and hence in terms of the discretisation onto $D^G$,


$$
\upsilon_i = \left(\frac{1}{|A_i|}\int_{A_i}\phib(\svec)\intd \svec\right)^\top \etab; \quad i = 1,\dots,N,
$$
\noindent so that $\upsilonb = \Smat\etab$, where $\Smat$ is the $N \times r$ matrix defined as follows:
\begin{equation}\label{eq:S_integral}
\Smat \equiv \left(\frac{1}{|A_i|}\int_{A_i}\phib(\svec)\intd\svec : i = 1,\dots,N\right)^\top.
\end{equation}

In \pkg{FRK}, we assume that $\etab$ is an $r$-dimensional Gaussian vector with mean zero and $r \times r$ covariance matrix $\Kmat$, and estimation of $\Kmat$ is based on likelihood methods; we denote this variant of FRK as FRK-V (where recall that `V' stands for `vanilla'). If some structure is imposed on $\var(\etab)$ in terms of parameters $\varthetab$, then $\Kmat = \Kmat_\circ(\varthetab)$ and $\varthetab$ needs to be estimated; we denote this variant as FRK-M (where recall that `M' stands for `model'). Frequently, the resolution of the BAUs is sufficiently fine, and the basis functions are sufficiently smooth, so that $\Smat$ can be approximated:
\begin{equation}\label{eq:Sapprox}
\Smat \approx \left(\phib(\svec_i) : i = 1,\dots,N\right)^\top,
\end{equation}
where $\{\svec_i: i = 1,\dots,N\}$ are the centroids of the BAUs. Since small BAUs are always assumed, this approximation is used throughout \pkg{FRK}.

In \pkg{FRK}, we do not directly model $\xi(\svec)$, since we are only interested in its discretised version. Rather, we assume that $\xi_i \equiv \frac{1}{|A_i|}\int_{A_i} \xi(\svec)\intd\svec$ has a Gaussian distribution with mean zero and variance
\begin{equation*}
\var(\xi_i) = \sigma^2_\xi \varv_{\xi,i},
\end{equation*}
where $\sigma^2_\xi$ is a parameter to be estimated, and the weights $\{\varv_{\xi,1},\dots,\varv_{\xi,N}\}$ are known and represent heteroscedasticity. These weights are typically generated from domain knowledge; they may, for example, correspond to topographical features such as terrain roughness \citep{Zammit_2015}. Since we specified $\xi(\cdot)$ to be `almost' spatially uncorrelated, it is reasonable to assume that the variables representing the discretised fine-scale variation, $\{\xi_i: i = 1,\dots,N\}$, are uncorrelated. From \eqref{eq:Yi1}, we can write
\begin{equation}\label{eq:SRE_Y}
\Yvec = \Tmat\alphab + \Smat\etab + \xib,
\end{equation}
where $\Tmat \equiv (\tvec_i: i = 1,\dots,N)^\top$, $\xib \equiv (\xi_i : i = 1,\dots,N)^\top$, and $\var(\xib) \equiv \sigma^2_\xi \Vmat_\xi$, for known $\Vmat_\xi \equiv \diag(\varv_{\xi,1},\dots,\varv_{\xi,N})$.

We now  assume that the hidden (or latent) process, $Y(\cdot)$, is observed with $m$ footprints (possibly overlapping) spanning one or more BAUs, where typically $m \gg r$ (note that both $m > N$ and $N \ge m$ are possible). We thus define the observation domain as $D^O \equiv \{ \cup_{i \in c_j} A_i : j = 1,\dots,m \}$, where $c_j$ is a non-empty set in $2^{\{1,\dots,N\}}$, the power set of $\{1,\dots,N\}$, and $m = |D^O|$.  For illustration, consider the simple case of  the discretised domain being made up of three BAUs. Then $D^G = \{A_1,A_2,A_3\}$ and, for example, $D^O = \{B_1, B_2\}$, where $B_1 = A_1 \cup A_2$ (i.e., $c_1 = \{1,2\}$) and $B_2 = A_3$ (i.e., $c_2 = \{3\}$). Catering for different footprints is important for remote sensing applications in which satellite-instrument footprints can widely differ \citep[e.g.,][]{Zammit_2015}.

Each $B_j \in D^O$ is either a BAU or a union of BAUs. Measurement of $\Yvec$ is imperfect: We define the measurement process as noisy measurements of the process averaged over the footprints
\begin{equation}\label{eq:meas_process}
Z_j \equiv Z(B_j) = \left(\frac{\sum_{i =1}^N Y_i w_{ij}}{\sum_{i=1}^N w_{ij}}\right) + \left(\frac{\sum_{i =1}^N \delta_i w_{ij}}{\sum_{i=1}^N w_{ij}}\right) + \epsilon_j; \quad B_j \in D^O,
\end{equation}
where the weights,
$$ w_{ij} = |A_i|\mathbb{I}(A_i \subset B_j); \quad i = 1,\dots,N;~~j = 1,\dots, m; ~~B_j \in D^O,$$
depend on the areas of the BAUs, and $\mathbb{I}(\cdot)$ is the indicator function. Currently, in \pkg{FRK}, BAUs of equal area are assumed, but we give \eqref{eq:meas_process} in its most general form.   The random quantities $\{\delta_i\}$ and $\{\epsilon_i\}$ capture the imperfections of the measurement. Better known is the measurement-error component $\epsilon_i$, which is assumed to be mean-zero Gaussian distributed. The component $\delta_i$ captures any bias in the measurement at the BAU level, which has the interpretation of an intra-BAU systematic error. These systematic errors are BAU-specific, that is, the $\{\delta_i\}$ are uncorrelated with mean zero and variance
\begin{equation*}
\var(\delta_i) = \sigma^2_\delta \varv_{\delta,i},
\end{equation*}
where $\sigma^2_\delta$ is a parameter to be estimated, and $\{\varv_{\delta,1},\dots,\varv_{\delta,N}\}$  represent known heteroscedasticity.

We assume that $\Yvec$ and $\deltab$ are independent. We also assume that the observations are conditionally independent, when conditioned on $\Yvec$ \emph{and} $\deltab$. Equivalently, we assume that the measurement errors $\{\epsilon_j: j = 1,\dots,m\}$ are independent with $\var(\epsilon_i) = \sigma^2_\epsilon\varv_{\epsilon,i}$.

We represent the data as $\Zvec \equiv (Z_j : j = 1,\dots,m)^\top$. Then, since each element in $D^O$ is the union of subsets of $D^G$, one can construct a matrix
$$
\Cmat_Z \equiv \left(\frac{w_{ij}}{\sum_{l=1}^N w_{lj}} : i = 1,\dots,N; j = 1,\dots,m\right),
$$
such that
\begin{equation*}
\Zvec = \Cmat_Z\Yvec + \Cmat_Z\deltab +  \epsilonb,
\end{equation*}
\noindent where the three components are independent, $\epsilonb \equiv (\epsilon_j : j = 1,\dots,m)^\top$, and $\var(\epsilonb) = \Sigmamat_\epsilon \equiv \sigma^2_\epsilon\Vmat_\epsilon \equiv \sigma^2_\epsilon \diag(\varv_{\epsilon,1},\dots,\varv_{\epsilon,m})$ is an $m \times m$ diagonal covariance matrix. The matrix $\Sigmamat_\epsilon$ is assumed known from the properties of the measurement. If it is not known, $\Vmat_\epsilon$ is fixed to $\Imat$ and $\sigma^2_\epsilon$ is estimated using variogram techniques \citep{Kang_2009}. Notice that the rows of  the matrix $\Cmat_Z$ sum to 1.


It will be convenient to re-write
\begin{equation}\label{eq:Z_collapsed}
\Zvec = \Tmat_Z\alphab + \Smat_Z\etab + \xib_Z + \deltab_Z + \epsilonb,
\end{equation}
where $\Tmat_Z \equiv \Cmat_Z \Tmat$, $\Smat_Z \equiv \Cmat_Z \Smat$, $\xib_Z \equiv \Cmat_Z \xib$, $\deltab_Z \equiv \Cmat_Z \deltab$, $\var(\xib_Z) = \sigma^2_\xi\Vmat_{\xi,Z} \equiv \sigma^2_\xi\Cmat_Z\Vmat_{\xi}\Cmat_Z^\top$, $\var(\deltab_Z) = \sigma^2_\delta\Vmat_{\delta,Z} \equiv \sigma^2_\delta\Cmat_Z\Vmat_\delta\Cmat_Z^\top$, and where $\Vmat_\delta \equiv \diag(\varv_{\delta,1},\dots,\varv_{\delta,N})$ is known. Then, recalling that $\expect(\etab) = \zerob$ and $\expect(\xib_Z) = \expect(\deltab_Z) = \expect(\epsilonb) = \zerob$,
\begin{align*}
\expect(\Zvec) &=\Tmat_Z\alphab,\\
\var(\Zvec) &= \Smat_Z\Kmat\Smat_Z^\top + \sigma^2_\xi\Cmat_Z\Vmat_{\xi}\Cmat_Z^\top +  \sigma^2_\delta\Cmat_Z\Vmat_\delta\Cmat_Z^\top + \sigma^2_\epsilon\Vmat_\epsilon.
\end{align*}
In practice, it is not always possible for each $B_j$ to include entire BAUs. For simplicity, in \pkg{FRK} we assume that the observation footprint overlaps a BAU if and only if the BAU centroid lies within the footprint. Frequently, point-referenced data is included in $\Zvec$. In this case, each data point is attributed to a specific BAU and it is possible to have multiple observations of the process defined on the same BAU.

We collect the unknown parameters in the set $\thetab \equiv \{\alphab, \sigma^2_\xi, \sigma^2_\delta, \Kmat\}$ for FRK-V and $\thetab_\circ \equiv \{\alphab, \sigma^2_\xi, \sigma^2_\delta, \varthetab\}$ for FRK-M for which $\Kmat = \Kmat_\circ(\varthetab)$; their estimation is the subject of Section \ref{sec:estimation}. If the parameters in $\thetab$ or $\thetab_\circ$ are known, an inversion that uses the Sherman--Woodbury identity \citep{Henderson_1981} allows spatial prediction at any BAU in $D^G$. Estimates of $\thetab$ are substituted into these spatial predictors to yield FRK-V. Similarly, estimates of $\thetab_\circ$ substituted into the spatial-prediction equations yield FRK-M.

In \pkg{FRK}, we allow the prediction set $D^P$ to be as flexible as $D^O$; specifically, $D^P \subset \{ \cup_{i \in \tilde{c}_k} A_i : k = 1,\dots,N_P \}$, where $\tilde{c}_k$ is a non-empty set in $2^{\{1,\dots,N\}}$ and $N_P$ is the number of prediction areas. We can thus predict both at the individual BAU level or averages over an area spanning multiple BAUs, and these prediction regions may overlap. This is an important change-of-support feature of \pkg{FRK}. We provide the FRK equations in Section \ref{sec:prediction}.

\subsection{Parameter estimation using an EM algorithm} \label{sec:estimation}

In all its generality, parameter estimation with the model of Section \ref{sec:SREModel} is problematic due to confounding between $\deltab$ and $\xib$. In \pkg{FRK}, the user thus needs to choose between modelling the intra-BAU systematic errors (in which case $\sigma^2_\xi$ is fixed to 0) or the process' fine-scale variation (in which case $\sigma^2_\delta$ is fixed to 0). 
We describe below the estimation procedure for the latter case; due to symmetry, the estimation equations of the former case can be simply obtained by replacing the subscript $\xi$ with $\delta$. However, which case is chosen by the user has a considerable impact on the prediction equations for $Y$ (Section \ref{sec:prediction}). Recall that the measurement-error covariance matrix $\Sigmamat_\epsilon$ is assumed known from measurement characteristics, or estimated using variogram techniques prior to estimating the remaining parameters described below. For conciseness, in this section we use $\thetab$ to denote the parameters in both FRK-V and FRK-M, only distinguishing when necessary.

We carry out parameter estimation using an expectation maximisation (EM) algorithm \citep[similar to][]{Katzfuss_2011,Nguyen_2014} with \eqref{eq:Z_collapsed} as our model. Define the \emph{complete-data} likelihood $L_c(\thetab) \equiv [\etab,\Zvec \mid \thetab]$ (with $\xib_Z$ integrated out), where $[~\cdot~]$ denotes the probability distribution of its argument.  The EM algorithm proceeds by first computing the conditional expectation (conditional on the data) of the complete-data log-likelihood at the current parameter estimates (the E-step) and, second, maximising this function with respect to the parameters (the M-step). In mathematical notation, in the E-step the function
\begin{equation*}
Q(\thetab \mid \thetab^{(l)}) \equiv \expect(\ln L_c(\thetab) \mid \Zvec,\thetab^{(l)}),
\end{equation*}
is found for some current estimate $\thetab^{(l)}$. In the M-step, the updated parameter estimates
\begin{equation*}
\thetab^{(l+1)} = \argmax_\thetab Q(\thetab \mid \thetab^{(l)}),
\end{equation*}
are found.

The E-step boils down to finding the conditional distribution of $\etab$ at the current parameter estimates. One can use standard results in Gaussian conditioning \citep[e.g., ][ Appendix A]{Rasmussen_2006}  to show from the joint distribution, $[\etab,\Zvec \mid \thetab^{(l)}]$, that
\begin{equation*}
\etab \mid \Zvec,\thetab^{(l)} \sim \Gau(\muvec_\eta^{(l)},\Sigmamat_\eta^{(l)}),
\end{equation*}
where
\begin{align*}
\muvec_\eta^{(l)} &= \Sigmamat_\eta^{(l)} \Smat_Z^\top\left(\Dmat_Z^{(l)}\right)^{-1}\left(\Zvec - \Tmat_Z\alphab^{(l)}\right), \\
\Sigmamat_\eta^{(l)} &= \left(\Smat_Z^\top\left(\Dmat_Z^{(l)}\right)^{-1}\Smat_Z + \left(\Kmat^{(l)}\right)^{-1}\right)^{-1},
\end{align*}
where $\Dmat_Z^{(l)} \equiv (\sigma^2_\xi)^{(l)} \Vmat_{\xi,Z} + \Sigmamat_\epsilon,$ and where $\Kmat^{(l)}$ is defined below.

The update for $\alphab$ is
\begin{equation}
\alphab^{(l+1)} = \left(\Tmat_Z^\top \left(\Dmat_Z^{(l+1)}\right)^{-1} \Tmat_Z\right)^{-1}\Tmat_Z^\top\left(\Dmat_Z^{(l+1)}\right)^{-1}\left(\Zvec - \Smat_Z \muvec_\eta^{(l)}\right). \label{eq:alpha}
\end{equation}
In FRK-V, the update for $\Kmat^{(l+1)}$ is

\begin{equation*}
\Kmat^{(l+1)} = \Sigmamat_\eta^{(l)} + \muvec_\eta^{(l)} \muvec_\eta^{(l)^\top},
\end{equation*}

\noindent while in FRK-M, where recall that $\Kmat = \Kmat_\circ(\varthetab)$, the update is
\begin{equation*}
\varthetab^{(l+1)} = \argmax_{\varthetab} \ln\left| \Kmat_\circ(\varthetab)^{-1}\right| - \tr\left(\Kmat_\circ(\varthetab)^{-1}\left(\Sigmamat_\eta^{(l)} + \muvec_\eta^{(l)} \muvec_\eta^{(l)^\top}\right)\right),
\end{equation*}
\noindent which is numerically optimised using the function \code{optim} with $\varthetab^{(l)}$ as the initial vector.

The update for $\sigma_\xi^2$ requires the solution to
\begin{equation} \label{eq:sigma2d}
\tr((\Sigmamat_{\epsilon} + (\sigma^2_\xi)^{(l+1)}\Vmat_{\xi,Z})^{-1}\Vmat_{\xi,Z}) = \tr((\Sigmamat_{\epsilon} + (\sigma^2_\xi)^{(l+1)}\Vmat_{\xi,Z})^{-1}\Vmat_{\xi,Z}(\Sigmamat_{\epsilon} + (\sigma^2_\xi)^{(l+1)}\Vmat_{\xi,Z})^{-1}\Omegab),
\end{equation}
where
\begin{equation} \label{eq:Omegab}
\Omegab \equiv \Smat_Z \Sigmamat_\eta^{(l)} \Smat_Z^\top + \Smat_Z \muvec_\eta^{(l)}\muvec_\eta^{(l)^\top} \Smat_Z^\top - 2\Smat_Z\muvec_\eta^{(l)}(\Zvec - \Tmat_Z\alphab^{(l+1)})^\top + (\Zvec - \Tmat_Z\alphab^{(l+1)})(\Zmat - \Tmat_Z\alphab^{(l+1)})^\top.
\end{equation}

\noindent The solution to \eqref{eq:sigma2d}, namely $(\sigma^2_\xi)^{(l+1)}$, is found numerically using \code{uniroot} after \eqref{eq:alpha} is substituted into \eqref{eq:Omegab}. Then $\alphab^{(l+1)}$ is found by substituting $(\sigma^2_\xi)^{(l+1)}$ into \eqref{eq:alpha}. Computational simplifications are possible when $\Vmat_{\xi,Z}$ and $\Sigmamat_\epsilon$ are diagonal, since then only the diagonal of $\Omegab$ needs to be computed. Further simplifications are possible when $\Vmat_{\xi,Z}$ and $\Sigmamat_\epsilon$ are proportional to the identity matrix, with constants of proportionality $\gamma_1$ and $\gamma_2$, respectively. In this case,
\begin{equation*}
(\sigma^2_\xi)^{(l+1)} = \frac{1}{\gamma_1} \left(\frac{\tr(\Omegab)}{m} - \gamma_2 \right),
\end{equation*}
where recall that $m$ is the dimension of the data vector $\Zvec$ and $\alphab^{(l+1)}$ is, in this special case, the ordinary-least-squares estimate given $\muvec_\eta^{(l)}$ (see \eqref{eq:alpha}). These simplifications are used by \pkg{FRK} whenever possible.

Convergence of the EM algorithm is assessed using the (\emph{incomplete-data}) log-likelihood function at each iteration,
\begin{equation*}
\ln \left[\Zvec \mid \alphab^{(l)}, \Kmat^{(l)}, (\sigma^2_\xi)^{(l)}\right] = -\frac{m}{2}\ln 2\pi -\frac{1}{2}\ln \left|\Sigmamat_Z^{(l)}\right| - \frac{1}{2}(\Zvec - \Tmat_Z\alphab^{(l)})^\top(\Sigmamat_Z^{(l)})^{-1}(\Zvec - \Tmat_Z\alphab^{(l)}),
\end{equation*}
where
\begin{equation*}
\Sigmamat_Z^{(l)} = \Smat_Z \Kmat^{(l)} \Smat_Z^\top + \Dmat_Z^{(l)},
\end{equation*}
and recall that $\Dmat_{Z}^{(l)} \equiv (\sigma_\xi^2)^{(l)}\Vmat_{\xi,Z} + \Sigmamat_\epsilon$. Efficient computation of the log-likelihood is facilitated through the use of the Sherman--Morrison--Woodbury matrix identity and a matrix-determinant lemma \citep[e.g.,][]{Henderson_1981}. Specifically, the operations
\begin{align*}
\left(\Sigmamat_Z^{(l)}\right)^{-1} &= \left(\Dmat_Z^{(l)}\right)^{-1} - \left(\Dmat_Z^{(l)}\right)^{-1} \Smat_Z \left[\left(\Kmat^{(l)}\right)^{-1} + \Smat^\top_Z \left(\Dmat_Z^{(l)}\right)^{-1}\Smat_Z\right]^{-1}\Smat_Z^\top\left(\Dmat_Z^{(l)}\right)^{-1},\\
\left| \Sigmamat_Z^{(l)}  \right| &= \left| \left(\Kmat^{(l)}\right)^{-1} + \Smat_Z^\top\left(\Dmat_Z^{(l)}\right)^{-1} \Smat_Z \right| \left|\Kmat^{(l)} \right| \left|\Dmat_Z^{(l)}\right|,
\end{align*}
ensure that we only deal with vectors of length $m$ and matrices of size $r \times r$, where typically the fixed rank $r \ll m,$ the dataset size.


\subsection{Prediction} \label{sec:prediction}

The prediction task is to make inference on the hidden $Y$-process over a set of prediction regions $D^P$. Consider the process $\{Y_P(\tilde{B}_k): k = 1,\dots,N_P\}$, which is derived from the $Y$ process and, similar to the observations, is constructed using the BAUs $\{A_i: i = 1,\dots,N\}$. Here, $N_P$ is the number of areas at which spatial prediction takes place, and is equal to $|D^P|$. Then,

\begin{equation*}
Y_{P,k} \equiv Y_{P}(\tilde{B}_k) = \left(\frac{\sum_{i =1}^N Y_i \tilde w_{ik}}{\sum_{i=1}^N \tilde w_{ik}}\right); \quad \tilde{B}_k \in D^P,
\end{equation*}
where the weights are
$$ \tilde w_{ik} = |A_i|\mathbb{I}(A_i \subset \tilde{B}_k); \quad i = 1,\dots,N;~~k = 1,\dots, N_P; ~~\tilde{B}_k \in D^P.$$

Define $\Yvec_P \equiv (Y_{P,k} : k = 1,\dots,N_P)^\top$. Then, since each element in $D^P$ is the union of subsets of $D^G$, one can construct a matrix,
\begin{equation}\label{eq:CP}
\Cmat_P \equiv \left(\frac{\tilde w_{ik}}{\sum_{l=1}^N \tilde w_{lk}} : i = 1,\dots,N; k = 1,\dots,N_P\right),
\end{equation}
the rows of which sum to 1, such that
\begin{equation*}
\Yvec_P = \Cmat_P\Yvec = \Tmat_P\alphab + \Smat_P\etab + \xib_P,
\end{equation*}
where $\Tmat_P \equiv \Cmat_P \Tmat$, $\Smat_P \equiv \Cmat_P \Smat$, $\xib_P \equiv \Cmat_P \xib$ and $\var(\xib_P) = \sigma^2_\xi\Vmat_{\xi,P} \equiv \sigma^2_\xi\Cmat_P\Vmat_\xi\Cmat_P^\top$.  As with the observations, the prediction regions $\{\tilde{B}_k \}$ may overlap. In practice, it may not always be possible for each $\tilde{B}_k$ to include entire BAUs. In this case, we assume that a prediction region contains a BAU if and only if the BAU centroid lies within the region.

Let $l^*$ denote the EM iteration number at which convergence is deemed to have been reached. The final estimates are then
$$\widehat\muvec_\eta \equiv \muvec_\eta^{(l^*)},~~ \widehat\Sigmamat_\eta \equiv \Sigmamat_\eta^{(l^*)},~~ \widehat\alphab \equiv \alphab^{(l^*)}, ~~\widehat\Kmat \equiv \Kmat^{(l^*)}, ~~\widehat\sigma^2_\xi \equiv (\sigma^2_\xi)^{(l^*)}, \textrm{~~and~~} \widehat\sigma^2_\delta \equiv (\sigma^2_\delta)^{(l^*)}.$$
Recall from Section \ref{sec:estimation} that the user needs to attribute fine-scale variation at the BAU level to either the measurement process or the hidden process $Y$. This leads to the following two cases.

{\bf Case 1:} $\sigma^2_\xi = 0$ and estimate $\sigma^2_\delta$. The prediction vector $\widehat\Yvec_P$ and covariance matrix $\Sigmamat_{Y_P\mid Z}$, corresponding to the first two moments from the predictive distribution $[\Yvec_P \mid \Zvec]$ when $\sigma^2_\xi = 0,$ are
\begin{align*}
\widehat\Yvec_P \equiv \expect(\Yvec_P \mid \Zvec) &= \Tmat_P\widehat\alphab + \Smat_P\widehat\muvec_\eta, \\
\Sigmamat_{Y_P \mid Z} \equiv \var(\Yvec_P\mid \Zvec) &= \Smat_P \widehat\Sigmamat_\eta\Smat_P^\top. 
\end{align*}
 Under the assumptions taken, $[\Yvec_P \mid \Zvec]$ is a $\Gau(\widehat\Yvec_P,\Sigmamat_{Y_P \mid Z})$ distribution. Note that all calculations are made after substituting in the EM-estimated parameters, and that $\widehat\sigma^2_\delta$ is present in the estimated parameters.

{\bf Case 2:} $\sigma^2_\delta = 0$ and estimate $\sigma^2_\xi$  (Default). To cater for arbitrary observation and prediction support, we predict $\Yvec_P$ by first carrying out prediction over the full vector $\Yvec$, that is, at the BAU level, and then transforming linearly to obtain $\widehat\Yvec_P$ through the use of the matrix $\Cmat_P$. It is easy to see that if $\widehat\Yvec$ is an optimal (squared-error-loss matrix criterion) predictor of $\Yvec$, then $\Amat\widehat\Yvec$ is an optimal predictor of $\Amat\Yvec$, where $\Amat$ is any matrix with $N$ columns.

Let $\Wvec \equiv (\etab^\top,\xib^\top)^\top$ and $\Pimat \equiv (\Smat,\Imat)$. Then \eqref{eq:SRE_Y} can be re-written as $\Yvec = \Tmat\alphab + \Pimat\Wvec$, and
\begin{align}
\widehat\Yvec \equiv \expect(\Yvec \mid \Zvec) &= \Tmat\widehat\alphab + \Pimat\widehat\Wvec, \nonumber \\
\Sigmamat_{Y \mid Z} \equiv \var(\Yvec \mid \Zvec) &= \Pimat \Sigmamat_W\Pimat^\top, \label{eq:Sigma_YZ}
\end{align}
for
\begin{align*}
\widehat\Wvec &\equiv \Sigmamat_W\Pimat^\top\Cmat_Z^\top\Sigmamat_\epsilon^{-1}(\Zvec - \Tmat_Z\widehat\alphab),\\
\Sigmamat_W &\equiv \left(\Pimat^\top \Cmat_Z^\top \Sigmamat_\epsilon^{-1} \Cmat_Z \Pimat + \Lambdamat^{-1}\right)^{-1},\\
\end{align*}
and the block-diagonal matrix $\Lambdamat \equiv \textrm{bdiag}(\widehat\Kmat,\widehat\sigma^{2}_\xi\Vmat_{\xi})$, where $\textrm{bdiag}(\cdot)$ returns a block diagonal matrix of its matrix arguments. Note that all calculations are made after substituting in the EM-estimated parameters.

For both Cases 1 and 2 it follows that $\widehat\Yvec_P = \expect(\Yvec_P \mid \Zvec) = \Cmat_P\widehat\Yvec$ and
\begin{equation}\label{eq:YvecP}
\Sigmamat_{Y_P\mid Z} = \var(\Yvec_P \mid \Zvec) = \Cmat_P \Sigmamat_{Y \mid Z} \Cmat_P^\top.
\end{equation}

Note that for Case 2 we need to obtain predictions for $\xib_P$ which, unlike those for $\etab$, are not a by-product of the EM algorithm of \ref{sec:estimation}. Sparse-matrix operations are used to facilitate the computation of \eqref{eq:YvecP} when possible.

\section[FRK-package structure and usage]{\pkg{FRK}-package structure and usage}\label{sec:usage}

In this section we discuss the layout and the interface of the package, and we show its use on the \code{meuse} dataset under `simple usage' and `advanced usage.' The former attempts to construct the SRE model automatically from characteristics of the data, while the latter gives the user more control through use of additional commands. The \code{meuse} dataset is not large and contains 155 readings of heavy-metal abundance in a region of The Netherlands along the river Meuse. For more details on the dataset see the vignette titled `gstat' in the package \pkg{gstat}.

\subsection{Usage overview}\label{sec:usage_overview}

By leveraging the flexibility of the spatial and spatio-temporal objects in the \pkg{sp} \citep{Bivand_2013} and \pkg{spacetime}  \citep{Pebesma_2012} packages, \pkg{FRK} provides a consistent, easy-to-use interface for the user, irrespective of whether the datasets have different spatial supports, irrespective of the manifold being used, irrespective of whether or not a temporal dimension needs to be included, and irrespective of the `prediction resolution.' 

\begin{figure}[!t]
\centering
\includegraphics[width=0.7\linewidth]{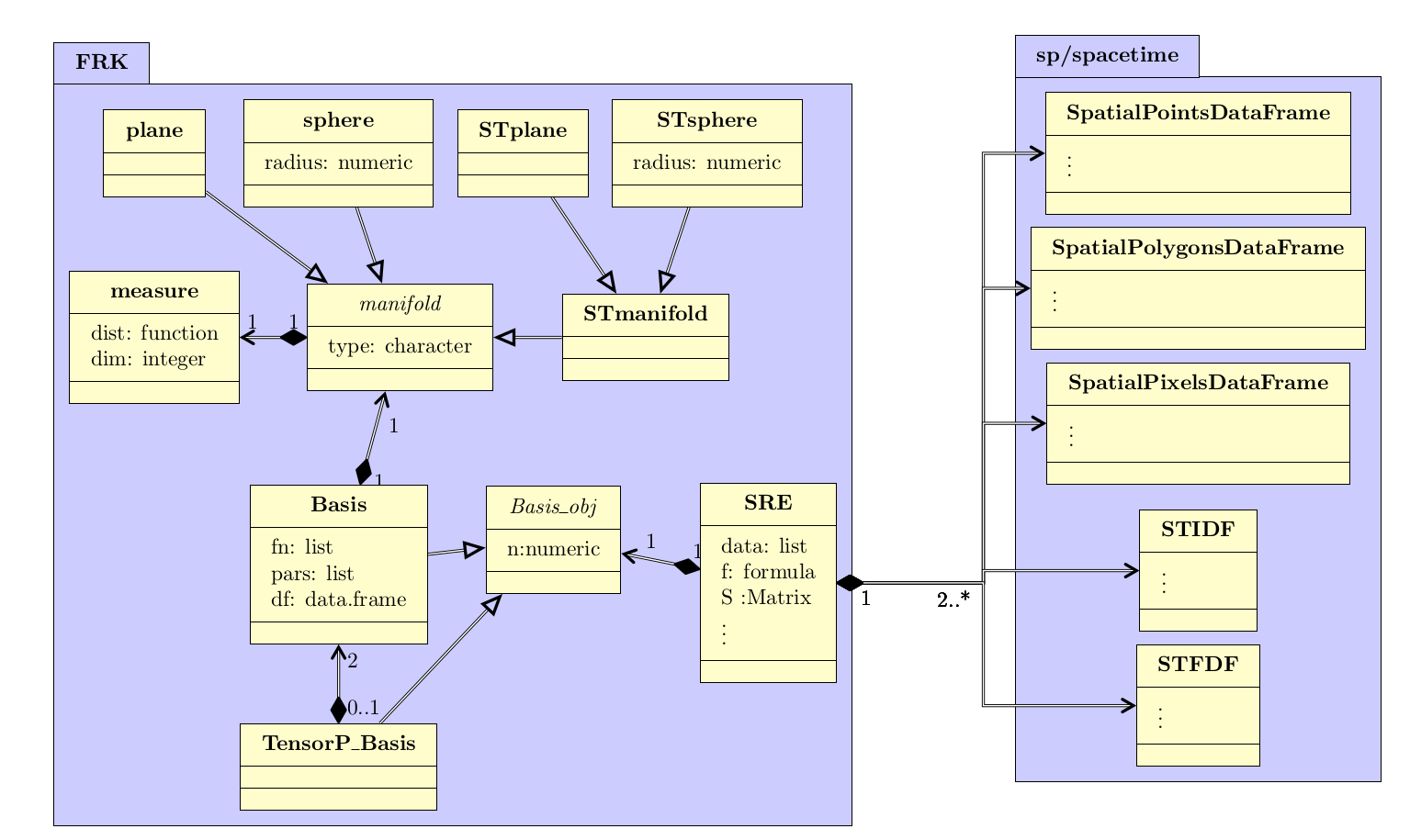}
\caption{Partial UML diagram showing the most important classes of \pkg{FRK} and their interaction to the relevant classes in \pkg{sp} and \pkg{spacetime}. See Table~\ref{tab:classes} for a brief description of these classes. For conciseness, in each class diagram (yellow box) only a few attributes are shown and no class operations are listed. Italicised class names indicate virtual classes, an arrow with an open arrowhead indicates inheritance, and a line with a diamond at one end and an arrowhead at another indicates a compositional (``has a'') relationship. The numbers on these lines indicate the number of instances involved in the relationship. For example, the \code{SRE} class always has two or more \pkg{sp} or \pkg{spacetime} instances  (BAUs and data), while a \code{TensorP\_Basis} object may or may not be needed when setting up the SRE model. In the former case we use the notation `2..*' to denote `two or more,' while in the latter we use `0..1' to note that the user may have 0 or 1  \code{TensorP\_Basis} objects when using \pkg{FRK}. \label{fig:UML}}
\end{figure}

In Figure~\ref{fig:UML} we provide a partial unified modelling language (UML) diagram summarising the important package classes and their interaction with the packages \pkg{sp} and \pkg{spacetime}, while in Table \ref{tab:classes} we provide a brief summary of these classes. BAUs should be \code{Spatial} or \code{ST} pixel or polygon objects, while the data can also be point objects (although they are subsequently mapped to BAUs by \pkg{FRK}). Each \code{Spatial} and \code{ST} object is equipped with a coordinate reference system (CRS), which needs to be identical across objects. The main class is the \code{SRE} class, the object of which incorporates all information about fitting and prediction using the data, BAUs, and basis functions.

\begin{table}
  \begin{tabular}{p{4cm}p{10cm}} \hline
    Class             &  Description \\\hline
    \code{Basis}      &  Defines basis functions on a specified manifold. \\
    \code{Basis\_obj} &  A virtual class that other basis classes inherit from.\\
    \code{manifold}   &  A virtual class that other manifold classes inherit from. \\
    \code{measure}    &  Defines objects that compute distances on a specified manifold. \\
    \code{plane, real\_line, sphere, STplane, STsphere, STmanifold }  &  Subclasses that inherit from the virtual class \texttt{manifold}. \\
    \code{SRE} & Defines the spatial-random-effects model, which is used to do FRK. \\
    \code{TensorP\_Basis} & Tensor product of two basis functions. \\ \hline
  \end{tabular}  
\caption{Important class definitions in \pkg{FRK}. \label{tab:classes}}
\end{table}

The basis functions are constructed on a manifold which, at the time of writing, can be $\mathbb{R}$ (\code{real\_line}), $\mathbb{R}^2$ (\code{plane}), $\mathbb{S}^2$ (surface of \code{sphere}), and their spatio-temporal counterparts (\code{STplane} and \code{STsphere}). Some consistency checks are made to ensure that the CRS in the BAUs and the data objects are compatible with the manifold on which the basis functions are constructed. As with \code{spDists} in the \pkg{sp} package, distances on the manifold are either Euclidean or great-circle. The function \code{spDists} in \pkg{sp} is not used, rather a function in an object of class \code{measure} is used for abstraction -- this redundant structure is intended to facilitate future implementation of \pkg{FRK} on arbitrary manifolds and with arbitrary distance functions. The package \pkg{FRK} has support for spatio-temporal data (see Section~\ref{sec:ST}); in this case, basis functions are of class \code{TensorP\_Basis} and, as the name implies, are constructed through the tensor product of spatial and temporal basis functions.

The package is built around a straightforward model (outlined in Section \ref{sec:theory}) and has the capability of handling large datasets (up to a few hundred thousand data points on a standard desktop machine, and a few million on a big memory machine). For linear algebraic calculations, it leverages routines from the \pkg{sparseinv} package \citep[][which is built from \texttt{C} code written by \citet{Sparseinv}]{sparseinv_2018} and the \proglang{R} package \pkg{Matrix} \citep{Matrix_2015}. The package \pkg{INLA} \citep{Lindgren_2015} is used for finding a non-convex hull of the data points and for placing basis functions irregularly in the domain (if desired).

 The user has two levels of control; for simple problems one can call the function \code{FRK}, in which case basis-function construction and BAU generation is done automatically based on characteristics of the data. Alternatively, for more (advanced) control, the user can follow the following six steps.

\begin{itemize}
\item {\bf Step 1:} Place the data into an object with class defined in \pkg{sp} or \pkg{spacetime}, specifically either \code{SpatialPointsDataFrame} or \code{STIDF} for point-referenced data, and either \\ \code{SpatialPolygonsDataFrame} or \code{STFDF} for polygon-referenced data \citep{Pebesma_2012}.
\item {\bf Step 2:} Construct a prediction grid of BAUs using \code{auto\_BAUs}, where each BAU is representative of the finest scale upon which we wish to carry out inference (the process is discretised at the BAU level). The BAUs are usually of class \code{SpatialPixelsDataFrame} for spatial problems (or they could also be of class \code{SpatialPolygonsDataFrame}), and they are of class \code{STFDF} for spatio-temporal problems.
\item {\bf Step 3:} Construct a set of regularly or irregularly spaced basis functions using \code{auto\_basis}. The basis functions can be of various types (e.g., bisquare, Gaussian, or exponential functions).
\item {\bf Step 4:} Construct an SRE model using \code{SRE} from an \proglang{R} formula that identifies the response variable, the covariates, the data, the BAUs, and the basis functions.
\item {\bf Step 5:} Estimate the parameters within the SRE model using \code{SRE.fit}. Estimation is carried out using the EM algorithm described in Section \ref{sec:estimation}.
\item \begin{sloppypar} {\bf Step 6:} Predict either at the BAU level or over arbitrary polygons specified as \code{SpatialPolygon}s or \code{SpatialPolygonDataFrame}s in the spatial case, or as \code{STFDF}s in the spatio-temporal case, using \code{predict}. \end{sloppypar}
\end{itemize}

In Table \ref{tab:functions} we provide some of the important methods and functions, together with brief descriptions, available to the user of \pkg{FRK}.

\begin{table}[h!]
  \begin{tabular}{llp{9cm}} \hline
    Group           &  Method/Function    & Use \\\hline
    Basis functions &  \code{auto\_basis} & Automatically constructs a set of basis functions on a given manifold based on a supplied dataset.      \\
                    & \code{local\_basis} & Manually constructs a set of `local' basis functions from a set of centroids and scale parameters. \\
                    & \code{eval\_basis}  & Evaluates basis functions over arbitrary points or polygons. \\
                    & \code{remove\_basis}& Removes basis functions from an object of class \code{Basis}.  \\
                    & \code{show\_basis}  & Visualises basis functions. \\
    BAUs            & \code{auto\_BAUs} & Automatically constructs a set of BAUs on a given manifold around a supplied dataset.\\
                    & \code{BAUs\_from\_points} & Constructs BAUs from point-level data.\\
    Information     & \code{coef} & Returns regression coefficients from a fitted SRE model. \\
                    & \code{info\_fit} & Returns information from the EM algorithm (e.g., information on convergence).\\
                    & \code{nbasis} & Returns the number of basis functions in a \code{Basis} or \code{SRE} object. \\
                    & \code{nres} & Returns the number of basis-function resolutions in a \code{Basis} or \code{SRE} object. \\
                    & \code{opts\_FRK\$get} & Returns current option settings. \\
                    & \code{opts\_FRK\$set} & Sets an option. \\
                    & \code{summary} & Returns information on the \code{Basis} or \code{SRE} object. \\
    FRK operations  & \code{FRK} & Constructs and fits an SRE model from a supplied \proglang{R} formula and dataset. \\
                    & \code{predict} & Predicts over BAUs or at \code{newdata} using a fitted SRE model. \\
                    & \code{SRE}     & Constructs an SRE model from an \proglang{R} formula, data, BAUs, and basis functions. \\
                    & \code{SRE.fit} & Fits (estimates parameters in) an SRE model.   \\            \hline
  \end{tabular}  
\caption{Important methods and functions in \pkg{FRK}. \label{tab:functions}}
\end{table}

\subsection{Simple usage}

In simple cases, the user constructs and fits the SRE model using the function \code{FRK}, and then prediction is carried out using the function \code{predict}. The main function \code{FRK} takes two compulsory arguments: A standard \proglang{R} formula \code{f} and a list of data objects \code{data}, and it returns an object of class \code{SRE}. Each of the data objects in the list must be of class \code{SpatialPointsDataFrame}, \code{SpatialPolygonsDataFrame}, \code{STIDF}, or \code{STFDF}, and each must contain the dependent variable defined in \code{f}. If there are covariates, then the user must supply the covariate data with all the BAUs, that is, at both the BAU measurement locations and at the BAU prediction locations. The BAUs should be of class \code{SpatialPolygonsDataFrame} or \code{SpatialPixelsDataFrame} (in the spatial case) or \code{STFDF} (in the spatio-temporal case). Note that, unlike conventional spatial modelling tools, covariate information should not be supplied with the data, but with the BAUs. Also note that the intersection of the data support and that of the BAUs should never be null.

When no basis functions or BAUs are supplied, then these are elicited automatically based on characteristics of the supplied dataset(s). The number of basis functions used depends on whether $\Kmat$ is unstructered or not, on whether the data are spatial only or are spatio-temporal, and on the number of data points. For details, see the package's manual \citep{FRK}. The number of BAUs depends on the domain boundary and on whether the dataset is spatial or spatio-temporal. Domain construction and basis-function placement may make use of geometric functions available in \pkg{INLA}. If \pkg{INLA} is unavailable, simple geometric methods are used instead.

\pkg{FRK} was not built for small datasets, for which standard exact kriging is fast and memory efficient. However, to illustrate the utility of \pkg{FRK}, we consider the \code{meuse} dataset in the package \pkg{sp}. We first consider a simple model with no covariates, in which we model the logarithm of zinc concentrations. Basis functions can either be arranged on a grid by setting \code{regular = 1} or as a function of data density (using the \pkg{INLA} mesher) by setting \code{regular = 0}.

The \code{meuse} dataset is first loaded and cast into a \code{SpatialPointsDataFrame}.
\begin{Schunk}
\begin{Sinput}
R> library("sp")
R> data("meuse")
R> coordinates(meuse) <- ~x + y
\end{Sinput}
\end{Schunk}

Then, \code{FRK} is invoked as follows.

\begin{Schunk}
\begin{Sinput}
R> library("FRK")
R> f <- log(zinc) ~ 1
R> S <- FRK(f = f, data = list(meuse), regular = 0)
\end{Sinput}
\end{Schunk}

\noindent The returned \code{SRE} object \code{S} contains all the information about the fitted SRE model, which can be displayed using the \code{summary} command.

If we wish to use covariate information, we need to consider BAUs that have covariate information attached to them. Such BAUs are available for this problem in the package \pkg{sp} in \code{meuse.grid}, which we first cast into a \code{SpatialPixelsDataFrame} using the function \code{gridded} before using them in the SRE model.

\begin{Schunk}
\begin{Sinput}
R> data("meuse.grid")
R> coordinates(meuse.grid) <- ~x + y
R> gridded(meuse.grid) <- TRUE
\end{Sinput}
\end{Schunk}

In this example, based on prior exploratory data analysis (see the vignette `gstat' in the package \pkg{gstat}), we consider the square root of the distance from the centroid of a BAU to the nearest point on the river Meuse as the covariate. Recall that all covariates need to be supplied with the BAUs and not with the data, and \pkg{FRK} will throw an error if the data and BAUs have fields in common. In the code below, we first set any common fields to \code{NULL} in the data object, before running \code{FRK} using the user-specified BAUs.

\begin{Schunk}
\begin{Sinput}
R> meuse$soil <- meuse$dist <- meuse$ffreq <- NULL
R> f <- log(zinc) ~ 1 + sqrt(dist)
R> S <- FRK(f = f, data = list(meuse), BAUs = meuse.grid, regular = 0)
\end{Sinput}
\end{Schunk}

The other core function, which is also needed for `advanced usage,' is \code{predict}, which is used to compute prediction and prediction standard errors at all prediction locations. This function takes as compulsory argument the SRE model \code{S}. If no polygons are specified, prediction is carried out at the BAU level (in space and/or time). An important argument is the flag \code{obs\_fs}, which acts as a choice between Case 1 (process' fine-scale variance $\sigma^2_\xi = 0$; \code{obs\_fs = TRUE}) and Case 2 (systematic intra-BAU variance $\sigma^2_\delta = 0$;  \code{obs\_fs = FALSE}) of Section \ref{sec:prediction}.

\begin{Schunk}
\begin{Sinput}
R> Pred <- predict(S, obs_fs = FALSE)
\end{Sinput}
\end{Schunk}

\noindent The function \code{Pred} returns the polygons (or, in this case, the BAUs) containing the prediction \code{mu} and the prediction variance \code{var}, which can be readily used for visualisation. The predictions and prediction standard errors of the model having \code{sqrt(dist)} as a covariate are depicted in Figure \ref{fig:meuse}. In this instance, Case 2 was used, and the estimate of the fine-scale variance $\sigma^2_\xi$ was positive. Hence, the prediction and prediction-error maps exhibit `bulls-eye' features, where the prediction standard errors are much lower in BAUs containing data than in neighbouring BAUs not containing data. 

\begin{Schunk}
\begin{figure}
\includegraphics[width=\maxwidth]{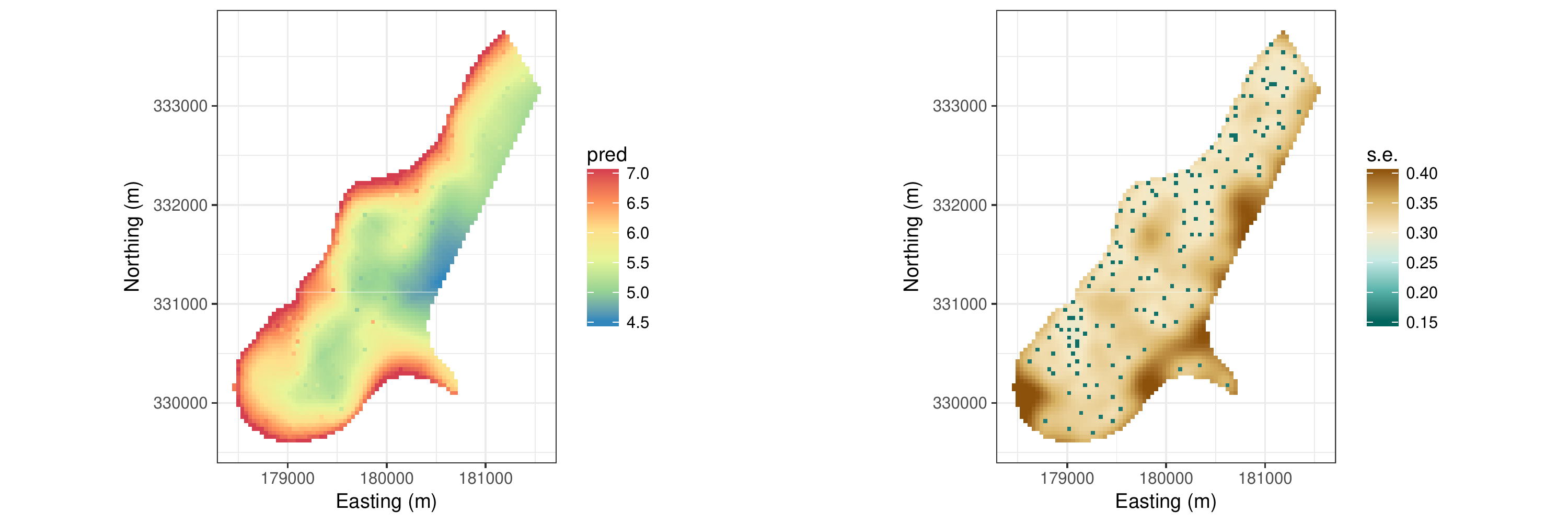} \caption{(Left panel) Prediction of log-zinc concentration obtained from \pkg{FRK} using the \code{meuse} dataset. (Right panel) Prediction standard errors for log-zinc concentration. Both quantities are in logs of ppm.\label{fig:meuse}}\label{fig:unnamed-chunk-7}
\end{figure}
\end{Schunk}

\subsubsection*{Point-level data and predictions}

In many cases, the user has one data object or data frame containing both observations and prediction locations with accompanying covariates. Missing observations are then usually denoted as \texttt{NA}. Since in \pkg{FRK} all covariates are associated with the BAUs and not the data, that one data object needs to be used to construct (i) a second data object where no data are missing and that does not contain missing covariates, and (ii) BAUs at both the observation and prediction locations supplied with their associated covariate data.

For example, assume that the first 10 log-zinc concentrations are missing in the \texttt{meuse} dataset.

\begin{Schunk}
\begin{Sinput}
R> data("meuse")
R> meuse[1:10, "zinc"] <- NA
\end{Sinput}
\end{Schunk}

\noindent Once the data frame is appropriately subsetted, it is then cast as a \code{SpatialPointsDataFrame} as usual.

\begin{Schunk}
\begin{Sinput}
R> meuse2 <- subset(meuse, !is.na(zinc))
R> meuse2 <- meuse2[, c("x", "y", "zinc")]
R> coordinates(meuse2) <- ~x + y
\end{Sinput}
\end{Schunk}

The BAUs, on the other hand, should contain all the data and prediction locations, but not the response variable itself. Their construction is facilitated by the function \code{BAUs\_from\_points} which constructs tiny BAUs around the data and prediction locations.

\begin{Schunk}
\begin{Sinput}
R> meuse$zinc <- NULL
R> coordinates(meuse) <- c("x", "y")
R> meuse.grid2 <- BAUs_from_points(meuse)
\end{Sinput}
\end{Schunk}

\noindent Once BAUs are constructed at both data and prediction locations, FRK may proceed as shown below. Predictions are by default made at both the observed and unobserved BAUs.
\begin{Schunk}
\begin{Sinput}
R> f <- log(zinc) ~ 1 + sqrt(dist)
R> S <- FRK(f = f, data = list(meuse2), BAUs = meuse.grid2, regular = 0)
R> Pred <- predict(S, obs_fs = FALSE)
\end{Sinput}
\end{Schunk}

\subsection{Advanced usage}\label{sec:advanced}

The package \pkg{FRK} provides several helper functions for facilitating basis-function construction and BAU construction when more control is needed. Harnessing the extra functionality requires following the six steps outlined in Section~\ref{sec:usage_overview}.

\noindent {\bf Step 1:} As before, we first load the data and cast it into a \code{SpatialPointsDataFrame}.

\begin{Schunk}
\begin{Sinput}
R> data("meuse")
R> coordinates(meuse) <- ~x + y
\end{Sinput}
\end{Schunk}

\vspace{0.1in}

\noindent {\bf Step 2:} Based on the geometry of the data we now generate BAUs. For this, we use the helper function \code{auto\_BAUs}, which takes several arguments (see \code{help(auto\_BAUs)} for details). In the code below, we instruct the helper function to construct BAUs on the plane, centred around the data in \code{meuse} with each BAU of size 100 $\times$ 100 m. The \code{type = "grid"} input indicates that we want a rectangular grid and not a hexagonal lattice (\code{type = "hex"}) and \code{convex = -0.05} is a parameter controlling the shape of the domain boundary when \code{nonconvex\_hull = TRUE} (see the help file of \code{INLA::inla.nonconvex.hull} and \citet{Lindgren_2015} for more details), and the extension of the convex hull of the data when \code{nonconvex\_hull = FALSE} (default).

\begin{Schunk}
\begin{Sinput}
R> GridBAUs1 <- auto_BAUs(manifold = plane(), type = "grid", cellsize = c(100, 
+    100), data = meuse, nonconvex_hull = TRUE, convex = -0.05)
\end{Sinput}
\end{Schunk}

For the $i$th BAU, we also need to supply the element $\varv_{\xi,i}$ (or $\varv_{\delta,i}$) that describes the hetereoscedascity of the fine-scale variation for that BAU. As described in Section \ref{sec:SREModel}, this component encompasses all process variation that occurs at the BAU scale and only needs to be known up to a constant of proportionality, $\sigma^2_\xi$ or $\sigma^2_\delta$ (depending on the chosen model); this constant is estimated using maximum likelihood with \code{SRE.fit}, which uses the EM algorithm of Section \ref{sec:estimation}. Typically, geographic features such as altitude are appropriate, but in this illustration of the package we just set this value to be 1 for all BAUs. This field is labelled \code{fs}, and \code{SRE} will throw an error if it is not set.

\begin{Schunk}
\begin{Sinput}
R> GridBAUs1$fs <- 1
\end{Sinput}
\end{Schunk}
\noindent The data and BAUs are illustrated using the \code{plot} function in Figure~\ref{fig:meuse2}.  At this stage, the BAUs only contain geographical information. To add covariate information to these BAUs from other \code{Spatial} objects, the function \code{sp::over} can be used.

\begin{Schunk}
\begin{figure}

{\centering \includegraphics[width=\linewidth]{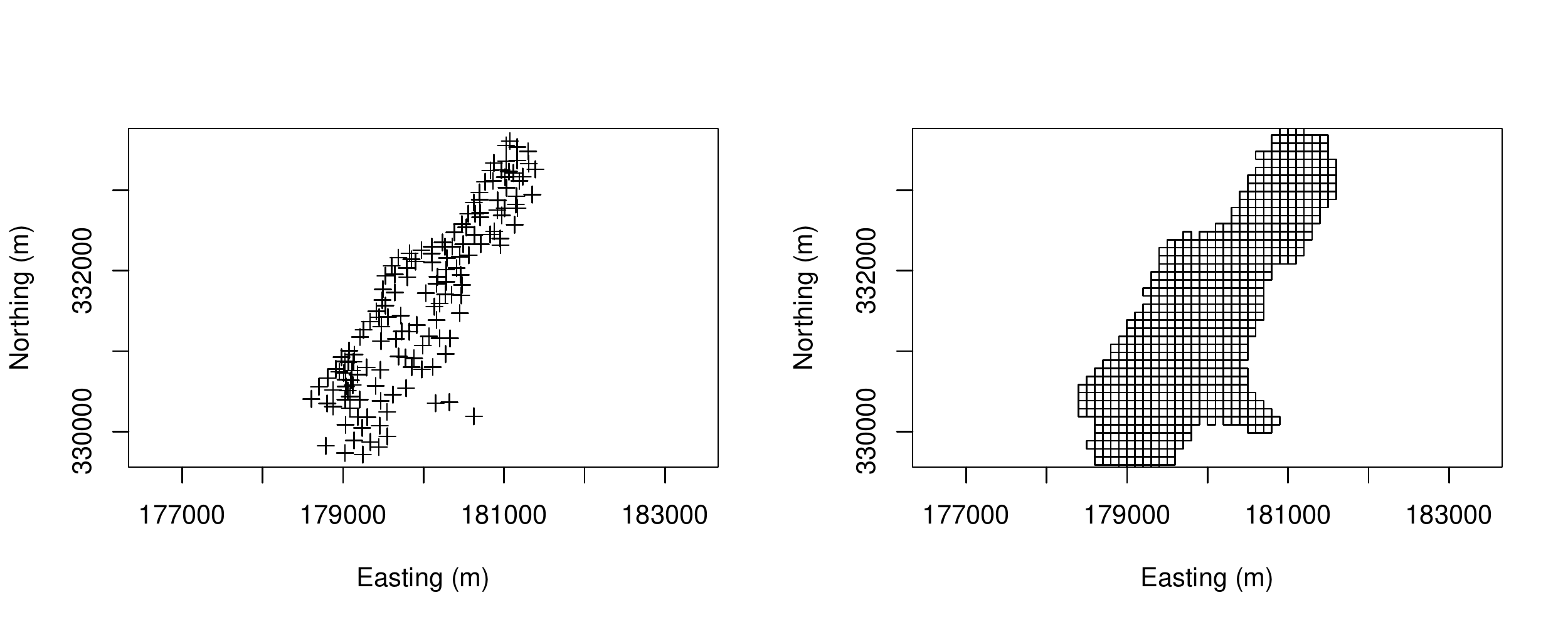} 

}

\caption{(Left panel) Locations of the \code{meuse} data. (Right panel) BAUs for FRK with the \code{meuse} dataset constructed using \code{auto\_BAUs}.\label{fig:meuse2}}\label{fig:unnamed-chunk-15}
\end{figure}
\end{Schunk}

\vspace{0.1in}

\noindent {\bf Step 3:} \pkg{FRK} decomposes the spatial process as a sum of basis functions that can be constructed using the helper functions \code{auto\_basis} as follows:

\begin{Schunk}
\begin{Sinput}
R> G <- auto_basis(manifold = plane(), data = meuse, regular = 0, 
+    nres = 3, type = "bisquare")
\end{Sinput}
\end{Schunk}

\noindent The argument \code{nres} indicates the number of basis-function resolutions to use, while \code{type} indicates the function to use, in this case the bisquare function,
$$
b(\svec_1,\svec_2) \equiv \left\{\begin{array}{ll} A\{1 - (\|\svec_2- \svec_1\|/r)^2\}^2; &\| \svec_2 -\svec_1  \| \le r \\
0; & \textrm{otherwise}, \end{array} \right.
$$
where $A$ is the amplitude and $r$ is the aperture. Other options are \code{"exp"} (the exponential covariance function) and \code{"Matern32"} (the Mat{\'e}rn covariance function with smoothness parameter equal to 1.5). The basis functions do not need to be positive-definite and users may define their own; see Section \ref{sec:custom_basis}. The argument \code{prune} (not used in this example) may be used to remove basis functions that are not influenced by data: \code{prune} should be used with care as, in general, basis functions are needed to represent variability in unobserved regions. However, it is useful when implementing FRK-V, where all the columns of $\Smat_Z$ in \eqref{eq:Z_collapsed} need to contain at least one non-zero element in order for $\etab$ to be  identifiable. See \code{help(auto\_basis)} for details.

The basis functions can be visualised using \code{show\_basis(G)}; see Figure~\ref{fig:meuse_basis}.

\begin{Schunk}
\begin{figure}[t]

{\centering \includegraphics[width=0.6\linewidth]{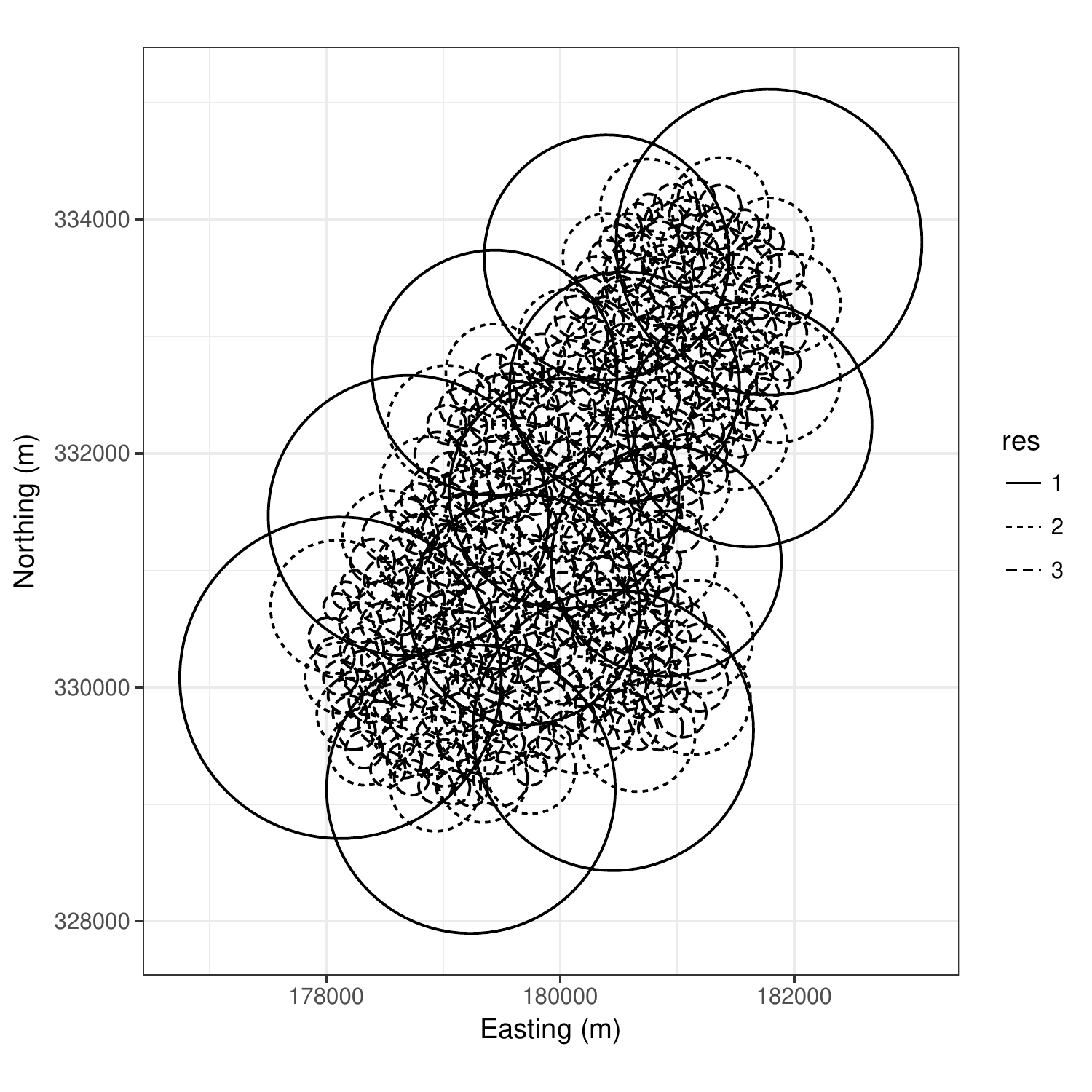} 

}

\caption{Basis functions automatically generated for the \code{meuse} dataset with 3 resolutions. The interpretation of the circles changes with the domain and basis-function type. For bisquare functions on the plane, each circle is centred at the basis-function centre, and has a radius equal to the function's aperture. See the help file of \code{show\_basis} for details.\label{fig:meuse_basis}}\label{fig:unnamed-chunk-17}
\end{figure}
\end{Schunk}

\vspace{0.1in}

\noindent {\bf Step 4:} The SRE model is constructed by supplying an \proglang{R} formula, the data, the BAUs, and the basis functions, to the function \code{SRE}. If the model contains covariates, one must make sure that they are specified at the BAU-level (and hence attributed to \code{GridBAUs1}). We use the following formula.

\begin{Schunk}
\begin{Sinput}
R> f <- log(zinc) ~ 1
\end{Sinput}
\end{Schunk}

\noindent The SRE model is then constructed using the function \code{SRE}, which essentially bins the data into the BAUs, constructs all the matrices required for estimation, and provides initial guesses for the parameters that need to be estimated. By default, \code{K\_type = "block-exponential"}, which signals the construction of the matrices $$\Kmat_n(\varthetab) = (\vartheta_{1n}\exp(-d_{ijn}/\vartheta_{2n}) : i,j = 1,\dots,r_n),$$ where $d_{ijn}$ is the distance between the centroids of the $i$th and $j$th basis functions at the $n$th resolution, $r_n$ is the number of basis functions at the $n$th resolution, $n = 1,\dots,n_{res}$, $n_{res}$ is the number of resolutions, $\vartheta_{1n}$ is the marginal variance at the $n$th resolution, and $\vartheta_{2n}$ is the e-folding length-scale  (i.e., the distance at which the correlation is $1/e$) at the $n$th resolution. Then the default is $\Kmat_\circ(\varthetab) = \textrm{bdiag}(\{\Kmat_n(\varthetab) : n = 1,\dots,n_{res}\})$, where $\varthetab \equiv (\vartheta_{11},\vartheta_{21}, \vartheta_{12}, \dots, \vartheta_{2n_{res}})^\top$ and $\textrm{bdiag}(\cdot)$ returns a block-diagonal matrix constructed from its arguments.

\begin{Schunk}
\begin{Sinput}
R> S <- SRE(f = f, data = list(meuse), BAUs = GridBAUs1, basis = G, 
+    est_error = TRUE, average_in_BAU = FALSE)
\end{Sinput}
\end{Schunk}

\noindent \code{K\_type = "unstructured"} can be used to invoke FRK-V.

\noindent When calling the function \code{SRE}, we supplied the formula \code{f} containing information on the dependent variable and the covariates; the data (as a list that can include additional datasets); the BAUs; the basis functions; a flag \code{est\_error}; and another flag \code{average\_in\_BAU}. The flag \code{est\_error = TRUE} is used to estimate the measurement-error variance $\sigma^2_\epsilon$ (where $\Sigmamat_\epsilon \equiv \sigma^2_\epsilon\Imat$) using semivariogram methods \citep{Kang_2009}. At the time of writing, \code{est\_error = TRUE} was only available for spatial data, not for spatio-temporal data. When not set to \code{TRUE}, each dataset needs to also contain a field \code{std}, the standard deviation of the measurement error (that can vary with the measurement).

\pkg{FRK} is built on the concept of a BAU, and hence the smallest spatial support of an observation has to be equal to that of a BAU. However, in practice, several datasets (such as the \code{meuse} dataset) are point-referenced. We reconcile this difference by assigning a support to every point-referenced datum equal to that of a BAU. Multiple point-referenced data falling within the same BAU are thus assumed to be noisy observations of the same random variable; see \eqref{eq:meas_process}.  As a consequence of this, when multiple observations fall into the same BAU, the matrices $\Vmat_{\xi,Z}$ and $\Vmat_{\delta,Z}$ will be sparse but \emph{not} diagonal (since $\Cmat_Z$ will contain more than one non-zero element per column). This can increase the computational time required for estimation considerably. For large point-referenced datasets, such as the \code{AIRS} dataset considered in Section \ref{sec:AIRS}, it is reasonable to summarise the data at the BAU level. Since \pkg{FRK} is designed for use with large datasets, the argument \code{average\_in\_BAU = TRUE} of the function \code{SRE} is defaulted to \code{TRUE}. In this default setting, all data falling into one BAU is averaged, and the standard deviation of the measurement error of the averaged data point is taken to be the average standard deviation of the measurement error of the individual data points. Consequently, the dataset is thinned. With large datasets and small BAUs, this thinning frequently does not cause performance degradation (see Section \ref{sec:AIRS}). Since the \code{meuse} dataset is relatively small, we set \code{average\_in\_BAU = FALSE}.

\vspace{0.1in}

\noindent {\bf Step 5:} The SRE model is fitted using the function \code{SRE.fit}. Maximum likelihood is carried out using the EM algorithm of Section \ref{sec:estimation}, which is assumed to have converged either when \code{n\_EM} is exceeded or when the log-likelihood across subsequent steps does not change by more than \code{tol}. In this example, the EM algorithm converged in about 30 iterations; see Figure~\ref{fig:EM}.

\begin{Schunk}
\begin{Sinput}
R> S <- SRE.fit(SRE_model = S, n_EM = 400, tol = 0.01, print_lik = TRUE)
\end{Sinput}
\begin{figure}

{\centering \includegraphics[width=\maxwidth]{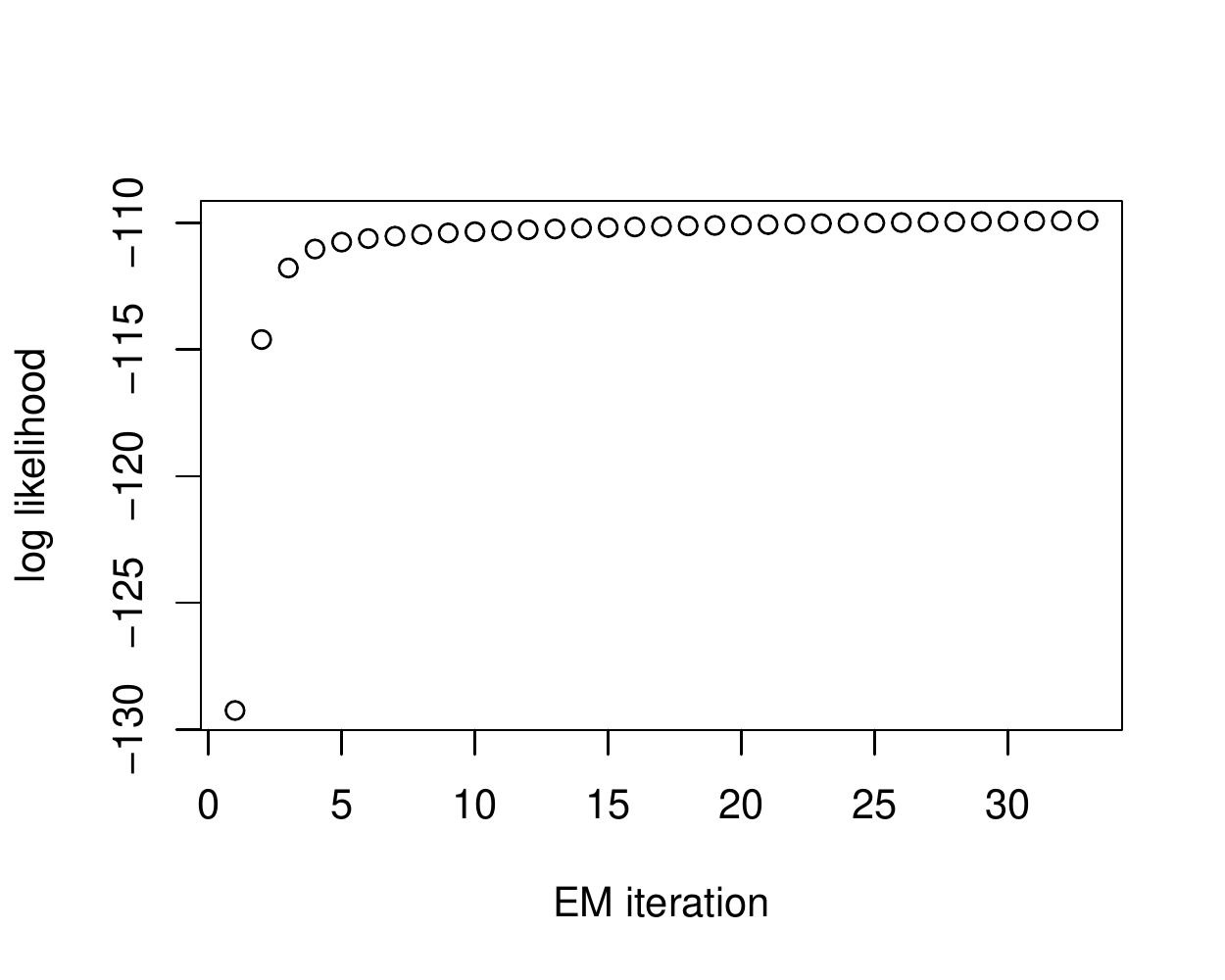} 

}

\caption{Convergence of the EM algorithm when using \pkg{FRK} with the \code{meuse} dataset.\label{fig:EM}}\label{fig:unnamed-chunk-19}
\end{figure}
\end{Schunk}

\vspace{0.1in}

\noindent {\bf Step 6:} Finally, we predict at all the BAUs with the fitted spatial model. This is done using the function \code{predict}. 
The argument \code{obs\_fs} dictates whether we attribute the fine-scale variation to intra-BAU systematic error (Case 1) or to the process model (Case 2). In the code below, we use the default setting and allocate it to the process model.

\begin{Schunk}
\begin{Sinput}
R> GridBAUs1 <- predict(S, obs_fs = FALSE)
\end{Sinput}
\end{Schunk}

\noindent The object \code{GridBAUs1} now contains the prediction vector, the prediction standard error, and the square of the prediction standard error at the BAU level in the fields \code{mu}, \code{sd}, and \code{var}, respectively. These can then be visualised using standard plotting commands.


\subsubsection*{Predicting over larger polygons/areas}

Now, assume that we wish to predict over regions encompassing several BAUs such that the matrix $\Cmat_P$ in \eqref{eq:CP} contains multiple non-zeros per row. We can create this larger regionalisation by using the function \code{auto\_BAUs} and specifying the cell size. This gives a `super-grid' shown in Figure~\ref{fig:PredictionPolygon}.

\begin{Schunk}
\begin{Sinput}
R> Pred_regions <- auto_BAUs(manifold = plane(), cellsize = c(600, 
+    600), type = "grid", data = meuse, convex = -0.05)
\end{Sinput}
\end{Schunk}

\noindent We carry out prediction on the super-grid by setting the \code{newdata} argument in the function \code{predict}, as given below.

\begin{Schunk}
\begin{Sinput}
R> Pred <- predict(S, newdata = Pred_regions, obs_fs = FALSE)
\end{Sinput}
\end{Schunk}

\noindent The predictions and the corresponding prediction standard errors on this super-grid are shown in Figure~\ref{fig:PredictionPolygon}.

\begin{Schunk}
\begin{figure}[t]
\includegraphics[width=\linewidth]{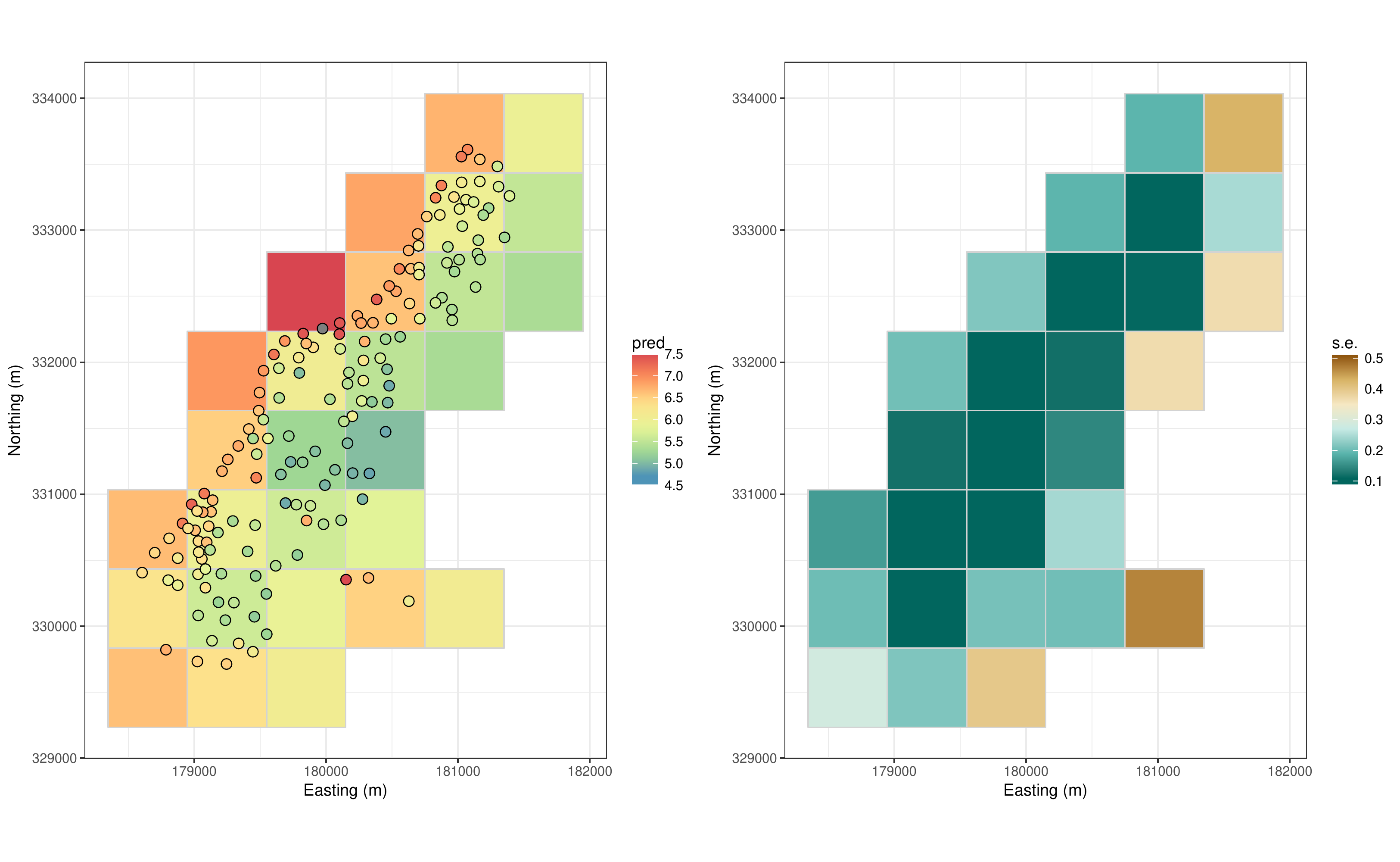} \caption{(Left panel) Prediction on a super-grid and overlayed observations for the \code{meuse} dataset.  (Right panel) Prediction standard errors obtained with \pkg{FRK} from the \code{meuse} dataset over the super-grid. Both quantities are in logs of ppm.\label{fig:PredictionPolygon}}\label{fig:unnamed-chunk-23}
\end{figure}
\end{Schunk}

\subsection{Computational considerations}

While FRK beats the curse of `data dimensionality' by dealing with matrices of size $r \times r$ instead of matrices of size $m \times m$, one must ensure that the number of basis functions, $r$, remains reasonably small. The reasons are two-fold. First, the computational time required to invert an $r \times r$ matrix increases cubicly with $r$, and several such inversions are required when running the EM algorithm. Second, it is likely that more EM-algorithm iterations are required when $r$ is large. In practice, $r$ should not exceed a few thousand. The number of basis functions $r$ can usually be controlled through the argument \code{nres}. The function \code{auto\_basis} also takes an argument \code{max\_basis} that automatically finds the number of resolutions required to not exceed the desired maximum number of basis functions.

The fitting and prediction algorithms scale linearly with the number of data points $m$ and the number of BAUs $N$. However, if one has millions of data points, then the number of BAUs must exceed this and a big-memory machine will probably be required. Irrespective of problem size, we have noted considerable improvements in speed when using the OpenBLAS libraries \citep{Wang_2013}.

Some of the operations in \pkg{FRK} can be run in parallel. To use a parallel back-end, one needs to set an option as follows:
\begin{Schunk}
\begin{Sinput}
R> opts_FRK$set("parallel", numcores)
\end{Sinput}
\end{Schunk}
\noindent where \code{numcores} is of class \code{integer} (e.g., \code{numcores = 4L} to use 4 cores). When this option is set, the \pkg{parallel} package is used to set up a parallel backend using \code{makeCluster}, which is subsequently used for parallel operations. Currently, parallelism is limited in \pkg{FRK} to
\begin{itemize}
  \item computing the integrals in \eqref{eq:S_integral} using Monte Carlo integration or, when appropriate, the approximation \eqref{eq:Sapprox};
  \item finding which data are influenced by which BAUs and computing the weights in \eqref{eq:meas_process}.   
\end{itemize}
\noindent Unfortunately the EM algorithm, which is the bottleneck in a spatial analysis using FRK, is serial in nature and difficult to parallelise. Hence, \code{SRE.fit} takes as argument \code{method}, in recognition that in the future other, possibly parallelisable, estimation methods might be implemented to speed up the fitting process.

\section{Comparison studies}\label{sec:comparison}

In this section we compare the utility of the SRE model in \pkg{FRK} to standard kriging using \pkg{gstat} \citep{Pebesma_2004}, and to two other popular models for modelling and predicting with large datasets in \proglang{R}: the LatticeKrig model that can be implemented with the package \pkg{LatticeKrig} \citep{LatticeKrig}, and the SPDE--GMRF model that can be implemented with the package \pkg{INLA} \citep{Lindgren_2015}. In both these models the spatial field is decomposed as
$$
\tilde{\lambda}(\svec) = \sum_{l = 1}^r\phi_l(\svec)\eta_l; \quad \svec \in D,
$$
and $\Kmat^{-1}$ is modelled in such a way that it is sparse. These two models allow for feasible computation with large $r$, however neither includes an extra fine-scale effect $\xi(\cdot)$. The SPDE--GMRF model has the added interpretable feature that, for a given set of basis functions, $\Kmat^{-1}$ is such that the resulting field approximates a Gaussian process with a stationary covariance function from the Mat{\'e}rn class.

In Section \ref{sec:spat_example}, we first analyse a 2D simulated dataset. We shall see that while \pkg{FRK} may sometimes perform less well in terms of prediction accuracy due to the practical limit on the number of basis functions it uses, it does not under-fit (i.e., it gives valid results) since fine-scale variation is taken into account. In fact, we see that the SRE model in \pkg{FRK} provides better coverage in terms of prediction intervals, even with large datasets, when compared to other models that use considerably more basis functions but that do not account for fine-scale variation. In the second case study (Section \ref{sec:AIRS}), we consider three days of column-averaged  carbon dioxide data from the Atmospheric InfraRed Sounder instrument on board the Aqua satellite \shortcites{Chahine_2006}.

\subsection{A 2D simulated dataset}\label{sec:spat_example}

Let $D = [0,1]\times [0,1] \subset \mathbb{R}^2$, and consider a process $Y(\cdot)$ with covariance function $\cov(Y(\svec),Y(\svec + \hvec)) \equiv \sigma^2\exp(-\| \hvec \| / \tau)$, where $\sigma^2$ is the marginal variance of the process and $\tau$ is the e-folding length-scale. Further, let $m$ be the number of observations and $\SNR$ be the signal-to-noise ratio, defined as the ratio of the marginal variance $\sigma^2$ to that of the measurement-error process, $\sigma^2_\epsilon$. In the inter-comparison, we consider cases where $m$ is either $1,000$ or $50,000$, $\SNR$ is 0.2, 1, or 5, and $\tau$ is either 0.015 or 0.15. These choices of parameters help highlight the strengths and weaknesses of FRK with respect to the other approaches. For example, due to the relatively small number $r$ of basis functions employed, we expect FRK to have lower prediction precision when the $\SNR$ is high and $\tau$ is low, but we expect the prediction intervals to be valid. We further split the domain into two side-by-side partitions, and we placed 95\% of the observations in the left half (LH) and 5\% in the right half (RH). This partitioning helps identify the different methods' capability of borrowing strength from a region with dense measurements to a region with sparse measurements. The measurement locations for the $m=1,000$ case are shown in Figure~\ref{fig:sim2}, left panel.

\begin{Schunk}
\begin{figure}
\includegraphics[width=\maxwidth]{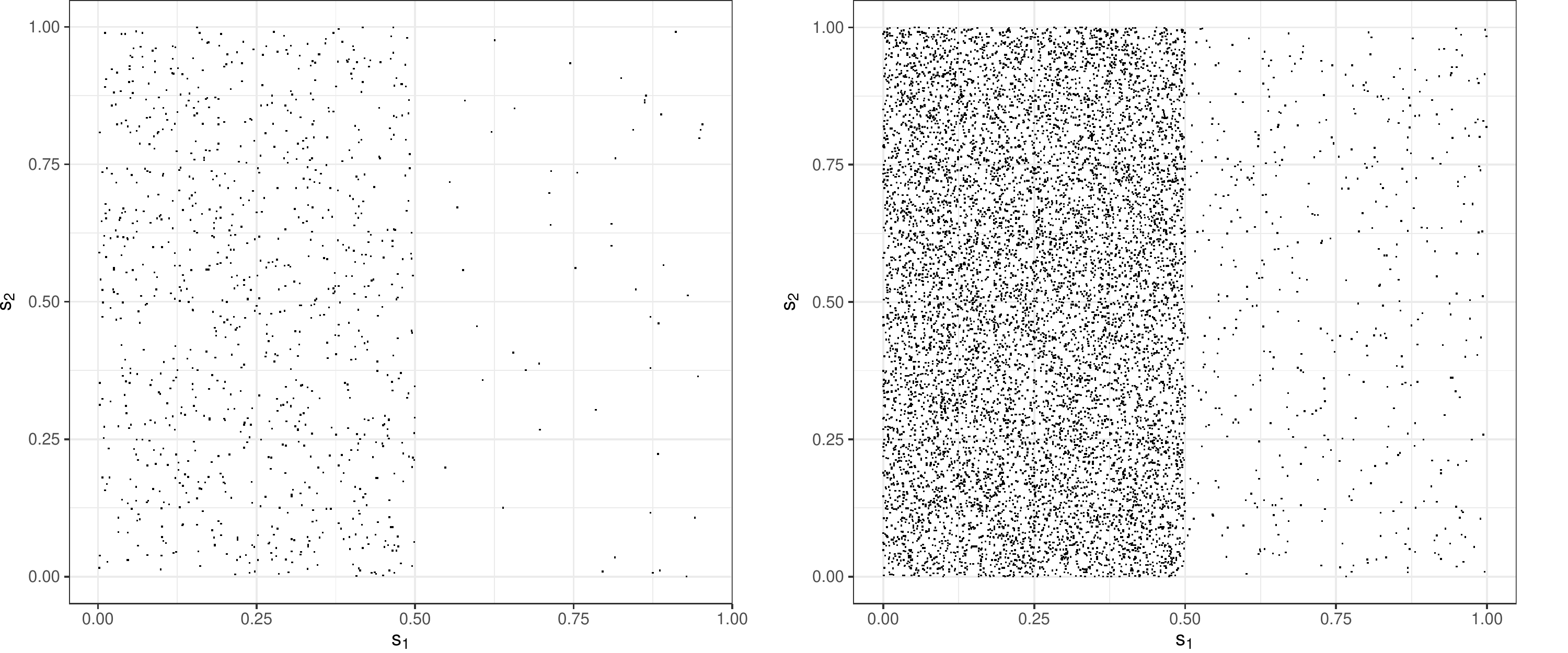} \caption{Measurement locations used in the experimental study. (Left panel) The number of locations $m=1,000$. (Right panel) The number of locations $m = 10,000$. These locations are fixed, for a given $m$, across simulations in the experiment.\label{fig:sim2}}\label{fig:unnamed-chunk-25}
\end{figure}
\end{Schunk}

We simulated the process on a 1,000 $\times$ 1,000 grid using the package \pkg{RandomFields} \citep{Schlather_2015}. We used the $10^6$ cells of the grid as our set of BAUs, $D^G$, and therefore each BAU was of size 0.001 $\times$ 0.001. One such spatial-process realisation for $\tau = 0.15$ and $\sigma^2 = 1$ is shown in Figure~\ref{fig:sim1}, left panel, while one with $\tau = 0.015$ and $\sigma^2 = 1$ is shown in Figure~\ref{fig:sim1}, right panel. With \pkg{gstat}, which we used to implement simple kriging (denoted $\gstat$), we assumed the true underlying covariance function was known. Hence, when available (for the $m=1,000$ case), the results of \pkg{gstat} should be taken as the gold standard. As $m$ gets larger, simple kriging quickly becomes infeasible, since it is $O(m^3)$ in computational complexity.

We implemented the LatticeKrig model (denoted $\LTK$) using the package \pkg{LatticeKrig}. We used \code{nlevel = 3} resolutions of Wendland basis functions, set the smoothness parameter \code{nu = 0.5}, and the number of grid points per spatial dimension at the coarsest resolution to \code{NC = 33}. The first resolution contained 1,849 basis functions, the second resolution contained 5,625, and the third resolution contained 19,321. In the case where $m=1,000$, we set \code{findAwght = TRUE} for the effective process range to be estimated by maximum likelihood methods. Setting \code{findAwght = TRUE} was prohibitive for $m=50,000$, but separate experiments showed that predictions from $\LTK$ were largely insensitive to this option for this value of $m$.

We implemented and fit the SPDE--GMRF model (denoted $\SPDE$) using the package \pkg{INLA}. We constructed a triangular mesh using \code{inla.mesh.2d} with \code{max.edge = c(0.05, 0.05)} and \code{cutoff = 0.012}. This gave a mesh with a higher density of basis function on the left-hand side of the domain and (as with the LatticeKrig model) a buffer to reduce edge effects. The basis functions are defined by the triangles, and their number was around 2,500 for $m = 1,000$, and 5,000 for $m = 50,000$, while the parameter $\alpha = 3/2$ was used to reproduce Gaussian fields with a Mat{\'e}rn  covariance function with smoothness parameter of 1/2 (i.e., an exponential covariance function). Unlike \pkg{LatticeKrig} and \pkg{FRK}, \pkg{INLA} uses an approximate Bayesian method for inference, and thus it requires the specification of prior distributions of the parameters, for which we use penalised complexity priors \citep{Fuglstad_2017}. For these simulation settings and our choice of prior distributions, we do not expect the inferential method to be a factor that largely influences the prediction and prediction errors.

For the SRE model implemented in \pkg{FRK} (denoted $\FRK$) we put a block-exponential covariance structure on $\Kmat_\circ(\varthetab)$ (\code{K\_type = "block-exponential"}), and we set \code{nres = 3}, yielding, in total, 819 basis functions regularly distributed in the domain $D$. In this study we used  \pkg{LatticeKrig} v6.4, \pkg{INLA} v17.06.20, and \pkg{FRK} v0.2.1.


For each configuration in the simulation experiment (i.e., the factorial design defined by $m \in \{1,000,50,000\}, \SNR \in \{0.2,1,5\},$ and $\tau \in \{0.015,0.15\}$), we simulated $L = 100$ datasets. For prediction locations we took $1,000$ locations at random on the left-hand side of the gridded domain $D^G$ that coincided with measurement locations, and another $1,000$ that did not; and we did the same for the right-hand side. When there were less than 1,000 measurement locations on a given side, all measurement locations were chosen as prediction locations for that side. The sets of locations are denoted as $D_{LH}^O$, $D_{LH}^M$, $D_{RH}^O$, and $D_{RH}^M$, respectively. These locations were kept constant across all simulation experiments for a given $m$.

In addition to the stationary, exponential, Gaussian process, we also simulated from the nonstationary process $Y^{\NS}(\cdot)$, where
\begin{equation}\label{eq:NS}
Y^{\NS}(\svec)= \frac{1}{2}\left\{Y_1(\svec)\sin(2\pi s_1)\cos(2\pi s_2) + Y_2(\svec)\sin(2\pi s_1)\right\}; \quad \svec = (s_1,s_2)^\top \in D,
\end{equation}
with $\cov(Y_1(\svec),Y_1(\svec + \hvec)) = \sigma_1^2\exp(-\| \hvec \| / \tau_1)$ and $\cov(Y_2(\svec),Y_2(\svec + \hvec)) = \sigma_2^2\exp(-\| \hvec \|^2 / \tau_2)$. For this additional experiment, we set $m = 10,000$, $\sigma_1 = \sigma_2 = 0.5$, and $\tau_1 = \tau_2 = 0.15,$ and we used all configurations in the original experiment as described above. The measurement locations for this case are shown in Figure~\ref{fig:sim2}, right panel.

As prediction-performance measures (`responses' of the experiment), we considered the following: \begin{itemize}
\item Root mean-squared prediction error: Let $\hat Y_X(\svec;l)$ denote the `model-$X$' predictor of $Y(\svec;l)$, where $Y(\svec;l)$ is the $l$th simulated process evaluated at location $\svec$ and $X = \gstat$, $\LTK$, $\SPDE$, $\FRK$. Then the model-$X$ predictor root-mean-squared prediction error for the $l$th simulation is
\begin{equation}\label{eq:RMSPE}
\RMSPE_X(l) \equiv \sqrt{\frac{1}{|D^*|}\sum_{\svec \in D^*} \left(\hat Y_X(\svec;l) - Y(\svec;l) \right)^2}; \quad l =1,\dots,L,
\end{equation}

where $D^* = D_{LH}^O, D_{LH}^M, D_{RH}^O,$ or $D_{RH}^M$. Since we are interested in benchmarking the model we use in \pkg{FRK}, we considered a measure of relative skill (RS), relative to $\FRK$:
\begin{equation*}
\RS_{X}(l) \equiv \RMSPE_{X}(l) / \RMSPE_{\FRK}(l);\quad l = 1,\dots,L,
\end{equation*}
where $X = \gstat, \LTK,\SPDE$. Hence, $\RS > 1$ $(< 1)$ indicates that $\FRK$ has better (worse) prediction accuracy.

\item Ninety-percent coverage: Let $\hat \sigma^2_X(\svec;l)$ denote the prediction variance under model $X$. Then, the ninety-percent coverage, $I^{90}_X(l)$, denotes the percentage of times $Y(\svec;l)$ lies in the interval $[\hat Y_X(\svec;l) - c\hat \sigma_X(\svec;l), \hat Y_X(\svec;l) + c\hat \sigma_X(\svec;l)],$ over $\svec \in D^*$. For the 90\% coverage measure, $c$ is the standard normal's inverse cumulative distribution function evaluated at $1 - (\frac{1}{2})(0.1) = 0.95$, which is 1.64 (rounded to two decimal places). That is,
\begin{equation}\label{eq:I90}
I^{90}_X(l) \equiv \frac{1}{|D^*|}\sum_{\svec \in D^*} \mathbb{I}(Y(\svec;l) \in [\hat Y_X(\svec;l) - c\hat \sigma_X(\svec;l), \hat Y_X(\svec;l) + c\hat \sigma_X(\svec;l)]); \quad l =1,\dots,L,
\end{equation}
where $D^* = D_{LH}^O, D_{LH}^M, D_{RH}^O, D_{RH}^M$, and where $X = \gstat, \LTK, \SPDE,\FRK$.
\end{itemize}
\noindent  We intentionally focus on coverage in order to highlight the strengths and weaknesses of the models in terms of uncertainty quantification. Other related measures, such as the Interval Score \citep{Gneiting_2007b}, penalise for both prediction interval width and poor coverage and are thus less suited to assess the issue of validity (i.e., whether the prediction intervals are correct, on average). The measures $\RS_{X}$ and $I^{90}_X$ were considered for $\{Y(\svec): \svec \in D^G\}$ simulated from the stationary process with exponential covariance function and from the nonstationary process $Y^{\NS}(\cdot)$ in \eqref{eq:NS}.


\begin{Schunk}
\begin{figure}
\includegraphics[width=\maxwidth]{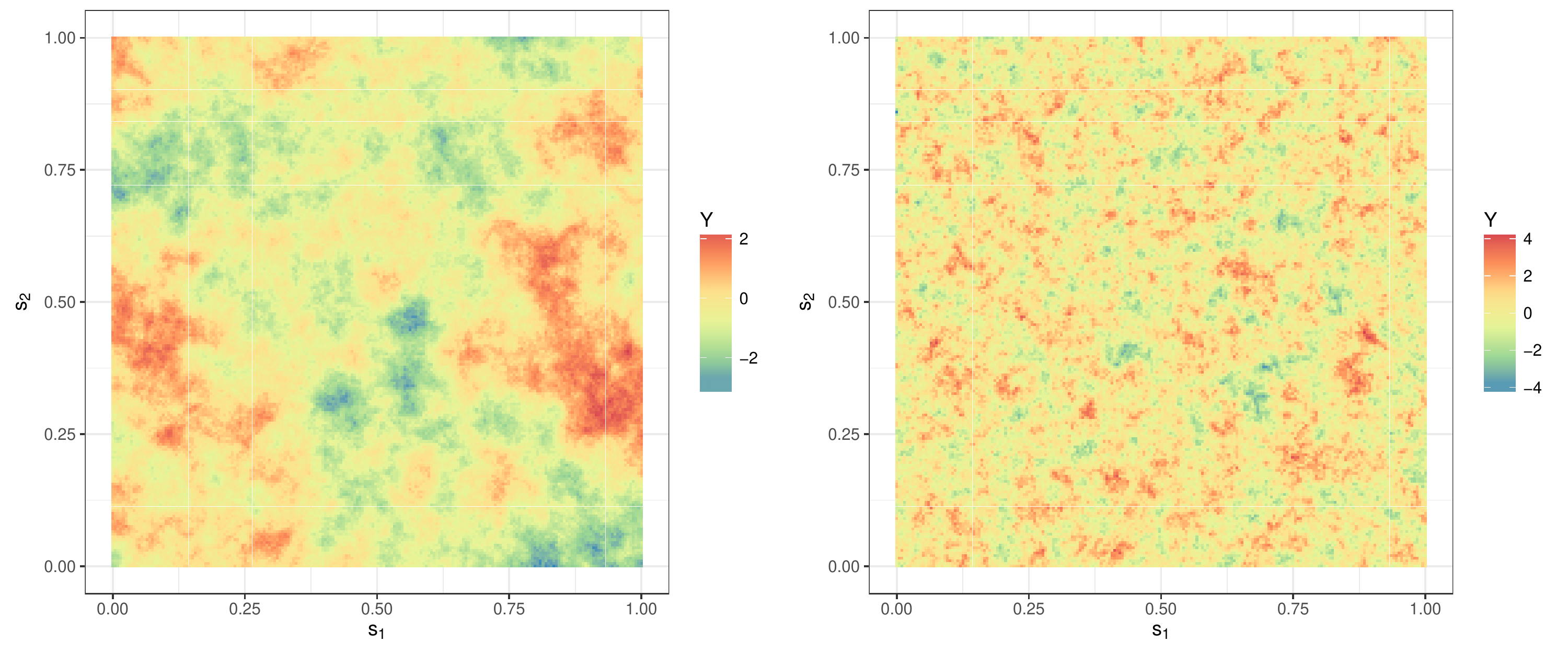} \caption{Simulations on a 1,000 $\times$ 1,000 grid $D^G$ from a stationary Gaussian process with exponential covariance function  $\cov(Y(\svec),Y(\svec + \hvec)) = \sigma^2\exp(-\| \hvec \| / \tau)$. (Left panel) The e-folding length scale $\tau = 0.15$.  (Right panel) The e-folding length scale $\tau = 0.015$. \label{fig:sim1}}\label{fig:unnamed-chunk-26}
\end{figure}
\end{Schunk}

Distributions of $\RS_X$ for the original experiment with $m=1,000$ and $m = 50,000$ are shown in Figures~\ref{fig:RMSPE1} and \ref{fig:RMSPE2}, respectively. While it is possible to proceed with an analysis of variance to analyse these results \citep[e.g.,][]{Zhuang_2014}, here we discuss their most prominent features. First, when there are few ($m=1,000$) data points (Figure~\ref{fig:RMSPE1}), there is little difference between the methods for low SNR but, for high SNR, $\SPDE$ and $\LTK$ perform better in terms of RS when $\tau$ is small ($\tau = 0.015$; see, for example, the bottom left two panels of Figure~\ref{fig:RMSPE1}). This was expected since the number of basis functions used begins to play an important role as the SNR increases \citep{Zammit_2018}. As expected, all prediction methods perform worse than or as well as, simple kriging with \pkg{gstat} (under a known covariance function).

The comparison in terms of RS is less clear when $m=50,000$ (Figure~\ref{fig:RMSPE2}). First, at unobserved locations, $\FRK$ is frequently outperformed in terms of RS by the other methods, since the relatively small number of basis functions is unable to adequately reconstruct the optimal (simple-kriging) predictor. On the other hand, at observed locations, the performance is $\SNR$ dependent and data-density dependent. In much of the design space, $\FRK$ performs worse (in terms of RS) than the other predictors at the measurement locations, but it begins to outperform $\SPDE$ and $\LTK$ as the $\SNR$ increases and when $\tau$ is small ($\tau = 0.015$). Now we turn to the question of `validity' of the predictors.

An equally important performance measure to RMSPE is coverage. Distributions of $I_X^{90}$ for $m=1,000$ and $m = 50,000$ are shown in Figures~\ref{fig:Cov1} and \ref{fig:Cov2}, respectively. In the small-data case $(m=1,000)$, all methods are over-confident (more so in the left-hand part of the domain) and by varying degrees. In the large-data case $(m=50,000)$, both \emph{SPDE} and \emph{LTK} perform poorly in terms of coverage, providing  over-confident predictions, especially when the $\SNR$ is large ($\SNR = 5$). This is a result of  these models relying on basis functions to reproduce the fine-scale variation and not attempting to separate out fine-scale variation from measurement error. The model implemented in \pkg{FRK} places a white-noise process at the BAU level to capture the fine-scale variation and can thus yield good coverage despite the use of a relatively low-dimensional manifold. It is worth nothing that it is straightforward in \pkg{INLA} to include an extra fine-scale-variation term and fix the standard deviation of the measurement error to some pre-specified value, although this is rarely done. Here we are illustrating that not doing this may lead to severe deleterious effects on coverage. The model used in \pkg{FRK} was also found to yield 90\% Interval Scores that were at least as good as, or better than, the other two models for the case with $m = 50,000$ (results not shown).

To further investigate this issue, we re-ran the simulations and generated coverage diagnostics for \emph{predicted data}, $\Yvec_P + \epsilonb_P$, rather than for just $\Yvec_P$. The coverage for all methods was very good (results not shown), indicating that all methods are able to correctly apportion \emph{total variability}. Consequently, these results show that inclusion of the fine-scale variation term is critical in reduced-rank approaches (irrespective of the number of basis functions) with large datasets when predicting the hidden process. (It is not critical when predicting missing data). The simple semivariogram method employed by \pkg{FRK} for estimating the measurement-error variance is a step in the right direction, and it appears to yield good results in the first instance. However, ideally, the standard deviation of the measurement error is known from the application and fixed \emph{a priori}.

Overall, all models have their own relative strengths and weaknesses, largely arising from the differences in (i) the type and number of basis functions employed, and (ii) the presence or otherwise of a fine-scale-process term. In this experiment we saw that the model employed by \pkg{FRK} produces predictions that are valid, on average. However for large-data situations, our experiment shows $\FRK$ predictions to be less efficient, as expected due to the restriction on the number of basis functions that can be used.

In the nonstationary case \eqref{eq:NS}, all methods performed similarly, with $\LTK$ being slightly overconfident and $\FRK$ being slightly underconfident; see Table \ref{tab:NS}. This similarity is not surprising since in \eqref{eq:NS} we set $\tau_1 = \tau_2 = 0.15$, which results in a process that is highly spatially correlated as well as rather smooth. The resulting process has a similar overall length scale and SNR to that simulated in the original experiment that yielded the results shown in the second row ($\SNR = 1$) of the third (LH, `10'), fourth (LH, `11'), seventh (RH, `10'), and eighth (RH, `11') columns of Figures~\ref{fig:RMSPE1} and \ref{fig:Cov1}. We see that all three methods performed similarly, and satisfactorily, in this case.

\begin{table}[ht]
\centering
\scalebox{0.7}{
\begin{tabular}{p{2cm}p{2cm}p{2cm}p{2cm}p{2cm}p{2cm}p{2cm}p{2cm}p{2cm}}
  \hline
 & RMSPE (LH obs) & RMSPE (LH unobs) & RMSPE (RH obs) & RMSPE (RH unobs) & 90\% cover. ~~~~~~~~~ (LH obs) & 90\% cover. ~~~~~~~~~ (LH unobs) & 90\% cover. ~~~~~~~~~ (RH obs) & 90\% cover. ~~~~~~~~~ (RH unobs) \\ 
  \hline
LTK & 0.13 & 0.12 & 0.24 & 0.25 & 0.86 & 0.87 & 0.88 & 0.87 \\ 
  SPDE & 0.13 & 0.12 & 0.24 & 0.25 & 0.90 & 0.89 & 0.91 & 0.88 \\ 
  FRK & 0.13 & 0.13 & 0.24 & 0.25 & 0.92 & 0.92 & 0.91 & 0.91 \\ 
   \hline
\end{tabular}
}
\caption{Root mean squared prediction error (RMSPE) and 90\% coverage for the case where $m = 10,000$ data were simulated from the nonstationary process $Y^{NS}(\cdot)$.\label{tab:NS}} 
\end{table}


\begin{Schunk}
\begin{figure}

{\centering \includegraphics[width=\maxwidth]{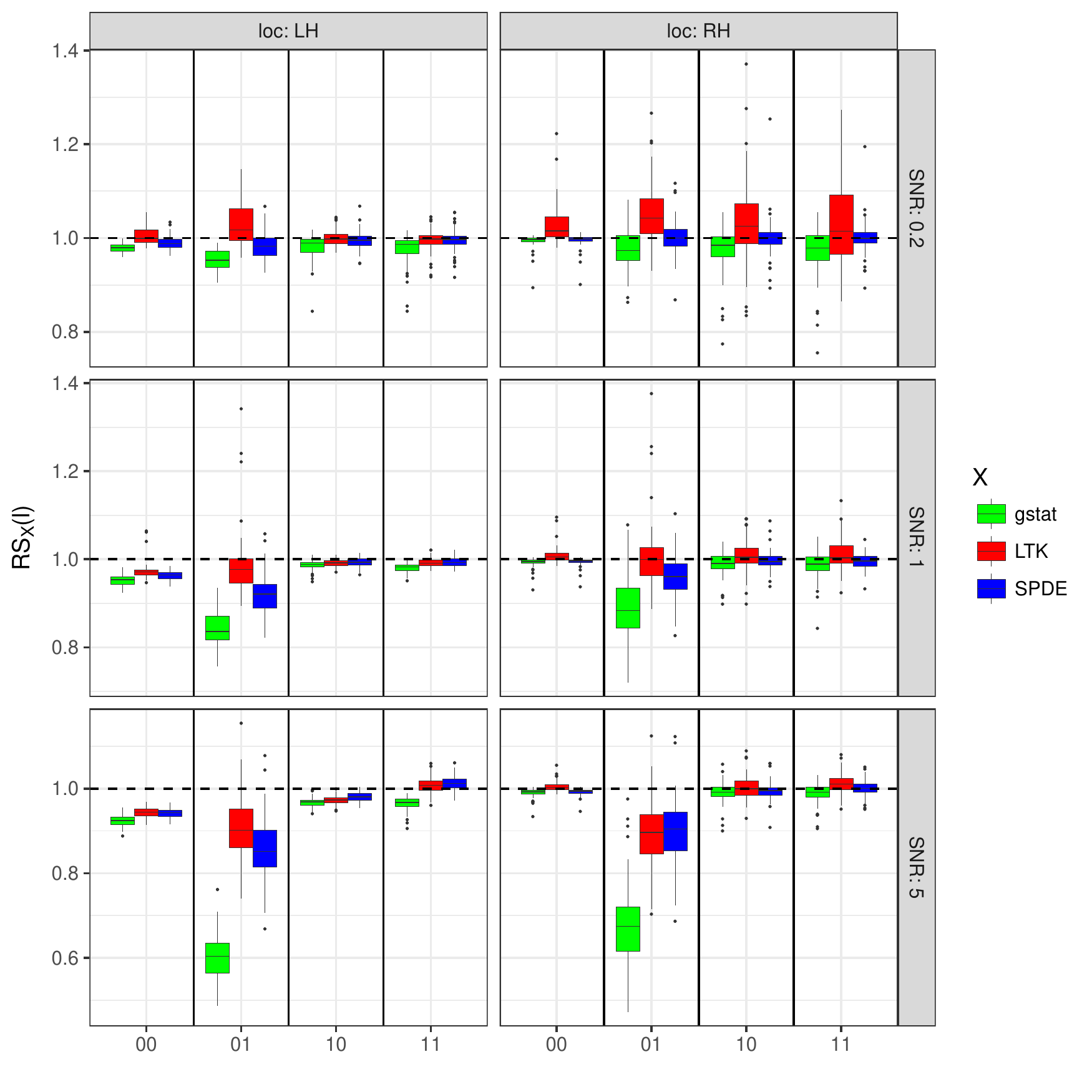} 

}

\caption{Boxplots of $\{\RS_X(l): l = 1,\dots,L\}$, $L = 100$, by model $X$ (box colour) for different SNR (signal-to-noise ratio), location (LH or RH), measurement-location coincidence, and process length scale, for the case where the number of data points $m = 1,000$. The $x$-axis labels are in the form $ab$ where $a$ is 0 if $\tau = 0.015$ and 1 if $\tau = 0.15$, and $b$ is 0 if the prediction locations do not coincide with the measurement locations and 1 if they do. The boxes denote the interquartile range, the whiskers extend to the last values that are within 1.5 times the interquartile range from the quartiles, and the dots show the values that lie beyond the end of the whiskers. Values of $RS_X$ larger (smaller) than 1 indicate superior (inferior) performance of $\FRK$; a dashed line is shown at $RS_X = 1$.\label{fig:RMSPE1}}\label{fig:unnamed-chunk-30}
\end{figure}
\end{Schunk}

\begin{Schunk}
\begin{figure}
\includegraphics[width=\maxwidth]{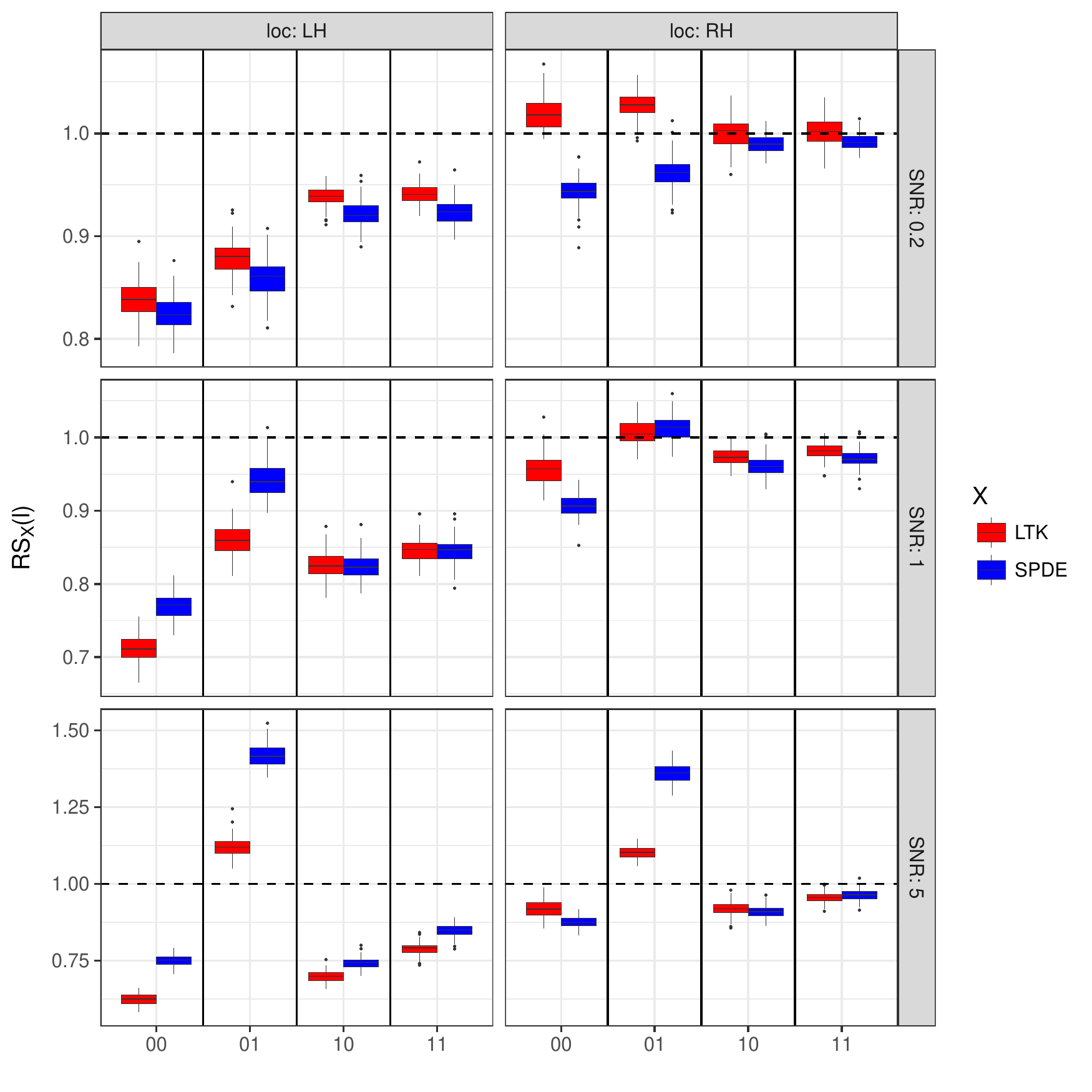} \caption{Same as Figure~\ref{fig:RMSPE1} but with the number of data points $m = 50,000$. Note that, in this case, simple kriging (with \gstat) is computationally prohibitive and hence does not appear in the figure. \label{fig:RMSPE2}}\label{fig:unnamed-chunk-31}
\end{figure}
\end{Schunk}


\begin{Schunk}
\begin{figure}

{\centering \includegraphics[width=\maxwidth]{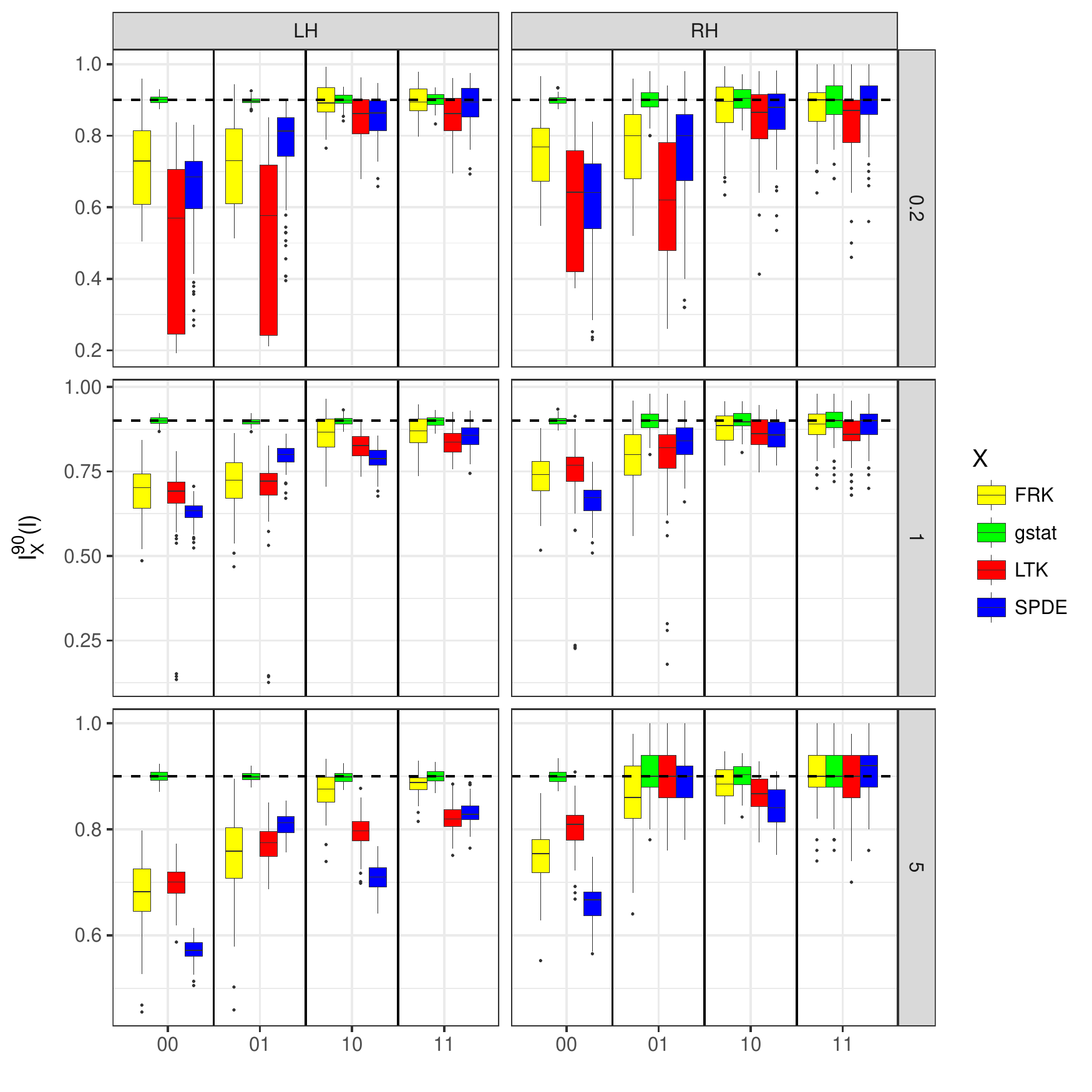} 

}

\caption{Boxplots of $\{I^{90}_X(l): l = 1,\dots,L\}$, $L = 100$, by model $X$ (box colour) for different SNR (signal-to-noise ratio), location (LH or RH), measurement-location coincidence, and process length scale, for the case where the number of data points $m = 1,000$. The $x$-axis labels are in the form $ab$ where $a$ is 0 if $\tau = 0.015$ and 1 if $\tau = 0.15$, and $b$ is 0 if the prediction locations do not coincide with the measurement locations and 1 if they do. The boxes denote the interquartile range, the whiskers extend to the last values that are within 1.5 times the interquartile range from the quartiles, and the dots show the values that lie beyond the end of the whiskers. The target value is 0.9 (dashed line). Values smaller than 0.9 indicate overconfidence of the prediction method. \label{fig:Cov1}}\label{fig:unnamed-chunk-32}
\end{figure}
\end{Schunk}

\begin{Schunk}
\begin{figure}
\includegraphics[width=\maxwidth]{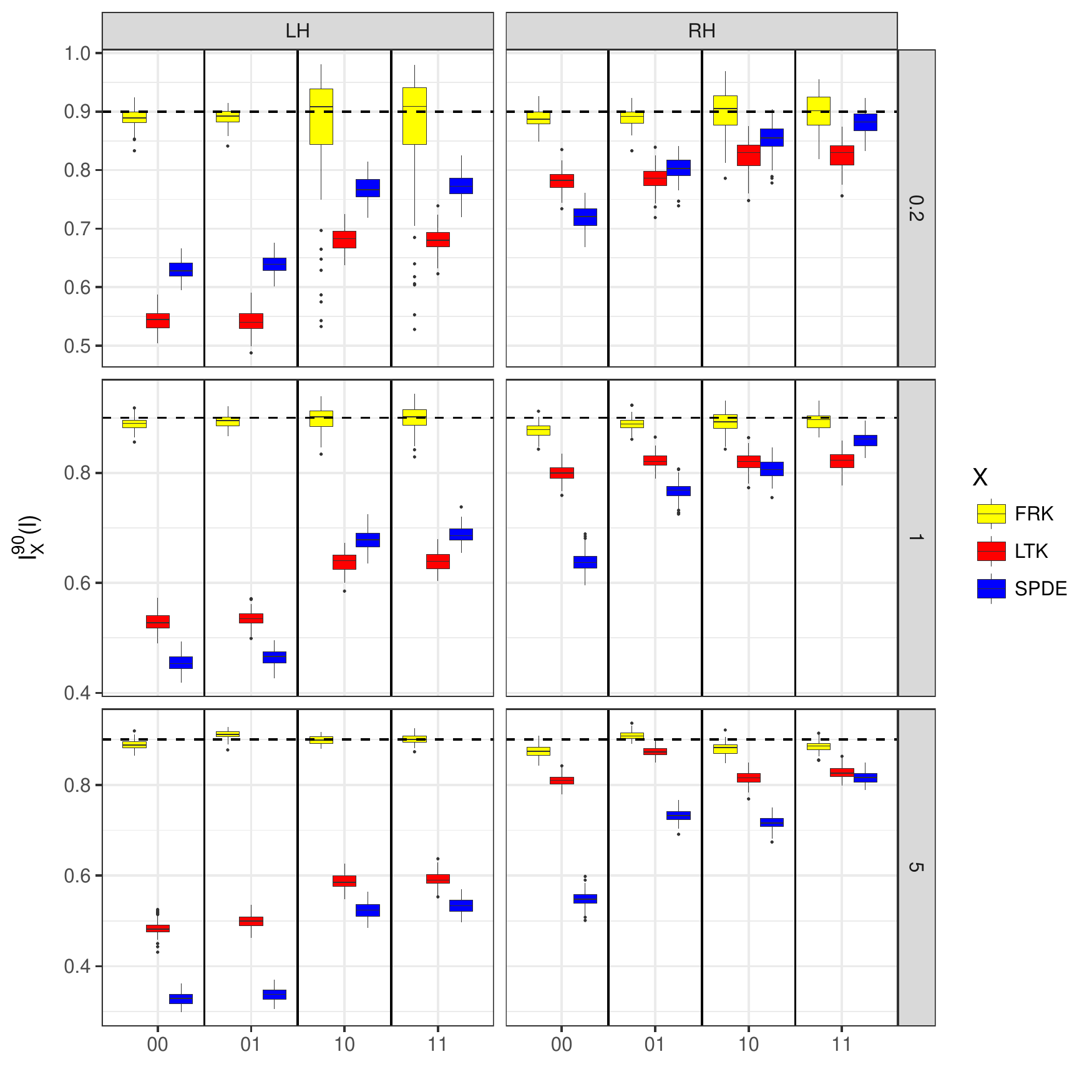} \caption{Same as Figure~\ref{fig:Cov1} but with the number of data points $m = 50,000$. Note that, in this case, simple kriging (with \gstat) is computationally prohibitive and hence does not appear in the figure. \label{fig:Cov2}}\label{fig:unnamed-chunk-33}
\end{figure}
\end{Schunk}

\subsection{Modelling and prediction with data from the AIRS instrument}\label{sec:AIRS}

The US National Aeronautics and Space Administration (NASA) launched the Aqua satellite on May 04 2002, with several instruments on board, including the Atmospheric Infrared Sounder (AIRS). AIRS retrieves column-averaged CO$_2$ mole fraction (in units of parts per million), denoted XCO$_2$ (with particular sensitivity in the mid-troposphere), amongst other geophysical quantities \citep{Chahine_2006}. The data we shall use consists of XCO$_2$ measurements taken between May 01 2003 and May 03 2003 (inclusive). These data are a subset of those available with \pkg{FRK}. We compare $\LTK, \SPDE$, and $\FRK$ on the three-day AIRS dataset, and we assess the utility of the methods on a validation dataset that we hold out.

Modelling on the sphere  with \pkg{FRK} proceeds in a very similar fashion to the plane, except that a coordinate reference system (CRS) on the surface of the sphere needs to be declared for the data. This is implemented using a \code{CRS} object with string \code{"+proj=longlat +ellps=sphere"}. We next outline the six steps required to fit these data using \pkg{FRK}.

\vspace{0.1in}

\noindent {\bf Step 1:} Fifteen days of XCO$_2$ data from AIRS (in May 2003) are loaded by using the command \code{data("AIRS\_05\_2003")}. In this case study, we subset the data to include only the first three days, which contains 43,059 observations of XCO$_2$ in parts per million (ppm). We subsequently divide the data into a training dataset of 30,000 observations, chosen at random (\code{AIRS\_05\_2003\_t}), and a validation dataset (\code{AIRS\_05\_2003\_v}) containing the remaining observations. To instruct \pkg{FRK} to fit the SRE model on the surface of a sphere, we assign the appropriate \code{CRS} object to the data as follows:

\begin{Schunk}
\begin{Sinput}
R> coordinates(AIRS_05_2003) <- ~lon + lat
R> proj4string(AIRS_05_2003) <- CRS("+proj=longlat +ellps=sphere")
\end{Sinput}
\end{Schunk}

\vspace{0.1in}

\noindent {\bf Step 2:} The next step is to create BAUs. This is done using the \code{auto\_BAUs} function but this time with the manifold specified to be the surface of a sphere with radius equal to that of Earth. We also specify that we wish the BAUs to form an ISEA Aperture 3 Hexagon (ISEA3H) discrete global grid (DGG) at resolution 9 (for a total of 186,978 BAUs). Resolutions 0--6 are included with \pkg{FRK}; higher resolutions are available in the package \pkg{dggrids} available from \code{https://github.com/andrewzm/dggrids}. By default, this will create a hexagonal grid on the surface of the sphere, however it is also possible to have the more traditional lon-lat grid by specifying \code{type = "grid"} and declaring a \code{cellsize} in units of degrees. An example of an ISEA3H grid, at resolution 5 (which would yield a total of 6,910 BAUs), is shown in Figure~\ref{fig:sphere_BAUs}, left panel, while a $5^\circ \times 5^\circ$ lon-lat grid using \code{type = "grid"} is  shown in Figure~\ref{fig:sphere_BAUs}, right panel.

\begin{Schunk}
\begin{Sinput}
R> isea3h_sp_poldf <- auto_BAUs(manifold = sphere(), isea3h_res = 9, 
+    type = "hex", data = AIRS_05_2003)
R> isea3h_sp_poldf$fs <- 1
\end{Sinput}
\end{Schunk}

\vspace{0.1in}

\noindent {\bf Step 3:} Now the basis functions are constructed, again of type \code{"bisquare"}, with three resolutions, to yield a total of 1,176 basis funcions; see Figure~\ref{fig:sphere_basis}, which was generated using the function \code{show_basis}.

\begin{Schunk}
\begin{Sinput}
R> G <- auto_basis(manifold = sphere(), data = AIRS_05_2003_t, nres = 3, 
+    isea3h_lo = 2, type = "bisquare")
\end{Sinput}
\end{Schunk}

\begin{Schunk}
\begin{figure}[t!]
\includegraphics[width=\maxwidth]{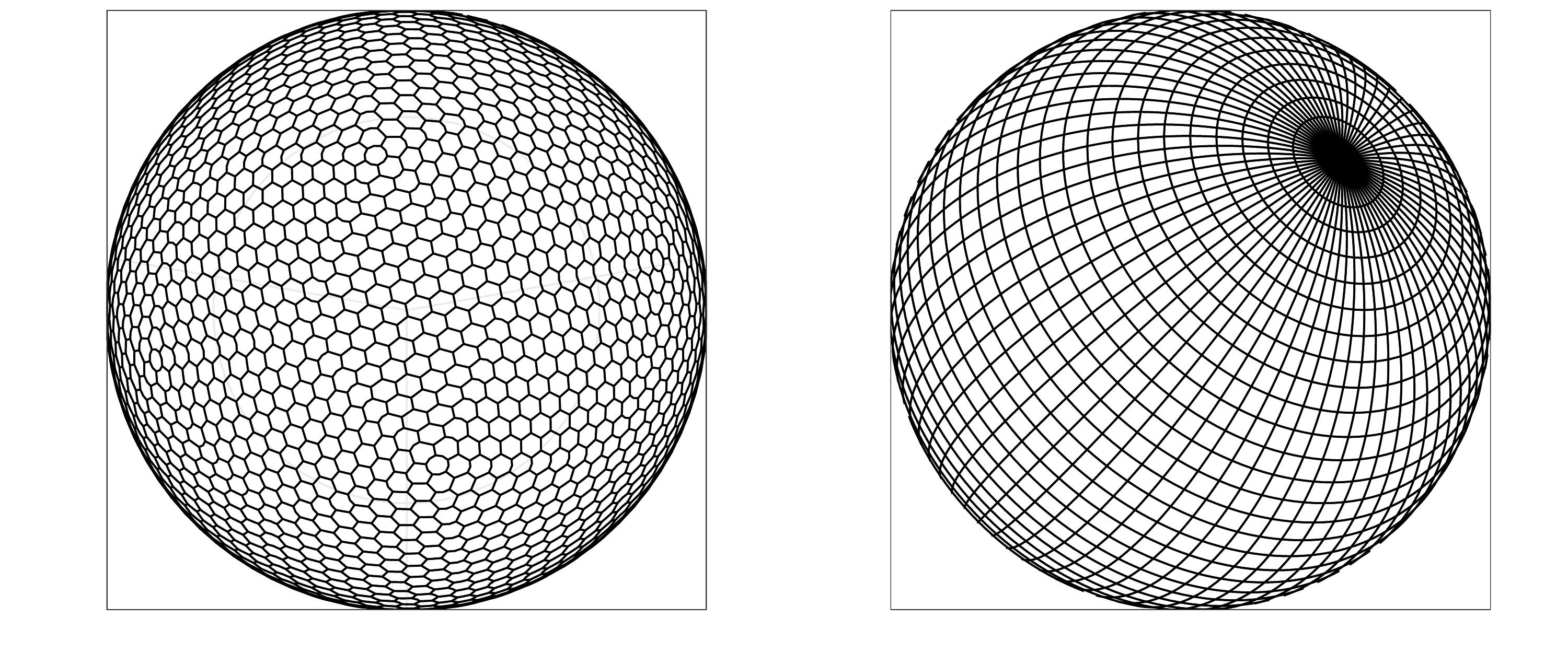} \caption{BAUs constructed on the surface of the sphere. (Left panel) BAUs are the ISEA3H hexagons (for visualisation purposes we show resolution 5). (Right panel) BAUs are a $5^\circ \times 5^\circ$ lon-lat grid.\label{fig:sphere_BAUs}}\label{fig:unnamed-chunk-41}
\end{figure}
\end{Schunk}

\begin{Schunk}
\begin{figure}[t!]
\includegraphics[width=\maxwidth]{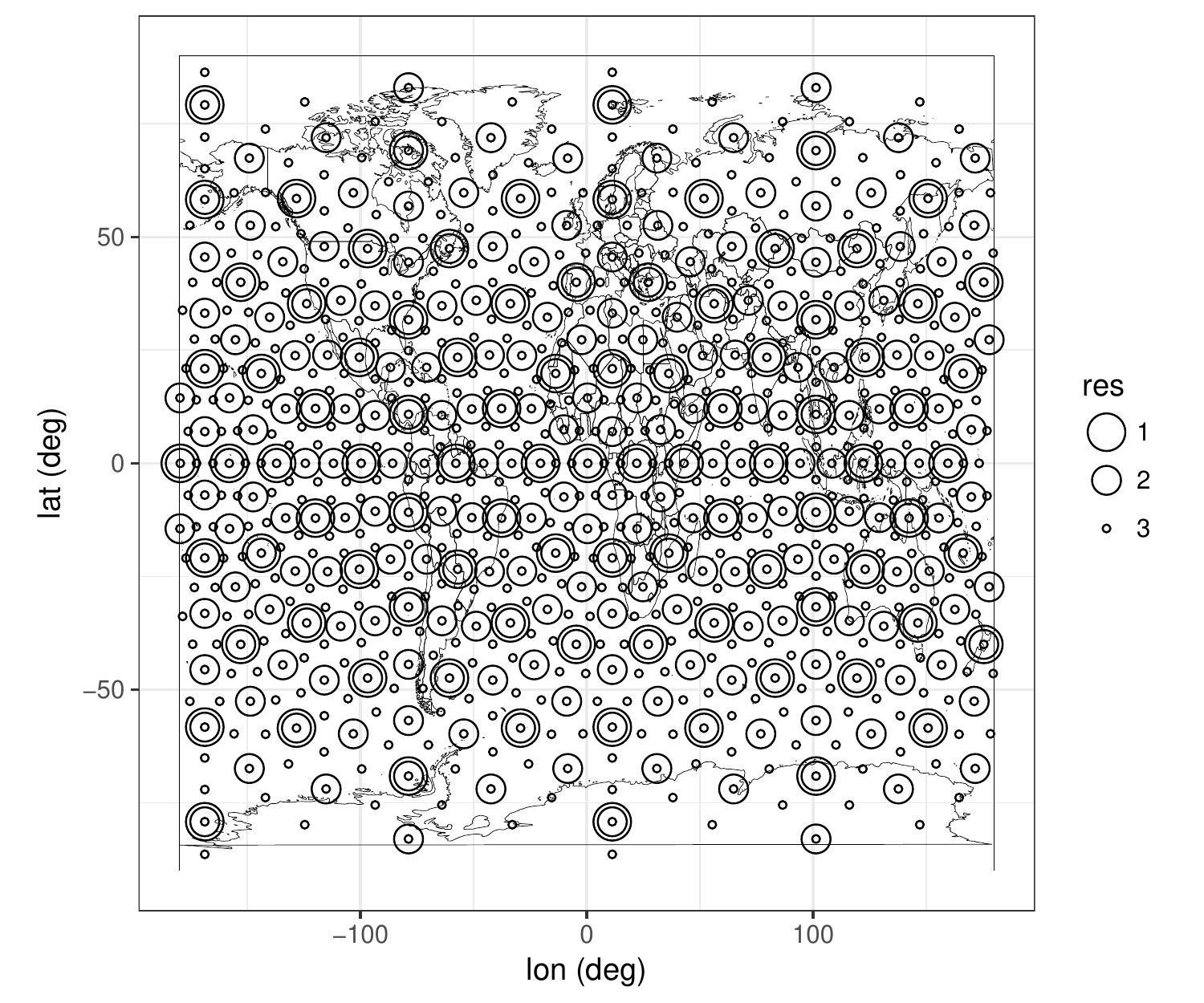} \caption{Basis functions used in modelling and predicting with the \code{AIRS} data. The basis-function centroids are constructed using the function \code{auto\_basis}. The diameter of the circles are indicative of the resolution the basis functions belong to.\label{fig:sphere_basis}}\label{fig:unnamed-chunk-42}
\end{figure}
\end{Schunk}
\vspace{0.1in}

\noindent {\bf Steps 4--5:} Since XCO$_2$, a column-averaged CO$_2$ mole fraction in units of ppm, has a latitudinal gradient, we use latitude as a covariate in our model. The SRE object is then constructed in the same way as in Section \ref{sec:advanced}. The AIRS footprint is approximately 50 km in diameter, which is smaller than the BAUs we use (approximately 100 km in diameter), and hence it is possible that multiple observations fall into the same BAU.  Recall from Section \ref{sec:advanced} (Step 4) that when multiple data points fall into the same BAU that these are correlated through either intra-BAU systematic error (Case 1) or fine-scale process variation at the BAU level (Case 2). However, recall also that when multiple observations fall into the same BAU, the matrices $\Vmat_{\xi,Z}$ and $\Vmat_{\delta,Z}$ are sparse but not diagonal, and this can increase computational time considerably. For large datasets in which each datum has relatively small (relative to the BAU) spatial support, such as the \code{AIRS} dataset, it is frequently reasonable to let the argument \code{average\_in\_BAU = TRUE} (as it is by default) to indicate that one wishes to summarise the data at the BAU level.

In the code below we implement FRK using the default Case (Case 2, where $\sigma^2_\delta = 0$ and $\sigma^2_\xi$ is estimated).

\begin{Schunk}
\begin{Sinput}
R> f <- co2avgret ~ lat
R> S <- SRE(f = f, list(AIRS_05_2003_t), basis = G, BAUs = isea3h_sp_poldf, 
+    est_error = TRUE, average_in_BAU = FALSE)
R> S <- SRE.fit(SRE_model = S, n_EM = 1000, tol = 0.1, print_lik = FALSE)
\end{Sinput}
\end{Schunk}


\vspace{0.1in}

\noindent {\bf Step 6:} To predict at the BAU level, we invoke the \code{predict} function.

\begin{Schunk}
\begin{Sinput}
R> pred_isea3h <- predict(S)
\end{Sinput}
\end{Schunk}

\begin{figure}[t!]
\begin{center}
\includegraphics[width=0.8\textwidth]{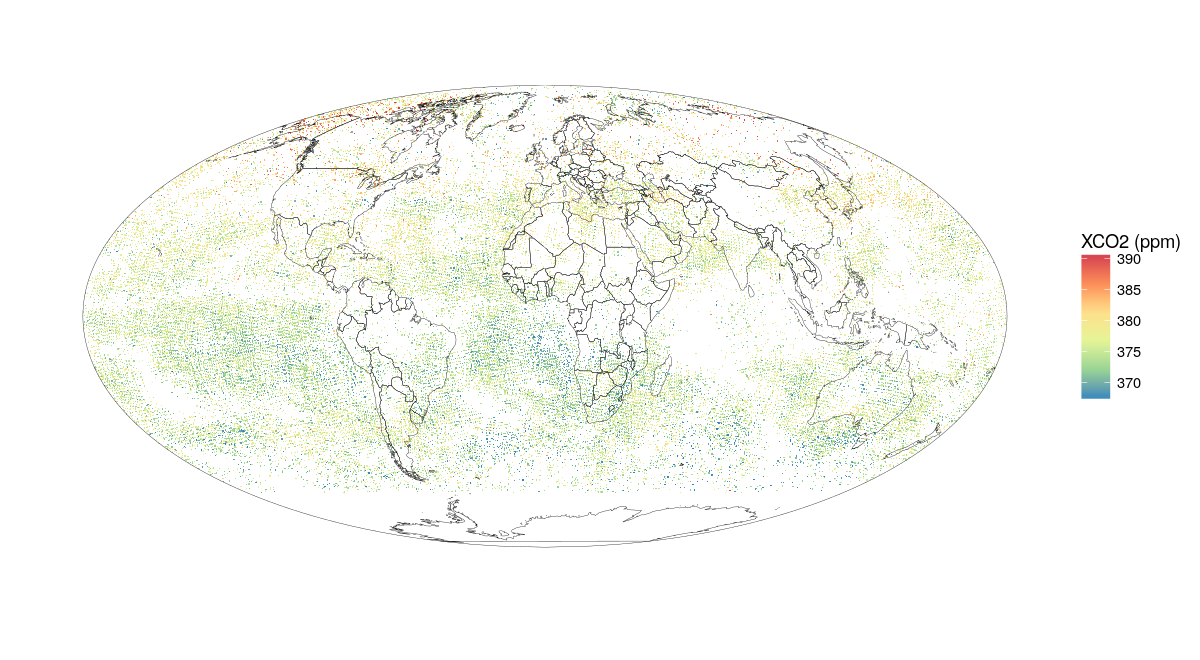}
\includegraphics[width=0.8\textwidth]{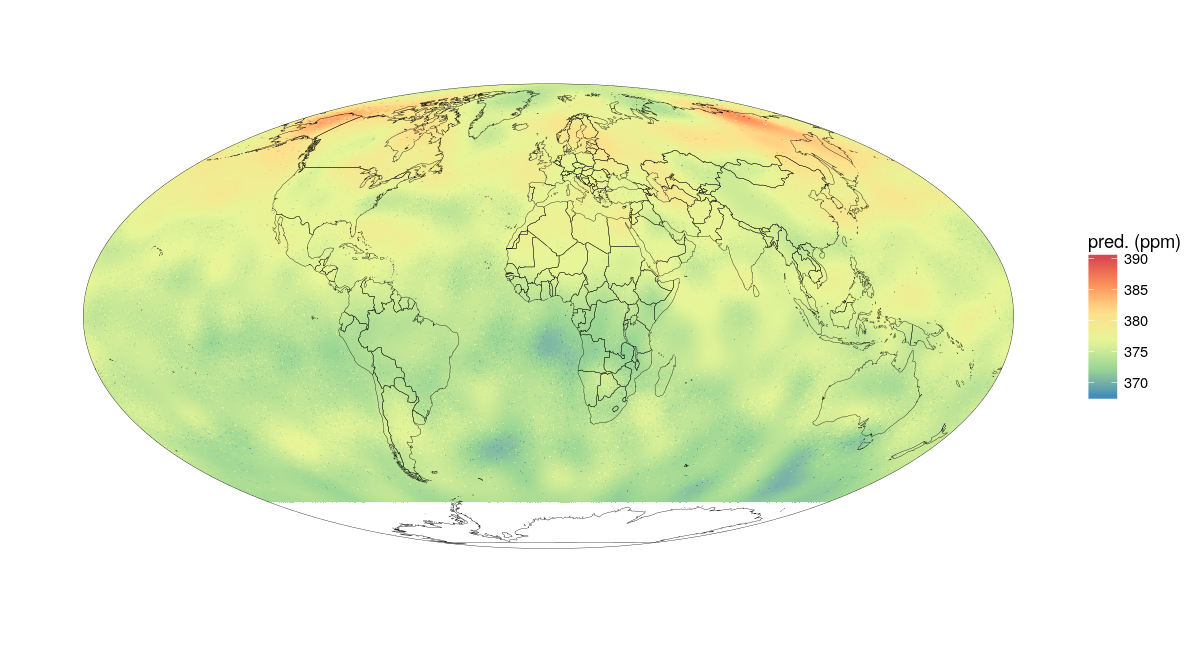}
\includegraphics[width=0.8\textwidth]{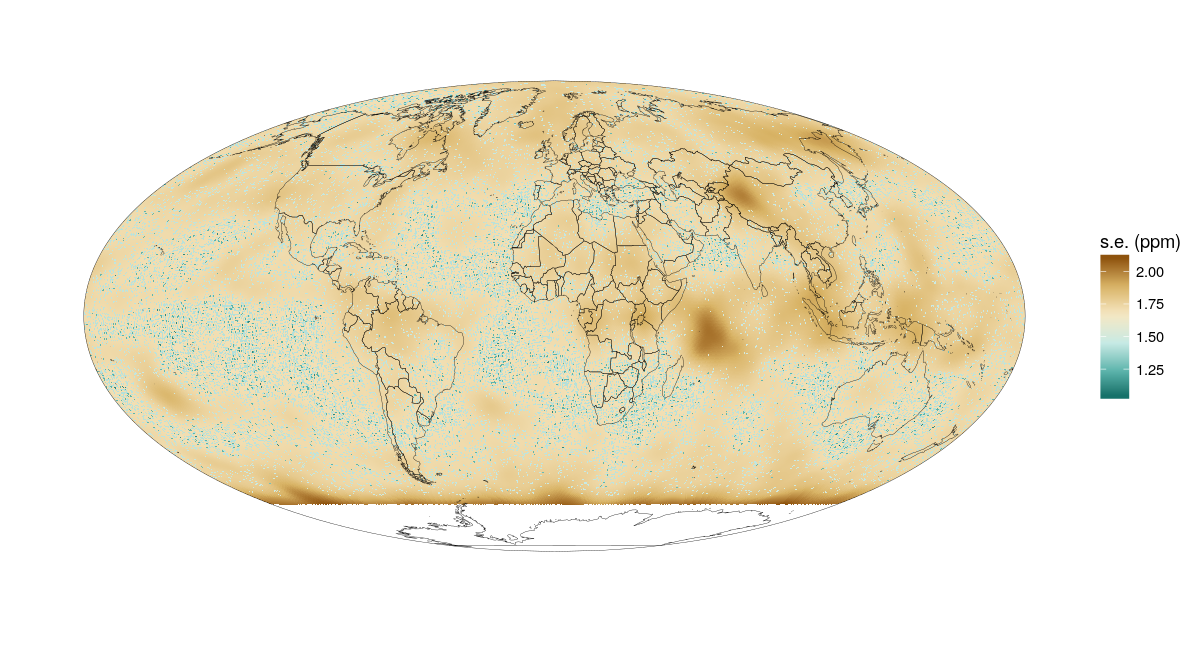}
\end{center}
\caption{(Top panel) XCO$_2$ data in parts per million (ppm) from the AIRS instrument between May 01 2003 and May 03 2003 (inclusive). (Middle panel) Prediction of $\Yvec_P$ in ppm using \pkg{FRK}. (Bottom panel) Prediction standard error of $\Yvec_P$ in ppm using \pkg{FRK}. Note that AIRS does not release data at latitudes below 60$^\circ$S. \label{fig:AIRSresults1}}
\end{figure}

\noindent The prediction and prediction standard error maps obtained using \pkg{FRK}, together with the observations, are shown in Figure~\ref{fig:AIRSresults1}. We denote the implementation above of FRK as FRK-Ma, where ``M'' denotes the case for the modelled $\var(\etab) = \Kmat_\circ(\varthetab)$ and ``a'' denotes the case for \code{average\_in\_BAU} set to \code{FALSE}.

    We evaluated the utility of the SRE model used in \pkg{FRK}, the LatticeKrig model (with \pkg{LatticeKrig}), and the SPDE--GRMF model (with {\bf INLA}), using out-of-sample prediction at the validation-data locations. We also re-ran \pkg{FRK} with \code{average\_in\_BAU} set to \code{TRUE} (denoted FRK-Mb) and \code{K\_type} = \code{"unstructured"} (FRK-V). With \pkg{INLA} we approximated an SPDE with $\alpha = 2$ on a global mesh of 6,550 basis functions and used penalised complexity prior distributions for the parameters \citep{Fuglstad_2017}. With \pkg{LatticeKrig} we used three resolutions with a total of 12,703 basis functions on $\mathbb{R}^2$ using the lon-lat coordinates to denote spatial locations. As comparison measures we used the RMSPE \eqref{eq:RMSPE} between the validation values and their respective predicted observations, the continuous ranked probability score \citep[CRPS,][]{Gneiting_2007}, and the actual coverage of a 90\% prediction interval \eqref{eq:I90} but obtained with respect to the validation data instead of the process $Y$ at the validation-data locations.

The results are summarised in Table \ref{tab:AIRSresults}.  In this example, we see that there is little practical difference in performance between the five methods despite the large difference in the number of basis functions and the form of the models; FRK performs about 2\% worse than the others in terms of RMSPE.  As expected, since we are validating against \emph{data} (and not against the true \emph{process}, which is unknown here), all methods perform acceptably in capturing total variation. However, the FRK methods gave prediction standard errors of the \emph{process} that were, on average, double those provided by $\LTK$ and $\SPDE$. This mirrors what was seen in Section \ref{sec:comparison}, where $\SPDE$ and $\LTK$ were generally overconfident, although in this case the true process is unknown and one can only speculate the reasons for the validation errors' behaviours. Nevertheless, the simulations in Section \ref{sec:spat_example} indicate FRK's accurate coverage of the process $Y$.

Computation time for all three packages were similar under the chosen configurations (except for FRK-Ma that assumes intra-BAU correlations). For {\bf FRK}, we computed the predictions and prediction standard errors directly using sparse-matrix operations, while we used \pkg{INLA}'s predictor functionality to obtain the prediction and prediction standard errors for the SPDE--GMRF model. We obtained \pkg{LatticeKrig}'s prediction errors using 30 conditional simulations.

\begin{table}[ht]
\centering
\begin{tabular}{rrrrrr}
  \hline
 & INLA & LTK & FRK-Ma & FRK-Mb & FRK-V \\ 
  \hline
RMSPE & 3.08 & 3.10 & 3.14 & 3.13 & 3.16 \\ 
  CRPS & 1.71 & 1.72 & 1.74 & 1.73 & 1.75 \\ 
  90\% coverage & 0.91 & 0.91 & 0.90 & 0.90 & 0.89 \\ 
  Time (s) & 74.00 & 336.00 & 1122.00 & 270.00 & 94.00 \\ 
   \hline
\end{tabular}
\caption{Root-mean-squared prediction error (RMSPE), continuous-ranked probability score (CRPS), and computational times for the different methods (see text for details). Computational times for \pkg{INLA} report the time taken to compute the \code{inla()} function; for \pkg{LatticeKrig} the time to fit the model, predict and generate 30 conditional simulations; and for \pkg{FRK} the time to generate the basis functions, the BAUs, the EM estimates of the SRE models (with an EM convergence tolerance of 0.1) and predictions. \label{tab:AIRSresults}} 
\end{table}

\section{Change-of-support, anisotropy, and custom basis functions}

Sections 1--4 introduced the core spatial functionality of \pkg{FRK}. The purpose of this section is to present additional functionality that may be of use to the spatial analyst.

\subsection{Multiple observations with different supports}

In \pkg{FRK}, one can make use of multiple datasets with different spatial supports with little difficulty. Consider the \code{meuse} dataset. We synthesise observations with a large support by changing the \code{meuse} object into a \code{SpatialPolygonsDataFrame}, where each polygon is a square of size 300 m $\times$ 300 m centred around the original \code{meuse} data point (see Figure \ref{fig:meuse_large}, left panel). For reference, the constructed BAUs are of size 50 m $\times$ 50 m. Once this object is set up, which we name \code{meuse\_pols}, we assign zinc values to the polygons by fitting a spherical semi-variogram model to the log zinc concentrations in the original \code{meuse} dataset, generating a realisation by conditionally simulating once at the BAU centroids, exponentiating the simulated values, aggregating accordingly, and adding on measurement error with variance 0.01. The analysis proceeds in precisely the same way as in Section \ref{sec:usage}, but with \code{meuse\_pols} used instead of \code{meuse}.

The predictions and the prediction standard errors using \code{meuse\_pols} are shown in Figure~\ref{fig:meuse_large}, centre and right panels, respectively. We note that the supports of the observations and the BAUs do not precisely overlap: Recall that, for simplicity, we assumed that an observation is taken to overlap a BAU if and only if the centroid of the BAU lies within the observation footprint. (A refinement of this simple method will require a more detailed consideration of the BAU and observation footprint geometry and is the subject of future work.)

\begin{Schunk}
\begin{figure}[t]
\includegraphics[width=\linewidth]{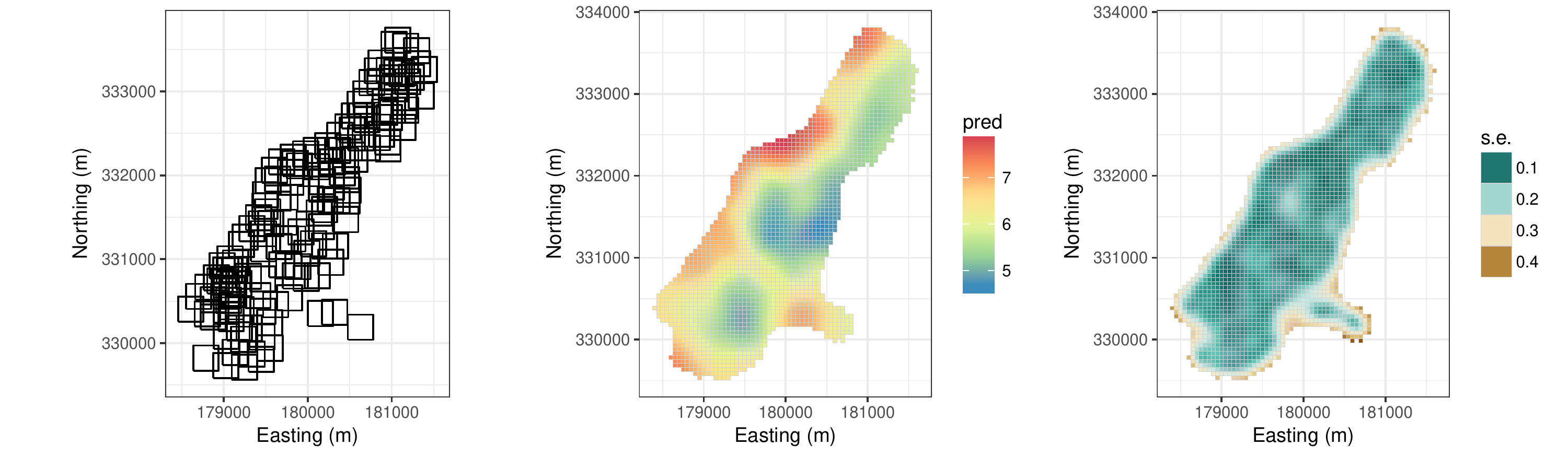} \caption{Data footprints, predictions and prediction standard errors obtained with \pkg{FRK} using the \code{meuse} dataset in logs of ppm, where each observation (black square) is assumed to have a spatial footprint of 300 m $\times$ 300 m. The BAUs are 50 m $\times$ 50 m in size. (Left panel) The spatial footprints of the synthesised data. (Centre panel) FRK predictions at the BAU level. (Right panel) FRK prediction standard errors at the BAU level. \label{fig:meuse_large}}\label{fig:unnamed-chunk-61}
\end{figure}
\end{Schunk}

\subsection{Anisotropy: Changing the distance measure}

Highly non-stationary and anisotropic fields may be easier to model on a deformed space on which the process is approximately stationary and isotropic \citep[e.g., ][]{Sampson_1992, Kleiber_2016}.  In \pkg{FRK}, a deformation can be introduced to capture geometric anisotropy by changing the distance measure associated with the manifold. As an illustration, we simulated an anisotropic, noisy, spatial process on a fine grid in $D = [0,1] \times [0,1]$ and assumed that 1,000 randomly-located grid points were observed. The process and the sampled data are shown in Figure~\ref{fig:aniso1}.

\begin{Schunk}
\begin{figure}[t]
\includegraphics[width=\linewidth]{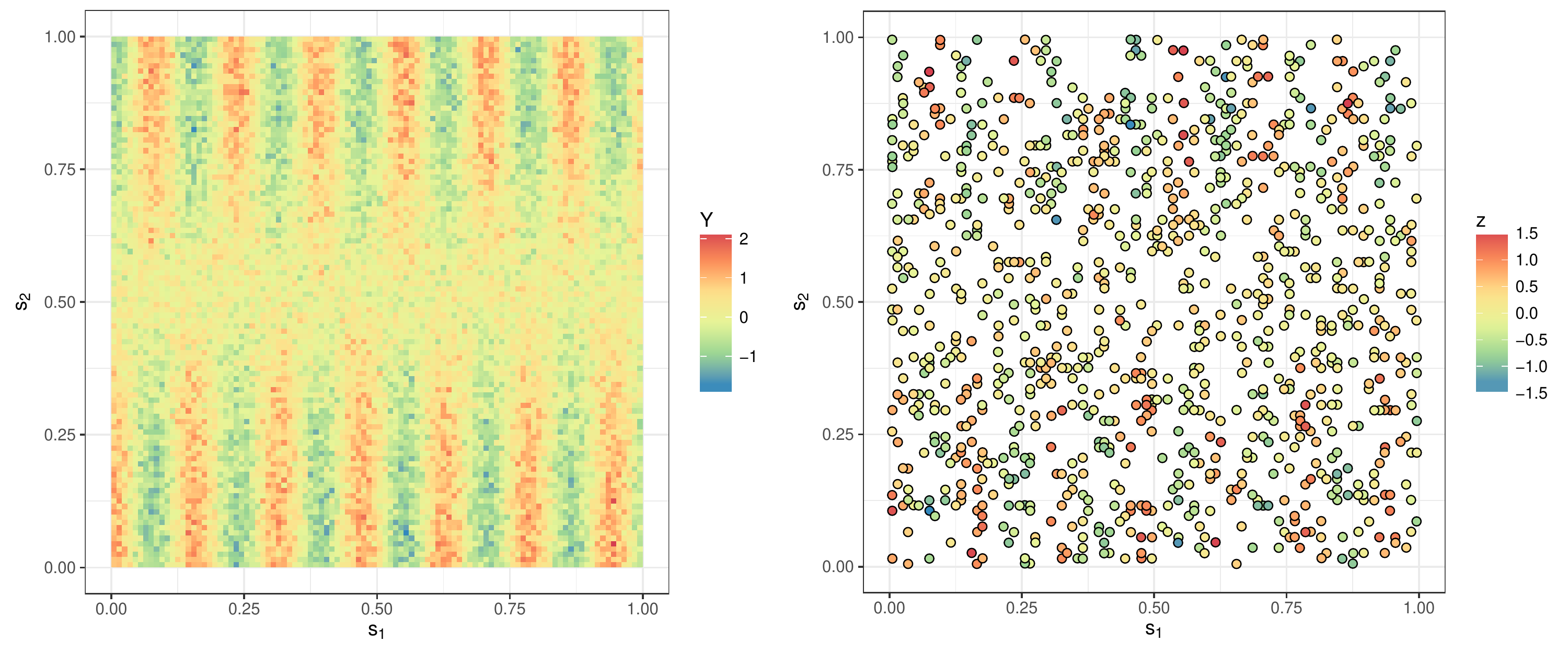} \caption{FRK with anisotropic fields. (Left panel) Simulated process. (Right panel) Observations at 1,000 randomly chosen grid points. \label{fig:aniso1}}\label{fig:unnamed-chunk-63}
\end{figure}
\end{Schunk}

In this simple case, to alter the modified distance measure we note that the spatial frequency in $x$ is approximately four times that in $y$. Therefore, in order to generate anisotropy, we use a measure that scales $x$ by 4. In \pkg{FRK}, a \code{measure} object requires a distance function and the dimension of the manifold on which it is used, and it is constructed as follows:
\begin{Schunk}
\begin{Sinput}
R> dist_fun <- function(x1, x2 = x1) {
+    scaler <- diag(c(4, 1))
+    distR(x1 %*% scaler, x2 %*% scaler)
+}
R> asymm_measure <- measure(dist = dist_fun, dim = 2L)
\end{Sinput}
\end{Schunk}

The distance function can be assigned to the manifold as follows.

\begin{Schunk}
\begin{Sinput}
R> TwoD_manifold <- plane(measure = asymm_measure)
\end{Sinput}
\end{Schunk}

We now generate a grid of basis functions (here at a single resolution) manually. First, we create a $5 \times 14$ grid on $D$, which we will use as centres for the basis functions. We then call the function \code{local\_basis} to construct bisquare basis functions centred at these locations with a range parameter (i.e., the radius in the case of a bisquare) of 0.4. Due to the scaling used, this implies a range of 0.1 in $x$ and a range of 0.4 in $y$. Basis-function number 23 is shown in Figure~\ref{fig:anisobasis}.

\begin{Schunk}
\begin{Sinput}
R> basis_locs <- expand.grid(seq(0, 1, length = 14), seq(0, 1, length = 5))
R> G <- local_basis(manifold = TwoD_manifold, loc = as.matrix(basis_locs), 
+    scale = rep(0.4, nrow(basis_locs)), type = "bisquare")
\end{Sinput}
\end{Schunk}

\begin{Schunk}
\begin{figure}[t]

{\centering \includegraphics[width=0.5\linewidth]{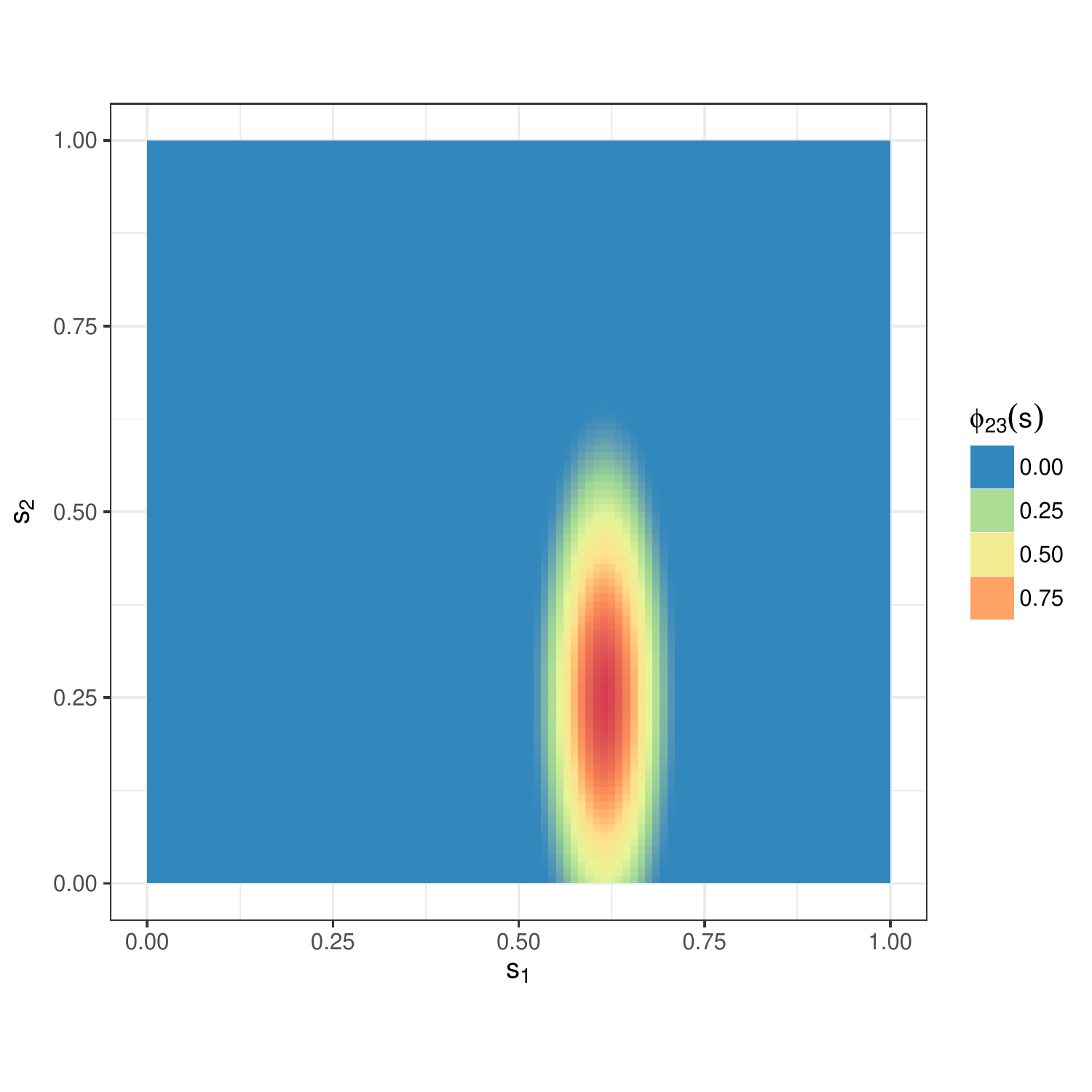} 

}

\caption{Basis-function number 23 of the 70 spatial basis functions constructed to fit an anisotropic spatial field.\label{fig:anisobasis}}\label{fig:unnamed-chunk-67}
\end{figure}
\end{Schunk}

From here on, the analysis proceeds in exactly the same way as given in the other examples. The predictions and prediction standard errors are shown in Figure~\ref{fig:aniso2}. Note that complicated distance functions need to be obtained offline, for instance using multi-dimensional scaling \citep{Sampson_1992}.

\begin{Schunk}
\begin{figure}[t]
\includegraphics[width=\linewidth]{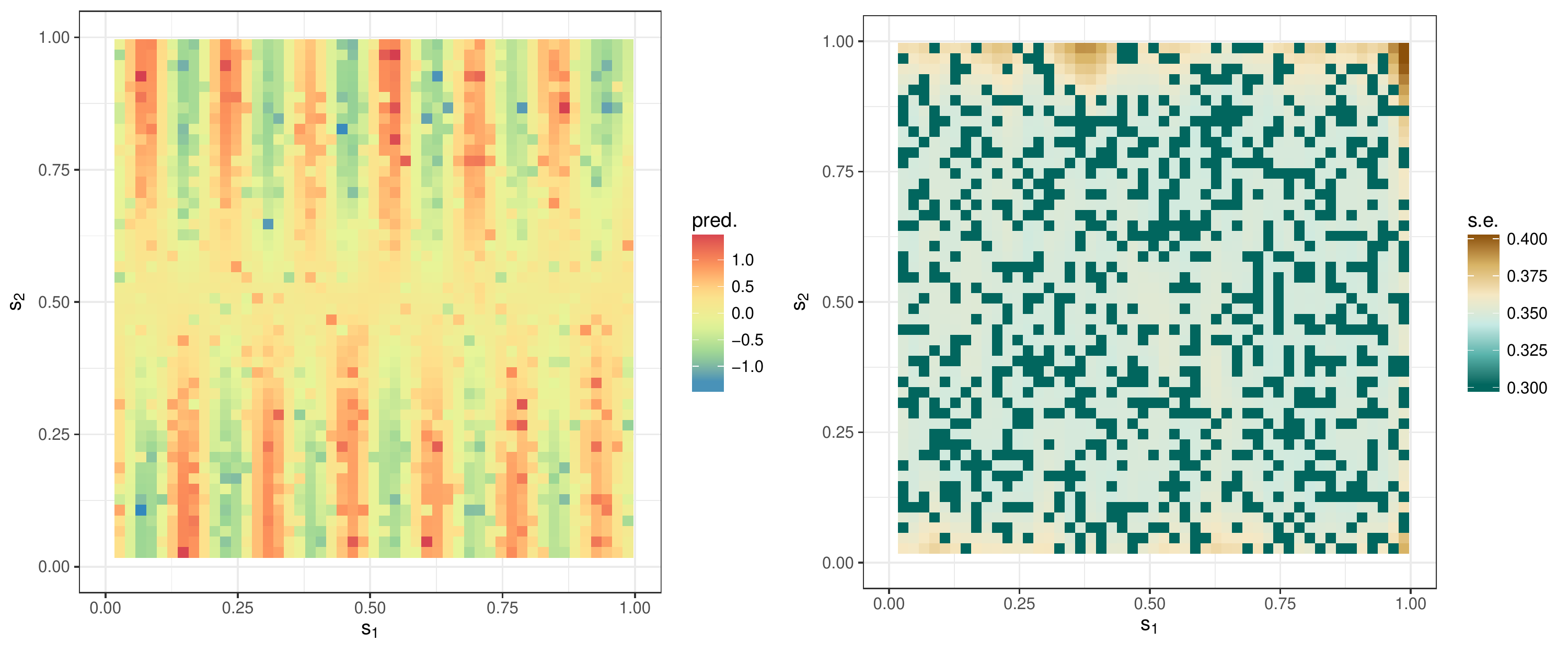} \caption{FRK using data shown in Figure~\ref{fig:aniso1} (generated by an anisotropic field). (Left panel) FRK predictions. (Right panel) FRK prediction standard errors.\label{fig:aniso2}}\label{fig:unnamed-chunk-69}
\end{figure}
\end{Schunk}

\subsection{Customised basis functions and BAUs} \label{sec:custom_basis}

The package \pkg{FRK} provides the functions \code{auto\_BAUs} and \code{auto\_basis} to help the user construct the BAUs and basis functions based on the data that are supplied. However, these could be done manually. The object containing the basis functions needs to be of class \code{Basis} that defines 5 slots:
\begin{itemize}
\item \code{dim}: The dimension of the manifold.
\item \code{fn}: A list of functions. By default, distances used in these functions (if present) are attributed to a manifold, but arbitrary distances can be used.
\item \code{pars}: A list of parameters associated with each basis function. For the local basis functions used in this paper (constructed using \code{auto\_basis} or \code{local\_basis}), each list item is another list with fields \code{loc} and \code{scale} where \code{length(loc)} is equal to the dimension of the manifold and \code{length(scale)} is equal to 1.
\item \code{df}: A data frame with number of rows equalling the number of basis functions and containing auxiliary information about the basis functions (e.g., resolution number).
\item \code{n}: An integer equal to the number of basis functions.
\end{itemize}
The constructor \code{Basis} can be used to instantiate an object of this class.

There are less restrictions for constructing BAUs; for spatial applications, they are usually either \code{SpatialPixelsDataFrame}s or \code{SpatialPolygonsDataFrame}s. In a spatio-temporal setting, the BAUs need to be of class \code{STFDF}, where the spatial component is usually either a \code{SpatialPixels} or a \code{SpatialPolygons} object. In either case, the \code{data} slot of the object must contain
\begin{itemize}
\item all covariates used in the model;
\item a field \code{fs} containing elements proportional to the fine-scale variation at the BAU level; and
\item fields that can be used to summarise the BAU as a point, typically the centroid of each polygon. The names of these fields need to equal those of the \code{coordnames(BAUs)} (typically \code{c("x", "y")} or \code{c("lon", "lat")}).
\end{itemize}

\section[Spatio-temporal FRK]{Spatio-temporal \pkg{FRK}}\label{sec:ST}

Fixed rank kriging in space and time is different from fixed rank filtering \citep{Cressie_2010}, where a temporal autoregressive structure is imposed on the temporally evolving basis-function weights $\{\etab_t \}$, and where Rauch--Tung--Striebel smoothing is used for inference on $\{\etab_t \}$. In \pkg{FRK}, the basis functions can also have a temporal dimension; then the only new aspect beyond the purely spatial analysis given above is specifying these spatio-temporal basis functions. We describe how this can be done in Section \ref{sec:STbasis}. In Section \ref{sec:OCO2} we show how this can be applied to modelling and prediction with data from the more recent Orbiting Carbon Observatory-2 (OCO-2) satellite that measures XCO$_2$.

\subsection{Basis-function construction}\label{sec:STbasis}

\pkg{FRK} allows for kriging in space and time through the use of spatio-temporal basis functions constructed through the tensor product of spatial basis functions with temporal basis functions. Specifically, consider a set of $r_s$ spatial basis functions $\{\phi_{p}(\svec): p = 1,\dots,r_s\}$ and a set of $r_t$ temporal basis functions $\{\psi_{q}(t): q = 1,\dots,r_t\}$. Then we construct the set of spatio-temporal basis functions as $\{\phi_{st,p_{st}}(\svec,t) : p_{st} = 1,\dots,r_sr_t\} = \{\phi_{p}(\svec)\psi_{q}(t) : p = 1,\dots,r_s; q = 1,\dots,r_t\}$.

To illustrate their construction, consider the following set of spatial basis functions.

\begin{Schunk}
\begin{Sinput}
R> centroids <- as.matrix(expand.grid(x = 1:3, y = 1:3))
R> G_spatial <- local_basis(manifold = plane(), loc = centroids, 
+    scale = rep(2, 9), type = "bisquare")
\end{Sinput}
\end{Schunk}

The function call above returns a set of bisquare basis functions centred at \code{loc} with aperture equal to \code{scale}. The same call, given below, can be used to construct temporal basis functions; note that now \code{manifold = real\_line()}, and we choose Gaussian basis functions instead (in which case \code{scale} represents the standard deviation). As in the spatial case, other basis functions (such as bisquare) can also be used.

\begin{Schunk}
\begin{Sinput}
R> G_temporal <- local_basis(manifold = real_line(), loc = matrix(seq(2, 
+    28, by = 4)), scale = rep(3, 7), type = "Gaussian")
\end{Sinput}
\end{Schunk}

\begin{Schunk}
\begin{figure}
\includegraphics[width=\maxwidth]{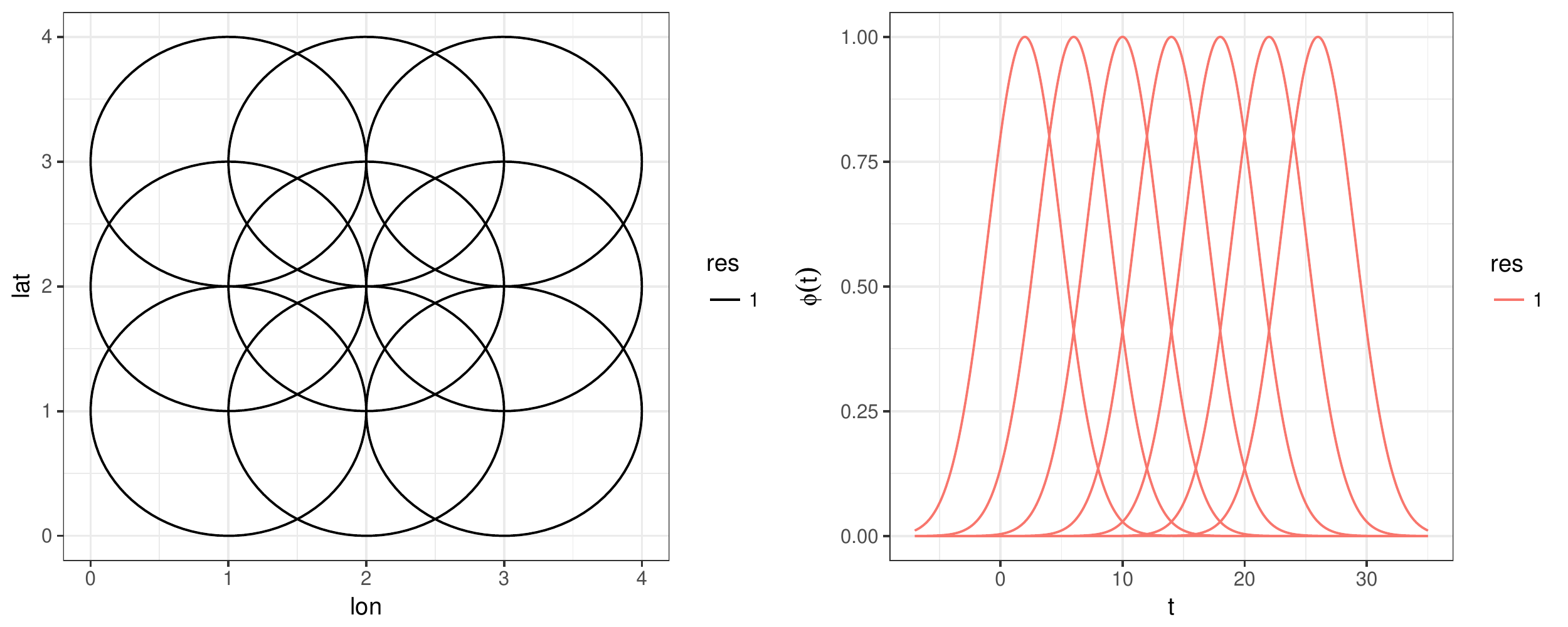} \caption{Basis functions used to construct the spatio-temporal basis functions. (Left panel) Locations of bisquare spatial basis functions (circles represent the apertures, here all equal). (Right panel) Temporal basis functions.\label{fig:basis}}\label{fig:unnamed-chunk-73}
\end{figure}
\end{Schunk}

The generated basis functions can be visualised using \code{show\_basis}; see Figure~\ref{fig:basis}. The spatio-temporal basis functions are then constructed using the function \code{TensorP} as follows:

\begin{Schunk}
\begin{Sinput}
R> G <- TensorP(G_spatial, G_temporal)
\end{Sinput}
\end{Schunk}

The object \code{G} can be subsequently used for constructing SRE models, as in the spatial case. Note that since we have nine spatial basis functions and seven temporal basis functions, we have 63 spatio-temporal basis functions in total. Care should be taken that the total number of basis functions does not become prohibitively large (say $> 4000$).

\subsection{Global prediction of column-averaged Carbon Dioxide from OCO-2}\label{sec:OCO2}

The NASA OCO-2 satellite was launched on July 02 2014, and it produces between 100,000 and 300,000 usable retrievals per day. Between the beginning of October 2014 and the end of February 2017, the satellite produced around 100 million retrievals. The specific data product we used was the OCO-2 Data Release 7R Lite File Version B7305Br \citep{OCO2_2017}. Following pre-processing, we reduced the number of data entries in the product to around 50 million. Each retrieval produces a number of variables; in this example, we consider the most commonly used one, XCO$_2$, the column-averaged mole fraction in ppm. Obtaining reliable global predictions of XCO$_2$ does not require consideration of all the data simultaneously. The atmosphere mixes quickly within hemispheres, and temporal correlation-length scales are on the order of days. Hence, we consider the OCO-2 data in a moving-window of 16 days, and for each 16 days we fit a spatio-temporal SRE model in order to obtain a global prediction of XCO$_2$ in the middle (i.e., the 8th day) of the window. We use this moving-window approach to obtain daily XCO$_2$ global prediction and prediction errors, between October 01 2014 and March 01 2017.

We first put the OCO-2 data into an \code{STIDF} object, before using the function \code{auto\_BAUs} to construct spatio-temporal voxels. The following code constructs gridded BAUs and instructs \pkg{FRK} to use 1 day as the smallest temporal unit and to limit the latitude grid to the minimum and maximum latitude of the OCO-2 data locations, rounded to the nearest degree.

\begin{Schunk}
\begin{Sinput}
R> BAUs <- auto_BAUs(manifold = STsphere(), data = STobj, type = "grid", 
+    tunit = "days", cellsize = c(1, 1, 1), xlim = c(-180, 180), 
+    ylim = c(floor(min(oco2data$lat)), ceiling(max(oco2data$lat))))
\end{Sinput}
\end{Schunk}

The code given above generated around 45,000 BAUs per day, for a total of around 720,000 BAUs per 16-day batch. We then constructed 396 spatial basis functions on the globe using

\begin{Schunk}
\begin{Sinput}
R> G_spatial <- auto_basis(manifold = sphere(), data = as(STobj, 
+    "Spatial"), nres = 3, type = "bisquare", isea3h_lo = 1)
\end{Sinput}
\end{Schunk}

\noindent and eight temporal basis functions, one basis function for each two-day period, using

\begin{Schunk}
\begin{Sinput}
R> G_temporal <- local_basis(manifold = real_line(), loc = matrix(seq(1, 
+    16, by = 2)), scale = rep(4, 8), type = "bisquare")
\end{Sinput}
\end{Schunk}

\noindent Finally we used \code{TensorP} to obtain a set of 3,168 spatio-temporal basis functions.

Following pre-processing, we had about one million usable soundings per 16-day window. However, several of these are in quick succession and thus also in close proximity to each other, so that they fall within the same spatio-temporal BAU. We therefore keep the flag \code{average\_in\_BAU} set to \code{TRUE} when calling \code{SRE} as in Section \ref{sec:AIRS}. Following averaging, the number of observations per 16-day window reduces by a factor of about 100. We did not need to estimate the measurement-error variance $\sigma^2_\epsilon$ in this case, since measurement-error variances are provided with the OCO-2 data. However, we forced all measurement-error standard deviations that were below 2 ppm to be equal to 2 ppm, since the reported values are likely to be optimistic. The total time needed to fit and predict with \pkg{FRK} in a single 16-day window was about 1 hour on a standard desktop computer.

\begin{figure}[t!]
\begin{center}
\includegraphics[width=0.8\textwidth]{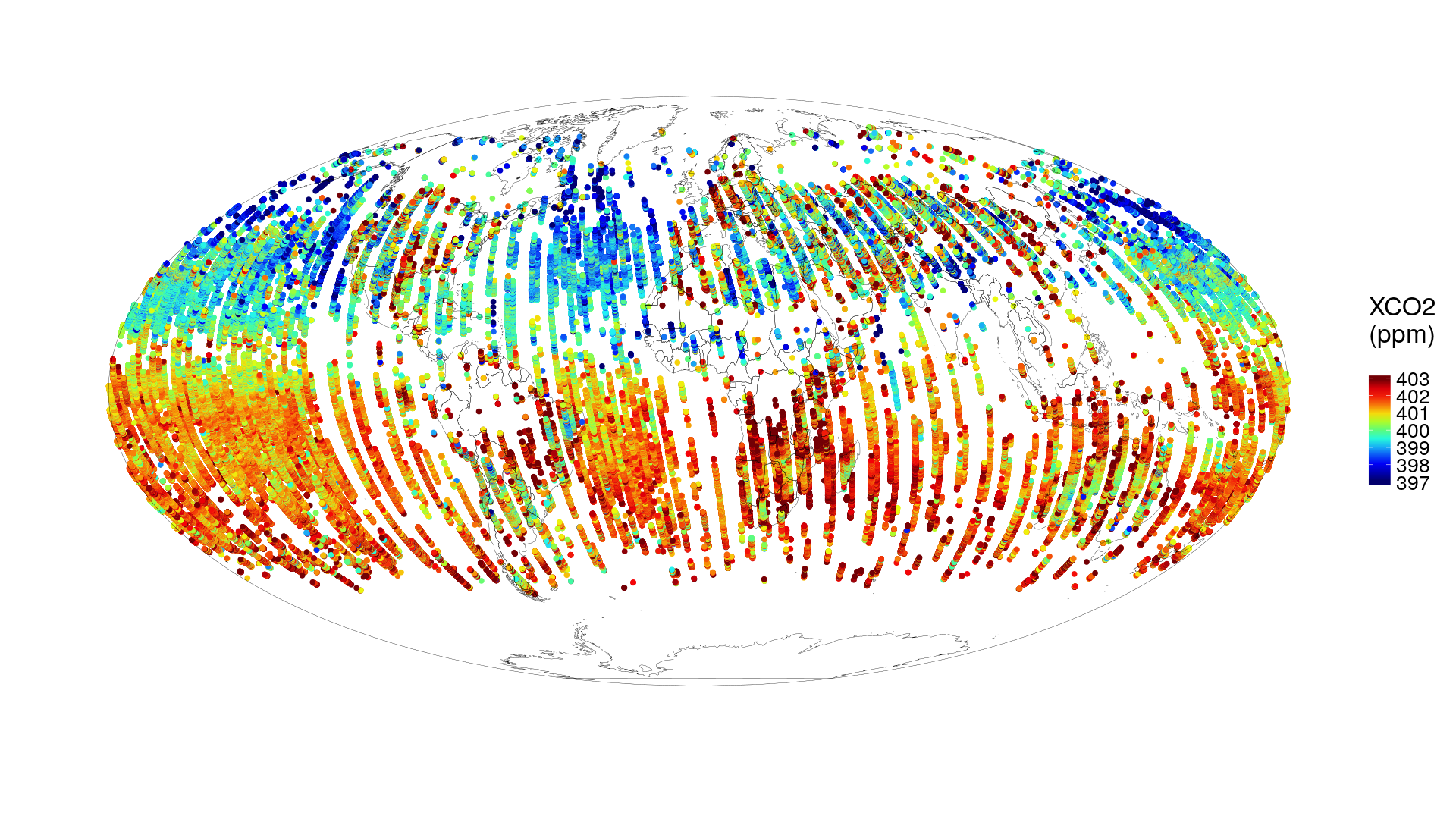}
\includegraphics[width=0.8\textwidth]{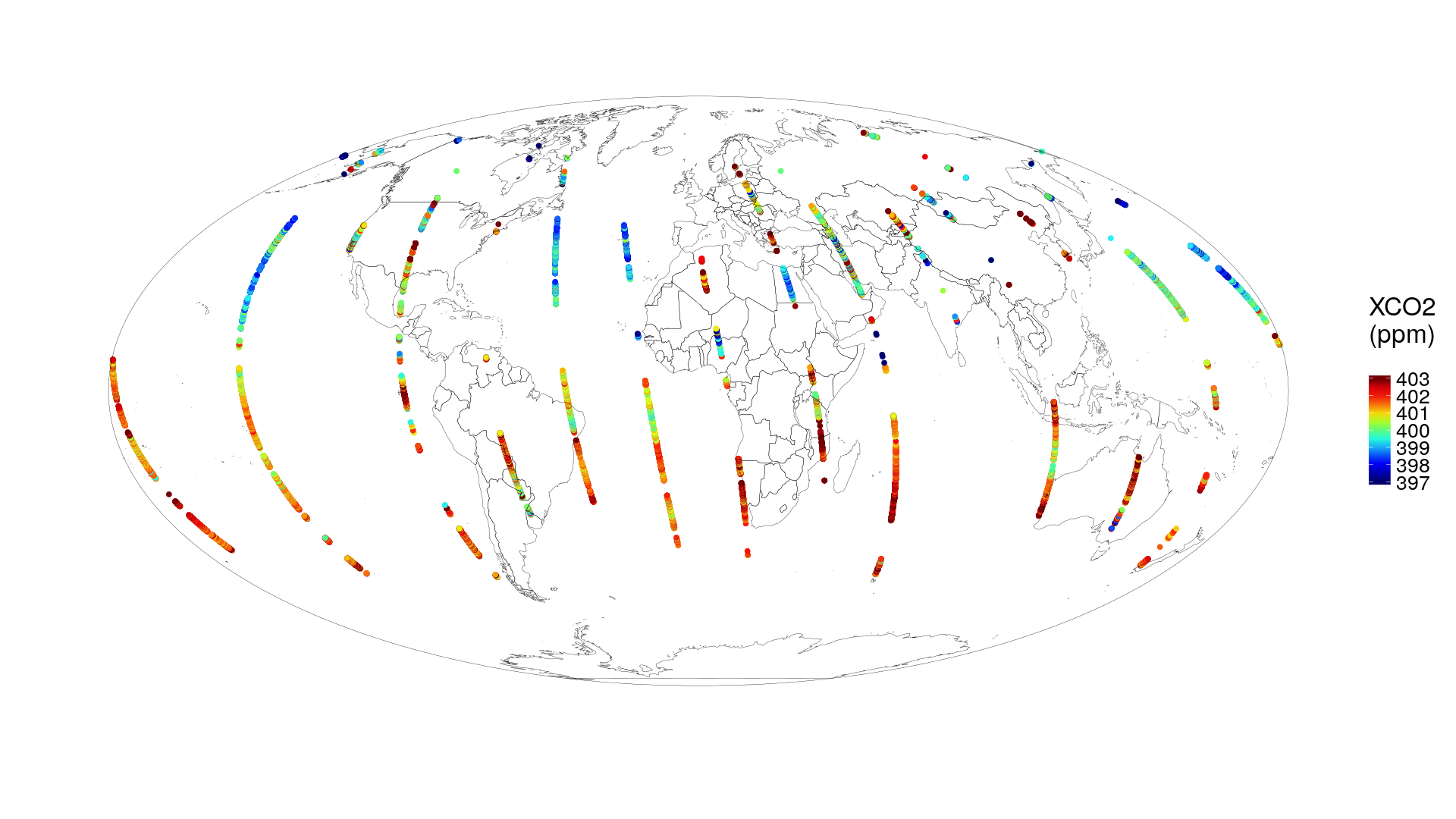}
\end{center}
\caption{OCO-2 measurements of column-averaged CO$_2$ (XCO$_2$) in ppm. (Top panel)  Data in the 16-day window from August 31 2016 to September 15 2016. (Bottom panel) Data from September 8 2016.\label{fig:OCO2data}}
\end{figure}

\begin{figure}[t!]
\begin{center}
\includegraphics[width=0.8\textwidth]{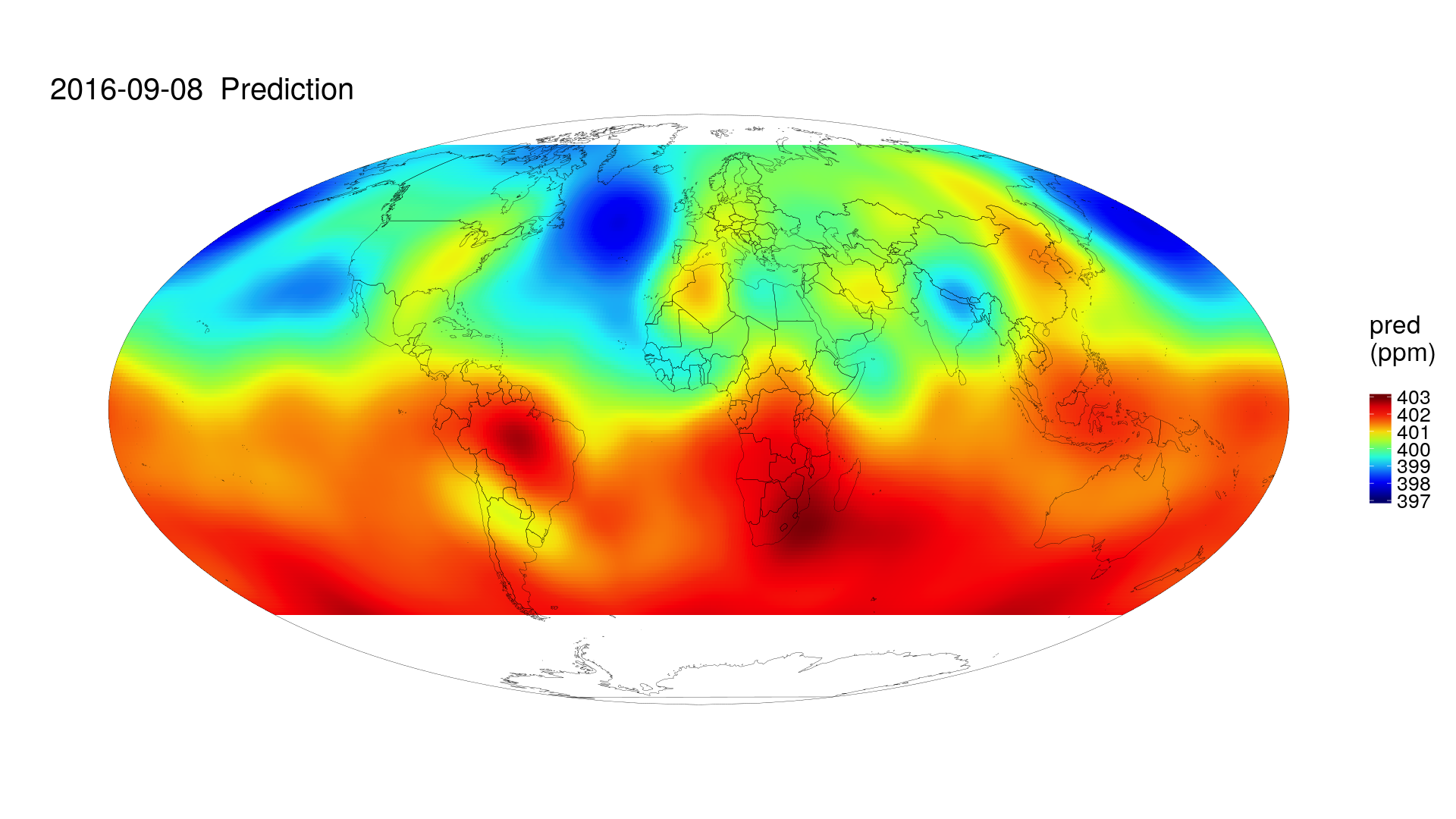}
\includegraphics[width=0.8\textwidth]{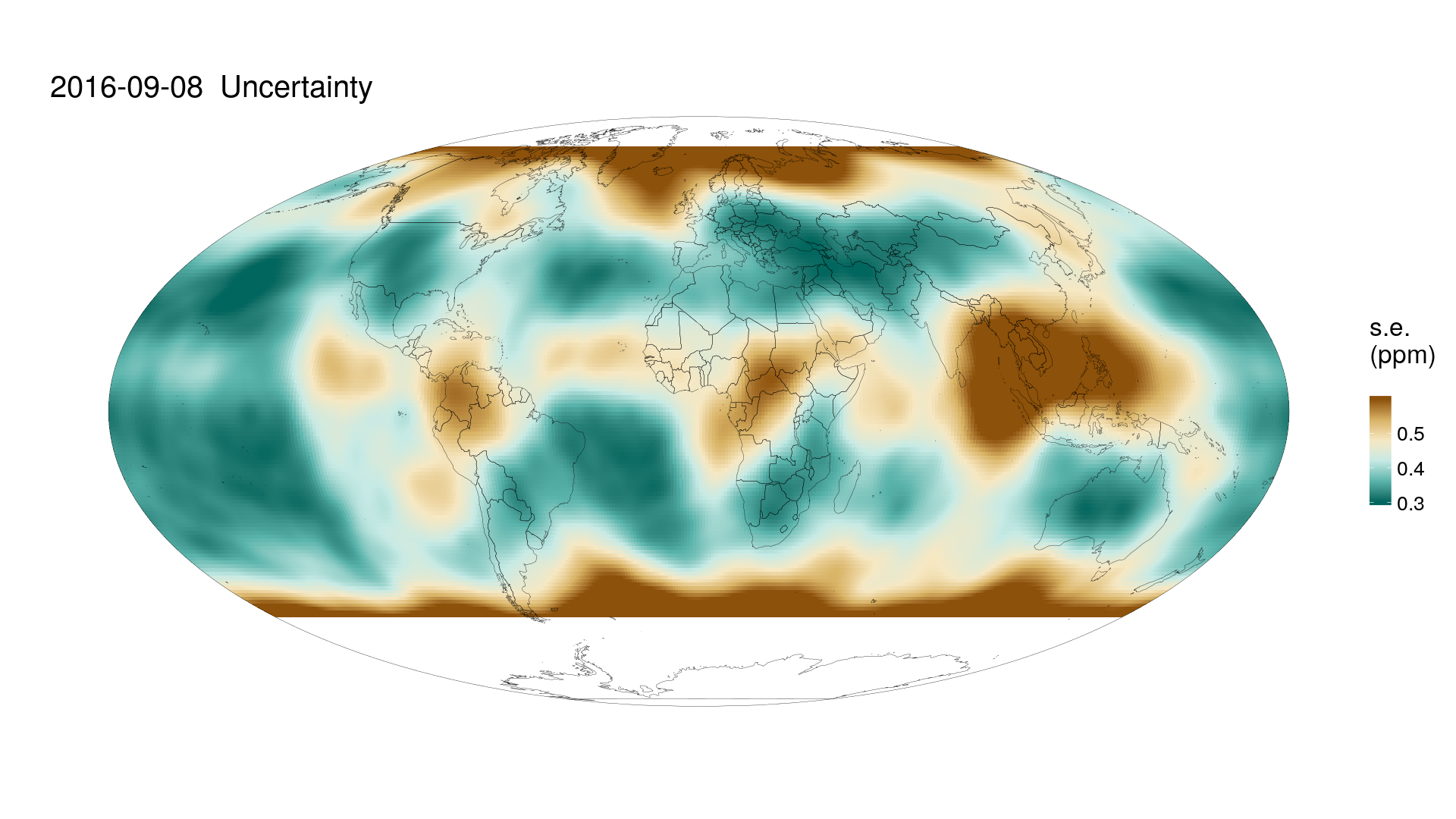}
\end{center}
\caption{Prediction and prediction standard errors of column-averaged CO$_2$ (XCO$_2$) on September 08 2016 obtained using FRK. (Top panel) Prediction of $\Yvec_P$ in ppm. (Bottom panel) Prediction standard errors in ppm. \label{fig:OCO2pred}}
\end{figure}

In Figure~\ref{fig:OCO2data} we show the measurement locations and values for the 16 days, and we show the central day of the 16-day window centred on September 08 2016. Predictions and prediction standard errors for the central day are shown in Figure~\ref{fig:OCO2pred}. Note how the error map reflects the data coverage over the entire 16-day window and not just the day at the centre of the window.  Prediction standard error maps on other days clearly show when the satellite is only taking readings over the ocean and when it is not taking any readings due to instrument reset or satellite manoeuvers. An animation showing these and other interesting features of predicted column-averaged CO$_2$ (i.e., XCO$_2$) between  October 01 2014 and March 01 2017 is available at \url{https://www.youtube.com/watch?v=wEws67WXvkY}.

\section{Future work} \label{sec:future}

There are a number of useful features that could be implemented in future versions of \pkg{FRK}, some of which are listed below:
\begin{itemize}

\item Currently, \pkg{FRK} is designed to work with local basis functions having an analytic form. However, the package could also accommodate basis functions that have no known functional form, such as empirical orthogonal functions (EOFs) and classes of wavelets defined iteratively; future work will attempt to incorporate the use of such basis functions. Vanilla FRK (FRK-V), where the entire positive-definite matrix $\Kmat$ is estimated, is particularly suited to the former (EOF) case where one has very few basis functions that explain a considerable amount of observed variability.

\item There is currently no component of the model that caters for sub-BAU process variation, and each datum that is point-referenced is mapped onto a BAU. Going below the BAU scale is possible, and intra-BAU correlation can be incorporated if the covariance function of the process at the sub-BAU scale is known \citep{Wikle_2005}.

\item Most work and testing in \pkg{FRK} has been done on the real line, the 2D plane, and the surface of the sphere ($\mathbb{S}^2$). Other manifolds can be implemented since the SRE model always yields a valid spatial covariance function, no matter the manifold. Some, such as the 3D hyperplane, are not too difficult to construct. Ultimately, it would be ideal if the package would allow the user to specify his/her own manifold, along with a function that can compute the appropriate distances on the manifold.

\item Currently, only the EM algorithm is implemented and hence the argument \code{method = "EM"} is implicit in the function \code{SRE.fit}. The EM algorithm has been seen to be slow to converge to a local maximum in other contexts \citep[e.g.,][]{McLachlan_2009}. Other methods for finding maximum, or restricted maximum, likelihood estimates for the SRE model \citep[e.g.,][]{Tzeng_2017} will be considered for future versions of \pkg{FRK}.

\item Although designed for very large data, \pkg{FRK} begins to slow down when several hundreds of thousands of data points are used. The flag \code{average\_in\_BAU} can be used to summarise the data and hence reduce the size of the dataset, however it is not always obvious how the data should be summarised (and whether one should summarise them in the first place). Future work will aim to provide the user with different options for summarising the data.

\item Currently, in \pkg{FRK}, all BAUs are assumed to be of equal area even if they are not. Unequal BAU area (see, for example, the lon-lat grid shown in Figure~\ref{fig:sphere_BAUs}) is important when aggregating the process or the predictions. In \pkg{FRK} there is the option to use equal-area icosahedral grids on the surface of the sphere, and regular grids on the real line and the plane for when areal data or large prediction regions are used. 
(Note that in Section \ref{sec:OCO2} an equal-area grid was not used, but also note that we did not spatially aggregate our results and that our data were point-referenced).

\end{itemize}

In conclusion, the package \pkg{FRK} is designed to provide core functionality for spatial and spatio-temporal prediction with large datasets. The low-rank model used by the package has validity (accurate coverage) in a big-data scenario when compared to high-rank models that do not explicitly cater for fine-scale variation.  However, it is likely to be less statistically efficient (larger root mean squared prediction errors) than other methods when data density is high and the basis functions are unable to capture the spatial variability.

\pkg{FRK} is available on the Comprehensive \proglang{R} Archive Network (CRAN). Its development page is \code{https://github.com/andrewzm/FRK}. Users are encouraged to report any bugs or issues relating to the package on this web page.

\section*{Acknowledgements}

Package development was facilitated with \pkg{devtools} \citep{Devtools}; this paper was compiled using \pkg{knitr} \citep{Xie_2015}; and package testing was carried out using \pkg{testthat} \citep{Wickham_2011} and \pkg{covr} \citep{Covr}. We would like to thank the reviewers for their feedback on the original submission and Esri staff with whom the classical factorial design of Section \ref{sec:spat_example} was discussed. We thank Jonathan Rougier for helpful comments on the manuscript; Doug Nychka and Finn Lindgren for advice on setting up \pkg{LatticeKrig} and \pkg{INLA}, respectively, to model data from a process with exponential covariance function; Clint Shumack for using \pkg{FRK} to analyse the OCO-2 data; and Enki Yoo and Behzad Kianian for providing valuable feedback on the package. AZM's research was partially supported by a Senior Research Fellowship from the University of Wollongong (2015--2017) and an Australian Research Council Discovery Early Career Research Award, DE180100203 (2018). NC's research was supported by an Australian Research Council Discovery Project, DP150104576.

\bibliography{FRK_bib}

\begin{thebibliography}{46}
\newcommand{\enquote}[1]{``#1''}
\providecommand{\natexlab}[1]{#1}
\providecommand{\url}[1]{\texttt{#1}}
\providecommand{\urlprefix}{URL }
\expandafter\ifx\csname urlstyle\endcsname\relax
  \providecommand{\doi}[1]{doi:\discretionary{}{}{}#1}\else
  \providecommand{\doi}{doi:\discretionary{}{}{}\begingroup
  \urlstyle{rm}\Url}\fi
\providecommand{\eprint}[2][]{\url{#2}}

\bibitem[{Banerjee \emph{et~al.}(2008)Banerjee, Gelfand, Finley, and
  Sang}]{Banerjee_2008}
Banerjee S, Gelfand A, Finley A, Sang H (2008).
\newblock \enquote{Gaussian Predictive Process Models for Large Spatial Data
  Sets.}
\newblock \emph{Journal of the Royal Statistical Society B}, \textbf{70}(4),
  825--848.

\bibitem[{Bates and Maechler(2015)}]{Matrix_2015}
Bates D, Maechler M (2015).
\newblock \emph{\pkg{Matrix}: Sparse and Dense Matrix Classes and Methods}.
\newblock \proglang{R}~package version~1.2-0,
  \urlprefix\url{http://CRAN.R-project.org/package=Matrix}.

\bibitem[{Bivand \emph{et~al.}(2013)Bivand, Pebesma, and
  Gomez-Rubio}]{Bivand_2013}
Bivand RS, Pebesma E, Gomez-Rubio V (2013).
\newblock \emph{{Applied Spatial Data Analysis with \proglang{R}}}.
\newblock 2nd edition. Springer-Verlag, New York, NY.

\bibitem[{Chahine \emph{et~al.}(2006)Chahine, Pagano, Aumann, Atlas, Barnet,
  Blaisdell, Chen, Divakarla, Fetzer, Goldberg, Gautier, Granger, Hannon,
  Irion, Kakar, Kalnay, Lambrigtsen, Lee, Marshall, McMillan, McMillin, Olsen,
  Revercomb, Rosenkranz, Smith, Staelin, Larrabee~Strow, Susskind, Tobin, Wolf,
  and Zhou}]{Chahine_2006}
Chahine M, Pagano T, Aumann H, Atlas R, Barnet C, Blaisdell J, Chen L,
  Divakarla M, Fetzer E, Goldberg M, Gautier C, Granger S, Hannon S, Irion F,
  Kakar R, Kalnay E, Lambrigtsen B, Lee SY, Marshall J, McMillan W, McMillin L,
  Olsen E, Revercomb H, Rosenkranz P, Smith W, Staelin D, Larrabee~Strow L,
  Susskind J, Tobin D, Wolf W, Zhou L (2006).
\newblock \enquote{AIRS: Improving Weather Forecasting and Providing New Data
  on Greenhouse Gases.}
\newblock \emph{Bulletin of the American Meteorological Society},
  \textbf{87}(7), 911--926.

\bibitem[{Cressie and Johannesson(2006)}]{Cressie_2006}
Cressie N, Johannesson G (2006).
\newblock \enquote{Spatial Prediction for Massive Data Sets.}
\newblock In \emph{Mastering the Data Explosion in the Earth and Environmental
  Sciences: Proceedings of the Australian Academy of Science Elizabeth and
  Frederick White Conference}, pp. 1--11. Australian Academy of Science,
  Canberra, Australia.

\bibitem[{Cressie and Johannesson(2008)}]{Cressie_2008}
Cressie N, Johannesson G (2008).
\newblock \enquote{Fixed Rank Kriging for Very Large Spatial Data Sets.}
\newblock \emph{Journal of the Royal Statistical Society B}, \textbf{70}(1),
  209--226.

\bibitem[{Cressie \emph{et~al.}(2010)Cressie, Shi, and Kang}]{Cressie_2010}
Cressie N, Shi T, Kang E (2010).
\newblock \enquote{Fixed Rank Filtering for Spatio-Temporal Data.}
\newblock \emph{Journal of Computational and Graphical Statistics},
  \textbf{19}(3), 724--745.

\bibitem[{Davis(2014)}]{Sparseinv}
Davis T (2014).
\newblock \emph{\pkg{sparseinv}: Sparse Inverse Subset}.
\newblock
  \urlprefix\url{https://au.mathworks.com/matlabcentral/fileexchange/33966-sparseinv--sparse-inverse-subset}.

\bibitem[{Finley \emph{et~al.}(2007)Finley, Banerjee, and Carlin}]{Finley_2007}
Finley A, Banerjee S, Carlin B (2007).
\newblock \enquote{\pkg{spBayes}: An \proglang{R} Package for Univariate and
  Multivariate Hierarchical Point-Referenced Spatial Models.}
\newblock \emph{Journal of Statistical Software}, \textbf{19}(4), 1--24.

\bibitem[{Fuglstad \emph{et~al.}(2018)Fuglstad, Simpson, Lindgren, and
  Rue}]{Fuglstad_2017}
Fuglstad GA, Simpson D, Lindgren F, Rue H (2018).
\newblock \enquote{Constructing Priors that Penalize the Complexity of
  {G}aussian Random Fields.}
\newblock \emph{Journal of the American Statistical Association}, \textbf{in
  press}.

\bibitem[{Gneiting \emph{et~al.}(2007)Gneiting, Balabdaoui, and
  Raftery}]{Gneiting_2007}
Gneiting T, Balabdaoui F, Raftery AE (2007).
\newblock \enquote{Probabilistic Forecasts, Calibration and Sharpness.}
\newblock \emph{Journal of the Royal Statistical Society B}, \textbf{69}(2),
  243--268.

\bibitem[{Gneiting and Raftery(2007)}]{Gneiting_2007b}
Gneiting T, Raftery AE (2007).
\newblock \enquote{Strictly Proper Scoring Rules, Prediction, and Estimation.}
\newblock \emph{Journal of the American Statistical Association},
  \textbf{102}(477), 359--378.

\bibitem[{Henderson and Searle(1981)}]{Henderson_1981}
Henderson H, Searle S (1981).
\newblock \enquote{On Deriving the Inverse of a Sum of Matrices.}
\newblock \emph{{SIAM} Review}, \textbf{23}(1), 53--60.

\bibitem[{Hester(2017)}]{Covr}
Hester J (2017).
\newblock \emph{\pkg{covr}: Test Coverage for Packages}.
\newblock \proglang{R}~package version~2.2.2,
  \urlprefix\url{https://CRAN.R-project.org/package=covr}.

\bibitem[{Kang and Cressie(2011)}]{Kang_2011}
Kang E, Cressie N (2011).
\newblock \enquote{Bayesian Inference for the Spatial Random Effects Model.}
\newblock \emph{Journal of the American Statistical Association},
  \textbf{106}(495), 972--983.

\bibitem[{Kang \emph{et~al.}(2009)Kang, Liu, and Cressie}]{Kang_2009}
Kang E, Liu D, Cressie N (2009).
\newblock \enquote{Statistical Analysis of Small-Area Data Based on
  Independence, Spatial, Non-Hierarchical, and Hierarchical Models.}
\newblock \emph{Computational Statistics \& Data Analysis}, \textbf{53}(8),
  3016--3032.

\bibitem[{Katzfuss and Cressie(2011)}]{Katzfuss_2011}
Katzfuss M, Cressie N (2011).
\newblock \enquote{Spatio-Temporal Smoothing and {EM} Estimation for Massive
  Remote-Sensing Data Sets.}
\newblock \emph{Journal of Time Series Analysis}, \textbf{32}(4), 430--446.

\bibitem[{Katzfuss and Hammerling(2017)}]{Katzfuss_2014}
Katzfuss M, Hammerling D (2017).
\newblock \enquote{Parallel Inference for Massive Distributed Spatial Data
  Using Low-Rank Models.}
\newblock \emph{Statistics and Computing}, \textbf{27}(2), 363--375.

\bibitem[{Kleiber(2016)}]{Kleiber_2016}
Kleiber W (2016).
\newblock \enquote{High Resolution Simulation of Nonstationary {G}aussian
  Random Fields.}
\newblock \emph{Computational Statistics and Data Analysis}, \textbf{101},
  277--288.

\bibitem[{Lindgren and Rue(2015)}]{Lindgren_2015}
Lindgren F, Rue H (2015).
\newblock \enquote{Bayesian Spatial Modelling with \proglang{R}-\pkg{INLA}.}
\newblock \emph{Journal of Statistical Software}, \textbf{63}(19), 1--25.

\bibitem[{McLachlan and Krishnan(2007)}]{McLachlan_2009}
McLachlan G, Krishnan T (2007).
\newblock \emph{The {EM} Algorithm and Extensions}.
\newblock John Wiley \& Sons, Hoboken, NJ.

\bibitem[{Nguyen \emph{et~al.}(2012)Nguyen, Cressie, and
  Braverman}]{Nguyen_2012}
Nguyen H, Cressie N, Braverman A (2012).
\newblock \enquote{Spatial Statistical Data Fusion for Remote Sensing
  Applications.}
\newblock \emph{Journal of the American Statistical Association},
  \textbf{107}(499), 1004--1018.

\bibitem[{Nguyen \emph{et~al.}(2014)Nguyen, Katzfuss, Cressie, and
  Braverman}]{Nguyen_2014}
Nguyen H, Katzfuss M, Cressie N, Braverman A (2014).
\newblock \enquote{Spatio-Temporal Data Fusion for Very Large Remote Sensing
  Datasets.}
\newblock \emph{Technometrics}, \textbf{56}(2), 174--185.

\bibitem[{Nychka \emph{et~al.}(2015)Nychka, Bandyopadhyay, Hammerling,
  Lindgren, and Sain}]{Nychka_2015}
Nychka D, Bandyopadhyay S, Hammerling D, Lindgren F, Sain S (2015).
\newblock \enquote{A Multiresolution {G}aussian Process Model for the Analysis
  of Large Spatial Datasets.}
\newblock \emph{Journal of Computational and Graphical Statistics},
  \textbf{24}(2), 579--599.

\bibitem[{Nychka \emph{et~al.}(2016)Nychka, Hammerling, Sain, and
  Lenssen}]{LatticeKrig}
Nychka D, Hammerling D, Sain S, Lenssen N (2016).
\newblock \emph{\pkg{LatticeKrig}: Multiresolution Kriging Based on Markov
  Random Fields}.
\newblock \proglang{R}~package version~6.2,
  \urlprefix\url{www.image.ucar.edu/LatticeKrig}.

\bibitem[{{OCO-2 Science Team} \emph{et~al.}(2015){OCO-2 Science Team}, Gunson,
  and Eldering}]{OCO2_2017}
{OCO-2 Science Team}, Gunson M, Eldering A (2015).
\newblock \emph{{OCO-2 Level 2 Bias-Corrected XCO2 and Other Select Fields from
  the Full-Physics Retrieval Aggregated as Daily Files, Retrospective
  Processing V7r}}.
\newblock Greenbelt, MD, USA, Goddard Earth Sciences Data and Information
  Services Center (GES DISC),
  \urlprefix\url{https://disc.gsfc.nasa.gov/datacollection/OCO2_L2_Lite_FP_7r.html}.

\bibitem[{Pebesma(2012)}]{Pebesma_2012}
Pebesma E (2012).
\newblock \enquote{\pkg{spacetime}: {S}patio-temporal Data in \proglang{R}.}
\newblock \emph{Journal of Statistical Software}, \textbf{51}(7), 1--30.

\bibitem[{Pebesma(2004)}]{Pebesma_2004}
Pebesma EJ (2004).
\newblock \enquote{Multivariable Geostatistics in \proglang{S}: The \pkg{gstat}
  Package.}
\newblock \emph{Computers \& Geosciences}, \textbf{30}(7), 683--691.

\bibitem[{Rasmussen and Williams(2006)}]{Rasmussen_2006}
Rasmussen CE, Williams CKI (2006).
\newblock \emph{{Gaussian Processes for Machine Learning}}.
\newblock The MIT Press, Cambridge, MA.

\bibitem[{{\proglang{R} Core Team}(2017)}]{R}
{\proglang{R} Core Team} (2017).
\newblock \emph{\proglang{R}: A Language and Environment for Statistical
  Computing}.
\newblock \proglang{R} Foundation for Statistical Computing, Vienna, Austria.
\newblock \urlprefix\url{https://www.R-project.org/}.

\bibitem[{Sampson and Guttorp(1992)}]{Sampson_1992}
Sampson PD, Guttorp P (1992).
\newblock \enquote{Nonparametric Estimation of Nonstationary Spatial Covariance
  Structure.}
\newblock \emph{Journal of the American Statistical Association},
  \textbf{87}(417), 108--119.

\bibitem[{Schlather \emph{et~al.}(2015)Schlather, Malinowski, Menck, Oesting,
  and Strokorb}]{Schlather_2015}
Schlather M, Malinowski A, Menck PJ, Oesting M, Strokorb K (2015).
\newblock \enquote{Analysis, Simulation and Prediction of Multivariate Random
  Fields with Package \pkg{RandomFields}.}
\newblock \emph{Journal of Statistical Software}, \textbf{63}(8), 1--25.

\bibitem[{Shi and Cressie(2007)}]{Shi_2007}
Shi T, Cressie N (2007).
\newblock \enquote{Global Statistical Analysis of {MISR} Aerosol Data: A
  Massive Data Product from {NASA's} Terra Satellite.}
\newblock \emph{Environmetrics}, \textbf{18}(7), 665--680.

\bibitem[{Stein(2008)}]{Stein_2008}
Stein M (2008).
\newblock \enquote{A Modeling Approach for Large Spatial Datasets.}
\newblock \emph{Journal of the Korean Statistical Society}, \textbf{37}(1),
  3--10.

\bibitem[{Tzeng and Huang(2018)}]{Tzeng_2017}
Tzeng S, Huang HC (2018).
\newblock \enquote{Resolution Adaptive Fixed Rank Kriging.}
\newblock \emph{Technometrics}, \textbf{60}(2), 198--208.

\bibitem[{Wang \emph{et~al.}(2013)Wang, Zhang, Zhang, and Yi}]{Wang_2013}
Wang Q, Zhang X, Zhang Y, Yi Q (2013).
\newblock \enquote{{AUGEM: {A}utomatically Generate High Performance Dense
  Linear Algebra Kernels on x86 CPUs}.}
\newblock In \emph{Proceedings of the International Conference on High
  Performance Computing, Networking, Storage and Analysis}, pp. 1--12. Denver,
  CO.

\bibitem[{Wickham(2011)}]{Wickham_2011}
Wickham H (2011).
\newblock \enquote{\pkg{testthat}: Get Started with Testing.}
\newblock \emph{The \proglang{R} Journal}, \textbf{3}(1), 5--10.
\newblock
  \urlprefix\url{http://journal.r-project.org/archive/2011-1/RJournal_2011-1_Wickham.pdf}.

\bibitem[{Wickham and Chang(2016)}]{Devtools}
Wickham H, Chang W (2016).
\newblock \emph{\pkg{devtools}: Tools to Make Developing \proglang{R} Packages
  Easier}.
\newblock \proglang{R}~package version~1.11.1,
  \urlprefix\url{https://CRAN.R-project.org/package=devtools}.

\bibitem[{Wikle and Berliner(2005)}]{Wikle_2005}
Wikle CK, Berliner LM (2005).
\newblock \enquote{Combining Information Across Spatial Scales.}
\newblock \emph{Technometrics}, \textbf{47}(1), 80--91.

\bibitem[{Xie(2015)}]{Xie_2015}
Xie Y (2015).
\newblock \emph{Dynamic {D}ocuments with \proglang{R} and \pkg{knitr}}.
\newblock 2nd edition. Chapman and Hall/CRC Press, Boca Raton, FL.

\bibitem[{Zammit-Mangion(2018{\natexlab{a}})}]{FRK}
Zammit-Mangion A (2018{\natexlab{a}}).
\newblock \emph{\pkg{FRK}: Fixed Rank Kriging}.
\newblock \proglang{R}~package version~0.2.1,
  \urlprefix\url{https://CRAN.R-project.org/package=FRK}.

\bibitem[{Zammit-Mangion(2018{\natexlab{b}})}]{sparseinv_2018}
Zammit-Mangion A (2018{\natexlab{b}}).
\newblock \emph{sparseinv: Computation of the Sparse Inverse Subset}.
\newblock \proglang{R}~package version 0.1.1,
  \urlprefix\url{https://CRAN.R-project.org/package=sparseinv}.

\bibitem[{Zammit-Mangion \emph{et~al.}(2018)Zammit-Mangion, Cressie, and
  Shumack}]{Zammit_2018}
Zammit-Mangion A, Cressie N, Shumack C (2018).
\newblock \enquote{On Statistical Approaches to Generate Level 3 Products from
  Satellite Remote Sensing Retrievals.}
\newblock \emph{Remote Sensing}, \textbf{10}(1), 155.

\bibitem[{Zammit-Mangion \emph{et~al.}(2015)Zammit-Mangion, Rougier, Sch{\"o}n,
  Lindgren, and Bamber}]{Zammit_2015}
Zammit-Mangion A, Rougier J, Sch{\"o}n N, Lindgren F, Bamber J (2015).
\newblock \enquote{Multivariate Spatio-Temporal Modelling for Assessing
  {A}ntarctica's Present-Day Contribution to Sea-Level Rise.}
\newblock \emph{Environmetrics}, \textbf{26}(3), 159--177.

\bibitem[{Zammit-Mangion \emph{et~al.}(2012)Zammit-Mangion, Sanguinetti, and
  Kadirkamanathan}]{Zammit_2012}
Zammit-Mangion A, Sanguinetti G, Kadirkamanathan V (2012).
\newblock \enquote{Variational Estimation in Spatiotemporal Systems from
  Continuous and Point-Process Observations.}
\newblock \emph{IEEE Transactions on Signal Processing}, \textbf{60}(7),
  3449--3459.

\bibitem[{Zhuang and Cressie(2014)}]{Zhuang_2014}
Zhuang L, Cressie N (2014).
\newblock \enquote{Bayesian Hierarchical Statistical {SIRS} Models.}
\newblock \emph{Statistical Methods \& Applications}, \textbf{23}(4), 601--646.

\end{thebibliography}

\end{document}